\documentclass[prd,twocolumn,amsmath,amssymb,floatfix,superscriptaddress,nofootinbib,preprintnumbers]{revtex4-1}

\usepackage{graphicx}
\usepackage{amssymb}
\usepackage{amsmath}
\usepackage{bm}
\usepackage{color}
\usepackage[utf8]{inputenc}
\usepackage{comment}
\usepackage{hhline}

\DeclareMathAlphabet\mathbfcal{OMS}{cmsy}{b}{n}
\definecolor{darkgreen}{RGB}{50,150,0}
\definecolor{purple}{cmyk}{0.5,0.75,0,0}

\newcommand{\refeq}[1]{Eq.~(\ref{eq:#1})}
\newcommand{\refeqs}[2]{Eqs.~(\ref{eq:#1})--(\ref{eq:#2})}
\newcommand{\refEq}[1]{Eq.~(\ref{eq:#1})}

\newcommand{\reffig}[1]{Fig.~\ref{fig:#1}}
\newcommand{\refFig}[1]{Fig.~\ref{fig:#1}}
\newcommand{\reffigs}[2]{Figs.~\ref{fig:#1}-\ref{fig:#2}}
\newcommand{\reftab}[1]{Tab.~\ref{tab:#1}}
\newcommand{\refTab}[1]{Tab.~\ref{tab:#1}}
\newcommand{\refsec}[1]{Sec.~\ref{sec:#1}}
\newcommand{\refapp}[1]{App.~\ref{app:#1}}

\def\be{\begin{equation}}
\def\ee{\end{equation}}
\def\ba#1\ea{\begin{align}#1\end{align}}
\newcommand{\vs}{\nonumber\\}

\renewcommand{\[}{\left[}
\renewcommand{\]}{\right]}
\renewcommand{\(}{\left(}
\renewcommand{\)}{\right)}

\newcommand{\Mpc}{~{\rm Mpc}}
\newcommand{\iMpc}{~{\rm Mpc}^{-1}}
\newcommand{\eV}{~{\rm eV}}

\newcommand\lsim{\mathrel{\rlap{\lower4pt\hbox{\hskip1pt$\sim$}}
        \raise1pt\hbox{$<$}}}
\newcommand\gsim{\mathrel{\rlap{\lower4pt\hbox{\hskip1pt$\sim$}}
        \raise1pt\hbox{$>$}}}
        
\newcommand{\Comment}[1]{{}}

\definecolor{ultramarine}{rgb}{0.07, 0.04, 0.56}
\definecolor{cadmiumgreen}{rgb}{0.0, 0.42, 0.24}
\definecolor{indigo(dye)}{rgb}{0.0, 0.25, 0.42}
\usepackage[linktocpage=true]{hyperref}
\hypersetup{
colorlinks=true,
citecolor=ultramarine,
linkcolor=cadmiumgreen,
urlcolor=indigo(dye),
pdfauthor={},
pdftitle={},
pdfsubject={}
}

\begin{document}

\preprint{YITP-SB-17-40}

\title{Scale-dependent bias and bispectrum in neutrino separate universe simulations}

\author{Chi-Ting Chiang}

\affiliation{C.N. Yang Institute for Theoretical Physics, Department of Physics \& Astronomy,
Stony Brook University, Stony Brook, NY 11794}

\author{Wayne Hu}

\affiliation{Kavli Institute for Cosmological Physics, Department of Astronomy \& Astrophysics,  Enrico Fermi Institute, University of Chicago, Chicago, IL 60637}

\author{Yin Li}

\affiliation{Berkeley Center for Cosmological Physics, Department of Physics
and Lawrence Berkeley National Laboratory, University of California, Berkeley, CA 94720}
\affiliation{Kavli Institute for the Physics and Mathematics of the Universe (WPI),
UTIAS, The University of Tokyo, Chiba 277-8583, Japan}

\author{Marilena LoVerde}

\affiliation{C.N. Yang Institute for Theoretical Physics, Department of Physics \& Astronomy,
Stony Brook University, Stony Brook, NY 11794}

\begin{abstract}
Cosmic background neutrinos have a large velocity dispersion, which causes the
evolution of long-wavelength density perturbations to depend on scale. This
scale-dependent growth leads to the well-known suppression in the linear theory
matter power spectrum that is used to probe neutrino mass. In this paper, we
study the impact of long-wavelength density perturbations on small-scale structure
formation. By performing separate universe simulations where the long-wavelength
mode is absorbed into the local expansion, we measure the responses of the cold
dark matter (CDM) power spectrum and halo mass function, which correspond to the
squeezed-limit bispectrum and halo bias. We find that the scale-dependent evolution
of the long-wavelength modes causes these quantities to depend on scale and provide
simple expressions to model them in terms of scale and the amount of massive neutrinos.
Importantly, this scale-dependent bias reduces the suppression in the linear halo
power spectrum due to massive neutrinos by 13 and 26\% for objects of bias $\bar{b}=2$
and $\bar{b} \gg1$, respectively. We demonstrate with high statistical significance
that the scale-dependent halo bias \emph{cannot} be modeled by the CDM and neutrino
density transfer functions at the time when the halos are identified. This reinforces
the importance of the temporal nonlocality of structure formation, especially when
the growth is scale dependent.
\end{abstract}

\maketitle

\section{Introduction}
\label{sec:intro}
Neutrinos are one of the most abundant particles in the universe, but are
the least explored species in the Standard Model of particle physics. The
solar \cite{Ahmad:2002jz,Fukuda:2002pe,Altmann:2005ix,Abdurashitov:2002nt}
and atmospheric oscillation experiments \cite{Fukuda:1998mi,Ashie:2005ik,Sanchez:2003rb}
have revealed the two mass-squared differences, but the individual masses
and their hierarchy are yet to be determined. Cosmological observables are primarily sensitive
to the sum of neutrino masses hence offer a complementary probe of neutrinos.
Thus, measuring the total neutrino mass is one of the most
important goals of future CMB \cite{Henderson:2015nzj,Benson:2014qhw,Abazajian:2016yjj}
and large-scale structure experiments
\cite{Aghamousa:2016zmz,Ellis:2012rn,Abate:2012za,Spergel:2015sza,Laureijs:2011gra}.

The effect of massive neutrinos on cosmology is extensively studied (see e.g.
Refs.~\cite{Lesgourgues:2006nd,Wong:2011ip} for reviews). At the background
level, the expansion history changes due to the presence of neutrinos, and
so the effect can, in principle, be probed by measuring the Hubble rate at
various redshifts. At the growth level, due to the high velocity dispersion,
massive neutrinos possess a free-streaming scale that acts as a Jeans scale
below which they no longer cluster with the cold dark matter (CDM). As a
result, the density perturbations become scale dependent, and the feature
is imprinted on the matter power spectrum \cite{Hu:1997vi,Eisenstein:1997jh}
as well as the halo bias in the spherical collapse model \cite{LoVerde:2014pxa}.
While the effect at the linear order is well understood, the nonlinear nature
of large-scale structure requires treatment beyond the leading order effect.
In the mildly nonlinear regime one can tackle the problem using perturbative approaches
\cite{Shoji:2010hm,Blas:2014hya,Fuhrer:2014zka,Dupuy:2015ega,Archidiacono:2015ota,Levi:2016tlf,Inman:2016qmg},
but $N$-body simulations are still necessary to capture the fully nonlinear
behavior, especially for accurately modeling the unprecedentedly precise
data from future observations. Moreover, even in the linear regime, the
biased tracers of large-scale structure such as galaxies and dark matter
halos are themselves nonlinear objects.

The most straightforward way to include massive neutrinos in $N$-body simulations
is to treat them as a different species of particles that have different mass
than CDM particles \cite{Bird:2011rb,Hannestad:2011td,Wagner:2012sw,VillaescusaNavarro:2012ag,Villaescusa-Navarro:2013pva,Castorina:2013wga,Costanzi:2013bha,Castorina:2015bma}.
Due to the large thermal motion, however, massive neutrinos occupy the
six-dimensional initial phase space (unlike CDM which occupy an effectively
three dimensional initial subspace), one either needs many more particles
for neutrinos than CDM to reduce the effect of Poisson shot noise or has
to start simulations at later times and approximate the neutrinos as cold
dark matter. There are many approaches put forward to bypass the difficulty.
Refs.~\cite{Agarwal:2010mt,Heitmann:2015xma} include the effect of massive
neutrinos in the background evolution and initial conditions, but simulate
only the dynamics of CDM+baryons. In Refs.~\cite{Inman:2015pfa,Yu:2016yfe,Emberson:2016ecv},
neutrinos are treated as a distinct species and injected into CDM simulations
at later redshift.  Hybrid approaches solve the coupled neutrino linear fluid
equation \cite{Brandbyge:2009ce} or Boltzmann equation \cite{AliHaimoud:2012vj}
with the nonlinear CDM evolution in $N$-body simulations. Recently, Ref.~\cite{Banerjee:2016zaa}
combined the particle and fluid descriptions to better estimate the properties
of the thermal species and reduce the effect of shot noise.

In this paper, we present a complementary approach to modeling neutrinos
in simulations, the separate universe (SU) approach. In the SU approach,
a long-wavelength perturbation changes the expansion history locally, and
the local observer would measure a different set of cosmological parameters
compared to the background universe that does not have any long-wavelength
perturbation. As a result, the small-scale structure formation in the SU would
be influenced, or respond, accordingly. Applying this technique in $N$-body
simulations, the response can be measured deep into the nonlinear regime
where perturbation theory breaks down \cite{McDonald:2001fe,Sirko:2005uz,Gnedin:2011kj,Li:2014sga,Wagner:2014aka}.
Specifically, the SU simulations have enabled studies on the power spectrum
covariance \cite{Li:2014sga,Barreira:2017kxd}, the squeezed-limit $n$-point function
\cite{Wagner:2015gva,Barreira:2017sqa}, the halo bias \cite{Li:2015jsz,Lazeyras:2015lgp,Baldauf:2015vio},
and the Lyman-$\alpha$ forest \cite{McDonald:2001fe,Cieplak:2015kra,Chiang:2017vsq,Chiang:2017qoh}.
The only limitations for SU simulations in $\Lambda$CDM are the usual ones
for any simulation: the resolution and the extent to which baryonic and
astrophysical effects are modeled in the deeply nonlinear regime.

In the universe with CDM and massive neutrinos, the growth of density
perturbations is scale dependent due to the free-streaming length of
massive neutrinos. Thus, for SUs with scale-dependent density perturbations,
the expansion histories are affected differently depending on the
wavelengths, and so is the response of the small-scale structure
formation. While it is nontrivial to find the corresponding densities
and curvature of the Friedmann equation in the SUs with additional components
that have Jeans scales \cite{Dai:2015jaa,Hu:2016ssz,Hu:2016wfa}, it is
easy to match the local Hubble expansion. Ref.~\cite{Chiang:2016vxa}
uses quintessence SU simulations to show that this approach provides
excellent agreement between the power spectrum response and the position-dependent
power spectrum (which is equivalent to the squeezed-limit bispectrum
\cite{Chiang:2014oga,Chiang:2015eza}) as well as the response and
clustering biases in the sub-Jeans limit as long as the Jeans scale
is much larger than the scales of interest. The goal of this paper
is to generalize SU simulations for massive neutrinos to study the
scale-dependent responses for long modes with different wavelengths
where the free-streaming scale plays the role of the Jeans scale.

The rest of the paper is organized as follows. In \refsec{theory} we construct
SUs with massive neutrinos for different long-wavelength density perturbations
and compute the responses of the linear growth. In \refsec{sims} we implement
the SU approach with massive neutrinos in $N$-body simulations. We present the
results of the neutrino SU simulations in \refsec{pk_resp} and \refsec{nh_resp}
for power spectrum response and response bias, respectively. In \refsec{observables}
we show how the linear halo power spectrum and the leading-order CDM squeezed-limit
bispectrum are changed by the scale-dependent response. We discuss the results
in \refsec{discussion}.
In \refapp{setup} we demonstrate the choices of initial and horizon entry
redshifts have minimum effects on solving the small-scale growth response.
In \refapp{2lpt} we derive the small-scale growths using the second-order
Lagrangian perturbation theory with long-wavelength density perturbations
in matter-radiation dominated universe.
In \refapp{caveat} we discuss the two main caveats of neutrino SU simulations,
that neutrino clustering is neglected within the simulations and that a separation
of scales is assumed between the SU observables and the long-wavelength mode.
In \refapp{sphericalcoll} we layout the detailed setup for numerically
evaluating the spherical collapse in neutrino SU, which is compared
against simulations in \refsec{nh_resp}.
In \refapp{clusteringbias}, we compare our predicted scale-dependent
linear bias to $N$-body simulations that contain massive neutrino particles.

Throughout the paper, unless otherwise stated, we adopt a spatially flat $\nu\Lambda{\rm CDM}$ cosmology
with a Hubble constant $h=0.7$, baryon density $\Omega_b=0.05$, CDM density
$\Omega_c=0.25$, the CMB temperature $T_{\rm cmb}=2.725$ K, helium
fraction $Y_{\rm He}=0.24$, and initial curvature power spectrum with
the spectral index $n_s=0.95$ and amplitude which sets $\sigma_8=0.83$
today for the power spectrum of CDM+baryons. We assume a degenerate neutrino
mass spectrum with each neutrino having $m_\nu=0.05\eV$. This choice
will produce a free-streaming scale consistent with that in the minimal
mass normal and inverted hierarchies. For this scenario, the free-streaming
scale is also still in the linear regime and the neutrino nonlinear clustering
can be neglected (see \refapp{caveat} for detailed discussion). Depending
on the number of massive neutrinos, $\Omega_\Lambda$ and the power spectrum
normalization would change accordingly. We choose to fix $\Omega_{bc}=\Omega_b+\Omega_c$
since the particle mass of the simulation is given by $\Omega_{bc}$ (see
\refsec{sims} for more details), and to fix $\sigma_8$ since the nonlinear scale
to the leading order is set by $\sigma_8$. In order to enhance the amplitude
of the neutrinos effects so that they can be measured with a small set of
simulations we shall study two cosmologies with $N_\nu=14$ and 28 massive
neutrinos, which can be converted to $f_\nu$ by
\be
 f_\nu=\frac{\Omega_\nu}{\Omega_{bc}+\Omega_\nu}
 =\frac{N_\nu\frac{m_\nu}{93\eV}}{\Omega_{bc}h^2+N_\nu\frac{m_\nu}{93\eV}} \,,
\label{eq:fnuNnu}
\ee
and the corresponding values are 0.049 and 0.093, respectively. Finally,
given our use of $N$-body techniques, in the following we will often refer to CDM+baryons as CDM.

\section{Neutrino separate universe}
\label{sec:theory}
The construction of the SU with components other than CDM that possess
Jeans scales has been studied extensively in Ref.~\cite{Hu:2016ssz}. Here we briefly
summarize the expansion history of the SU in \refsec{expansion}, and focus on
the discussion of the small-scale growth in the neutrino SU in \refsec{growth}.

\subsection{Expansion History}
\label{sec:expansion}
In the separate universe (SU) picture, an observer sitting in a long-wavelength
density perturbation $\delta_c$ would measure the \emph{local}
mean density $\bar{\rho}_{cW}(a)$ as
\be
 \bar{\rho}_{cW}(a)=\bar{\rho}_c(a)\[1+\delta_c(a)\] \,,
\ee
where $\bar{\rho}_c$ is the \emph{global} mean density, and the subscript
$W$ denotes the windowed average across the scale much smaller than that
of $\delta_c$. Note that the subscript $c$ denotes CDM+baryons under the assumption
that baryons trace the CDM at large scales.  While the SU picture does not require this assumption to be valid
on small scales, in our $N$-body simulations we do combine them into a single CDM-like component.   We therefore
refer to this component as CDM in the following for simplicity.
Since the total amount of CDM is conserved, this observer
would find the local scale factor of the SU as
\be
 a_W=a\[1+\delta_c(a)\]^{-1/3}\approx a\[1-\frac{1}{3}\delta_c(a)\] \,,
\label{eq:aW}
\ee
which leads to the local Hubble expansion of the SU
\be
 H_W=\frac{\dot{a}_W}{a_W}=H-\frac{1}{3}\dot{\delta}_c=H\(1-\frac{1}{3}\delta'_c\) \,,
\label{eq:HW}
\ee
with $'\equiv d/d\ln a$. At early times
\be
 \lim_{a\to 0}\delta_c(a)\to 0 \,, \quad
 \lim_{a\to 0}a_W\to a \,, \quad
 \lim_{a\to 0}H_W\to H \,,
\ee
and the physical conditions in local and global cosmology coincide. Notice
that we implicitly assume that there is a universal time coordinate between
the local SU and global universe, hence in the relativistic limit $\delta_c$
is the synchronous-gauge density perturbation \cite{Hu:2016ssz}.

As we can see above, the construction of the SU only requires $\delta_c(a)$.
While other components affect the evolution of $\delta_c(a)$, they do not enter explicitly
into $a_W$. If these components influence the small-scale structure formation
only through the local expansion, then the effect should be completely characterized
by $\delta_c(a)$. The local scale factor $a_W$ does not even have to follow
a Friedmann equation with the corresponding local densities and curvature
which is only guaranteed above the Jeans scale \cite{Gnedin:2011kj,Hu:2016ssz}.
Using quintessence as an example, it has been shown in Ref.~\cite{Chiang:2016vxa}
that the effect of the Jeans length on the small-scale observables can indeed be
modeled accurately by $N$-body simulations with the SU expansion even below
the Jeans scale where the SU technique might naively be supposed to fail.

In general $\delta_c(a)$ depends on wavelength, so $\delta_c(a)$ with
different wavelengths would correspond to different SUs. Therefore,
even if $\delta_c(a)$ of different wavelengths have the same value at
the final time, their evolutionary histories are still distinct.
This indicates the importance of the temporal nonlocality in structure
formation. As a result, the response of the small-scale observable will
become \emph{scale dependent}, with the scale being the wavelength of
the long mode. We shall demonstrate these scale-dependent responses
in \refsec{pk_resp} and \refsec{nh_resp}.

\subsection{Small-scale linear growth}
\label{sec:growth}
\begin{figure}[h]
\centering
\includegraphics[width=0.498\textwidth]{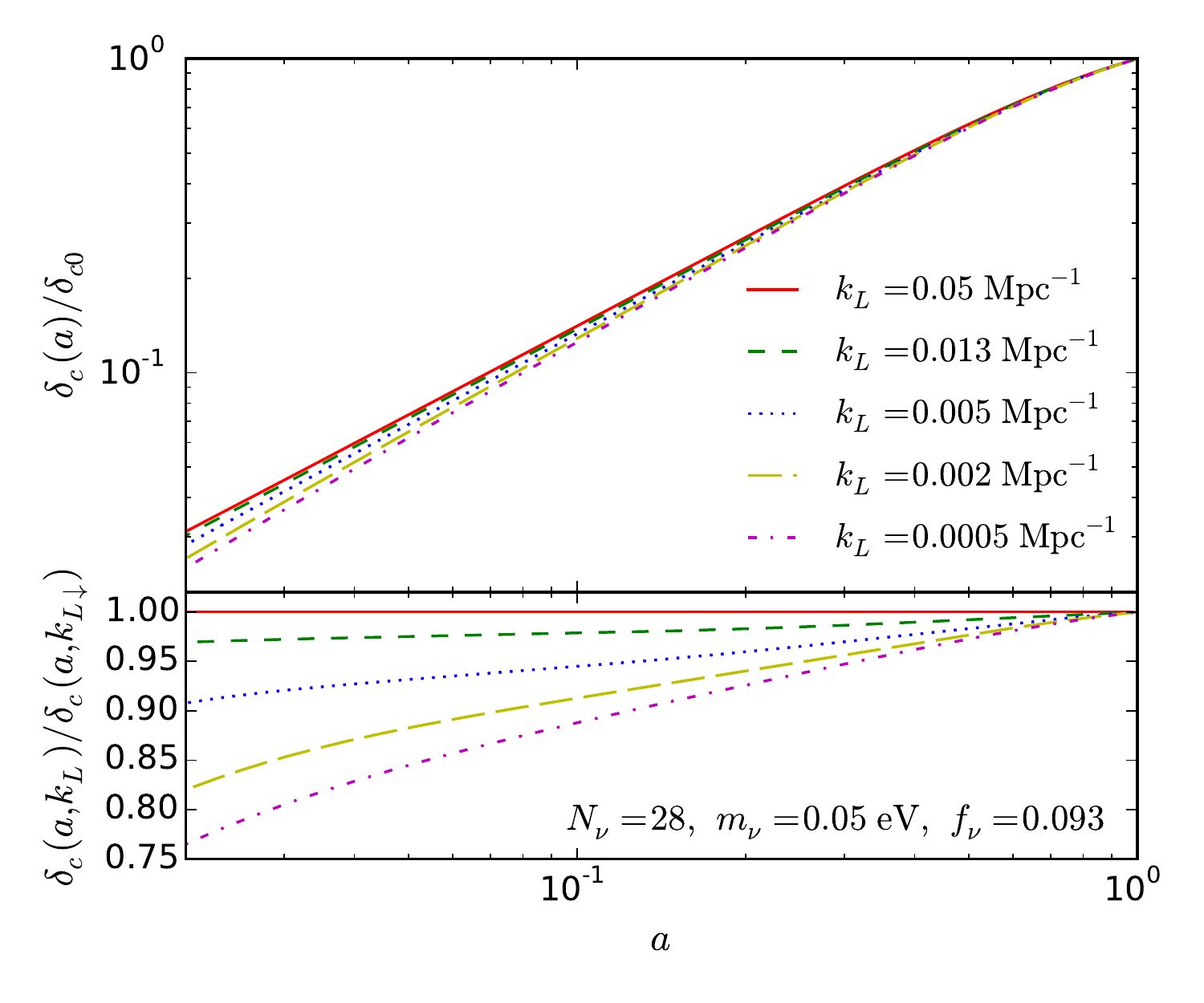}
\caption{(Top) Evolution of $\delta_c(a)$, including both CDM and baryons,
normalized to $\delta_{c0}=\delta_c(a=1)$ for different long-wavelengths
$k_L$ with $N_\nu=28$. (Bottom) Ratios of $\delta_c(a)$ to that of
$k_{L\downarrow}=5\times10^{-2}\iMpc$.}
\label{fig:deltac}
\end{figure}

Massive neutrinos provide a perfect arena to explore the scale dependence
of the response of the small-scale observables to the large-scale density
perturbation. Due to their high velocity dispersion, neutrinos do not 
cluster with CDM on scales smaller than the free-streaming scale
\cite{Lesgourgues:2006nd}
\be
 k_{\rm fs}(z)=0.0287\frac{\sqrt{\Omega_\Lambda+\Omega_m(1+z)^3}}{(1+z)^2}\iMpc \,,
\label{eq:kfs}
\ee
for our choice of $m_\nu$. Below this scale the fluctuations are washed out
by free-streaming. As a result, the evolution of $\delta_c$ becomes scale
dependent, and so does the corresponding SU. We use \texttt{CLASS}
\cite{Blas:2011rf,Lesgourgues:2011rh} to compute $\delta_c$ (including both
CDM and baryons) as a function of the scale factor $a$ and the large-scale
wavenumber $k_L$. \refFig{deltac} shows $\delta_c(a)/\delta_{c0}$ for different
$k_L$ with $N_\nu=28$, where $\delta_{c0}=\delta_c(a=1)$. Modes with smaller
$k_L$ grow faster than those of larger $k_L$ since neutrinos cluster with CDM
on scales larger than the free-streaming scale, and the free-streaming scale
decreases with time. With the evolution of $\delta_c(a)$, we can straightforwardly
construct the SU expansion using \refeqs{aW}{HW} for various $k_L$.

Since in the SU approximation only CDM clusters, the small-scale linear growth
in the SU, $D_W$, is given by
\be
\label{eq:DW}
 \frac{d^2D_W}{d\ln a_W^2}+\(2+\frac{d\ln H_W}{d\ln a_W}\)\frac{dD_W}{d\ln a_W}
 =\frac{3}{2}\frac{H_{0W}^2}{H_W^2}\frac{\Omega_{cW}}{a_W^3}D_W \,,
\ee
where $\Omega_{cW}H_{0W}^2=\Omega_cH_0^2$ is the background physical CDM
energy density. Rewriting in terms of the global scale factor, we have
\ba
\label{eq:D}
 \:&D''+\(2+\frac{H'}{H}\)D'-\frac{3}{2}\Omega_c(a)D=0 \,, \\
 \:&\epsilon''+\(2+\frac{H'}{H}\)\epsilon'-\frac{3}{2}\Omega_c(a)\epsilon
 =\frac{2}{3}\delta_c'D'+\frac{3}{2}\Omega_c(a)\delta_cD \,,
\label{eq:epsiloneom}
\ea
where $D$ is the linear growth for sub-Jeans scale perturbations in the global universe, $\epsilon=D_W-D$ is the
perturbation on $D$ due to $\delta_c$, and $\Omega_c(a)=(\Omega_c H_0^2)/(H^2 a^3)$
is the CDM energy density in units of the critical energy density as a function
of time.
Note that the scale independence of both $D$ and $D_W$ is due to the SU
approximation, since the full growth in cosmology with massive neutrinos is
scale dependent. In other words, we only consider the growth with scale much
smaller than the neutrino free-streaming scale ($k\gg k_{\rm fs}$), and in this
limit it is scale independent.
The challenge for solving \refeq{epsiloneom} is to set up the initial
condition for $\epsilon$. Specifically, as we have $f_\nu\gtrsim0.05$ ($N_\nu=14$
and 28), at higher redshift massive neutrinos have even larger contribution to
the total energy density compared to CDM, and so one cannot assume the SU being
matter dominated to set up the initial condition of $\epsilon$.

Instead, let us consider setting up the initial condition at $a_i$ with
$a_{\rm eq}\gg a_i$, and so the universe was radiation dominated and
neutrinos were relativistic. During this epoch, $2+H'/H\propto a_i/a_{\rm eq}\to0$
and $\Omega_c(a)\propto a_i/a_{\rm eq}\to0$, so the solution of the
background growth is
\be
 D=C_1\ln\frac{a}{a_H} \,,
\label{eq:D_RD}
\ee
where $C_1$ and $a_H$ are integration constants. When solving the full
perturbation equations from super to sub horizon $a_H$ is fixed to be
approximately the scale factor at horizon-entry for the small-scale mode
(see Ref.~\cite{Hu:1995en} Eq.~B12) and we have assumed $a_i \gg a_H$ in
order to approximate these equations in \refeq{D} by dropping radiation
clustering.  Plugging \refeq{D_RD} into \refeq{epsiloneom}, we require
\be
 \epsilon''=\frac{2}{3}\delta_c'D'=\frac{2}{3}C_1\delta_c'
\ee
in the radiation-dominated universe. Assuming that the long-wavelength
mode is super-horizon during this time, we also have $\delta_c\propto a^2$,
which leads to
\be
 \epsilon=\frac{1}{3}C_1\delta_c+C_2+C_3\ln a \,,
\ee
where $C_2$ and $C_3$ are integration constants. As $\delta_c$ grows as
$a^2$, the $C_2$ and $C_3$ terms should be negligible for any reasonable
choice of parameters, and we only keep the $C_1$ term. Furthermore, $C_1$
is equivalent to an overall normalization which drops out once normalized
to the final conditions, hence it is sufficient to describe the initial
conditions as
\be
 D_i=\ln\frac{a_i}{a_H} \,, \quad
 \epsilon_i=\frac{1}{3}\delta_c(a_i)D'_i \,.
\label{eq:Di_ei_RD}
\ee
In \refapp{setup}, we show that the results are insensitive to the choices
of $a_i$ and $a_H$, hence we fix $a_i=10^{-6}$ and $a_H=10^{-10}$, which
satisfies $a_{\rm eq}\gg a_i\gg a_H$.

\begin{figure}[h]
\centering
\includegraphics[width=0.498\textwidth]{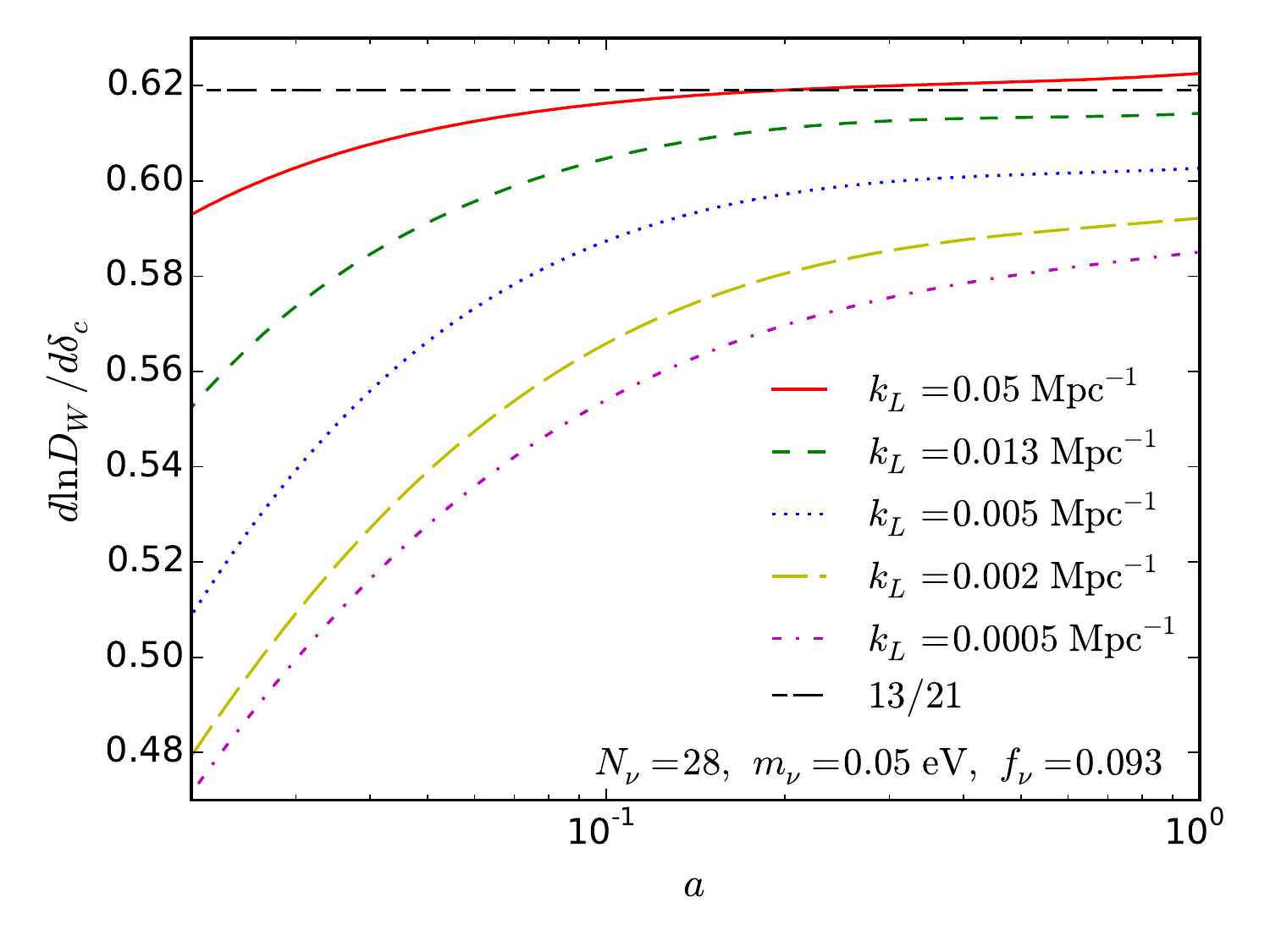}
\caption{Responses of the small-scale linear growth to the long-wavelength
$\delta_c$ for different $k_L$ as a function of global scale factor for
$N_\nu=28$. The horizontal dashed line represents the response in the
matter-dominated SU, i.e. 13/21.}
\label{fig:dlnDW}
\end{figure}

\refFig{dlnDW} shows the response of the linear growth for different
$k_L$ as a function of the global scale factor for $N_\nu=28$. The
horizontal dashed line represents the response in the matter-dominated
SU, i.e. 13/21. As we can see, the smaller the $k_L$, the smaller the
response. This is because on large scales neutrinos cluster with CDM,
hence the SU is closer to the fiducial universe. We also find that in
no case is the matter-dominated response appropriate since even at
high redshift the radiation plays a role in the response which cannot
be neglected when considering the effect of neutrinos.

\section{Separate universe simulations}
\label{sec:sims}
To extend the treatment of the response of small-scale structure to
the long-wavelength density perturbation into the nonlinear regime,
we perform $N$-body simulations in SUs with different long-wavelength
modes. Ref.~\cite{Chiang:2016vxa} presented in detail the technique
of running SU simulations with components other than CDM. In short,
one first computes a table of $(a_W,H_W)$, passes it to the $N$-body
code, and interpolates $H_W(a_W)$ when necessary. Note that $H_W$
contains the energy density of the relativistic components, i.e.
photons and neutrino.

The rest of the setup follows the standard procedure of SU simulations.
Specifically, for constructing the initial conditions, we choose the
initial power spectrum in the SU to be
\be
 P_W(k,a_{Wi})=P(k,a_0)\[\frac{D_W(a_{Wi})}{D(a_0)}\]^2 \,,
\ee
where we set $a_{Wi}=0.02$ to be the initial scale factor of the SU and
$a_0=1$ is the scale factor today in the global universe. Since CDM+baryons
is the only component that clusters in SU approximation, we use the linear
CDM+baryon transfer function from \texttt{CLASS} to calculate the linear
power spectrum for the initial conditions, in cosmology with the
corresponding numbers of massive neutrinos. To avoid confusion, in this
paper we use units of comoving $[{\rm Mpc}^{-1}]$, and convert for code
purposes as necessary. We then generate the initial conditions using realizations
of Gaussian random fields for the primordial fluctuations, and evolve to
$a_{Wi}$ using the second-order Lagrangian perturbation theory (2LPT)
\cite{Crocce:2006ve}. In \refapp{2lpt}, we derive the second-order growth
under 2LPT with long-wavelength perturbations in matter-radiation dominated
SUs, and the results are used to set up the initial conditions of SUs.
The simulations are carried out by \texttt{Gadget-2} \cite{Springel:2005mi}
from $a_{Wi}$ to the final scale factor $a_{W0}=a_0(1-\delta_{c0}/3)$,
corresponding to the same \emph{physical} time as $a_0$.

We identify halos using the Amiga Halo Finder \cite{Gill:2004km,Knollmann:2009pb},
and use its dark energy feature to input the table of $(a_W,H_W,\Omega_{cW})$
in the corresponding SU. To account for the fact that in overdense and underdense
SUs the thresholds for forming halos decrease and increase respectively, we set
the density threshold in the SU to be
\be
 \Delta_W=\frac{\Delta}{1+\delta_c(a)}\approx\Delta\[1-\delta_c(a)\] \,,
\ee
with $\Delta=200$. We set the minimum number of particles for halos to be
100 for our halo catalogs, but to be conservative we only report the result
with halos having more than 400 particles.

We run two sets of cosmologies, $N_\nu=14$ and 28 (corresponding to $f_\nu=0.049$
and 0.093), for the neutrino SU simulations. For both cosmologies we fix the
box size $L$ and the number of particles $N_p^3$. Since the particle mass of
the simulation is proportional to $\Omega_{bc}L^3/N_p^3$, fixing $\Omega_{bc}=0.3$
results in the identical mass resolution for the two $N_\nu$. We choose $L=700\Mpc$,
which is large enough so that the response in the linear regime can be obtained
at $z=0$. Note that this box size is larger than the free-streaming scale of
massive neutrinos of $0.05\eV$, which is roughly $R_{\rm fs}=200\Mpc$ today
\cite{Lesgourgues:2006nd}. We set the number of particles to be $N_p^3=640^3$,
hence the minimum halo mass we report in this paper is $2\times10^{13}~M_\odot$.

For the long-wavelength perturbations, we set $\delta_{c0}=0$ and $\pm0.01$.
In order to sample the scale dependence of the response well, we run five
different $k_L$ ($0.05-0.0005\iMpc$) for $N_\nu=28$; for $N_\nu=14$ we only
run two limiting $k_L$ to quantify the dependence of the response on $N_\nu$,
or equivalently $f_\nu$. For each choice of $N_\nu$ and $k_L$ we have $N_s=40$
sets of SU simulations for $\delta_{c0}=0,\pm0.01$. To better quantify the
scale dependence of the halo bias, which does not require simulations of $\delta_{c0}=0$
with our analysis method (see \refsec{bh_meas} for detail), we additionally
run 40 sets of SU simulations for $N_\nu=14$ and $28$ for $\delta_{c0}=\pm0.01$.
For each set of SU simulations (with or without $\delta_{c0}=0$), we adopt the
same Gaussian realization so the cosmic variance largely cancels. The details
of the neutrino SU simulations are summarized in \reftab{sims}. To simplify the
notation, hereafter we denote $k_{L\uparrow}=0.0005\iMpc$ and $k_{L\downarrow}=0.05\iMpc$.

\begin{table}[h]
\centering
\begin{tabular}{l l l l l l}
\hhline{======}
$N_\nu$ & $\delta_{c0}$ & $k_L\,[{\rm Mpc}^{-1}]$ & \!$L\,{ [{\rm  Mpc}]}$\! & $N_p$ & $N_s$ \\
\hline
14 & 0 & $0.05, 0.0005$ & 700 & $640^3$ & 40 \\
28 & 0 & $0.05, 0.013, 0.005, 0.002, 0.0005$ & 700 & $640^3$ & 40 \\
14 & $\pm0.01$ & $0.05, 0.0005$ & 700 & $640^3$ & 80 \\
28 & $\pm0.01$ & $0.05, 0.013, 0.005, 0.002, 0.0005$ & 700 & $640^3$ & 80 \\
\hhline{======}
\end{tabular}
\caption{Summary of the neutrino SU simulations.}
\label{tab:sims}
\end{table}

We point out that there are two main caveats in our neutrino SU simulations.
First, within the SU, only CDM clusters, and the other components are smooth.
Our simulation box, however, is larger than the neutrino free-streaming scale,
so the neutrino clustering is missing on scales larger than the free-streaming
scale of \refeq{kfs}. Second, we assume that the wavelength of $\delta_c$ is
much larger than the simulation box size, hence the curvature of $\delta_c$ is
ignored. This is a good approximation when $k_L\ll\pi/L\approx0.0045\iMpc$,
but will be violated for larger $k_L$. In \refapp{caveat}, we discuss these
two systematics in detail, and argue that neutrino clustering can be ignored
for $k\gtrsim0.05\iMpc$, while the corrections due to ignoring the curvature
of $\delta_c$ are $\mathcal{O}(k_L^2R_M^2)$ for halo bias with $R_M$ being
the Lagrangian radius of halo with mass $M$ and $\mathcal{O}(k_L^2/k^2)$ for
power spectrum response with $k$ being the wavenumber of the small-scale power
spectrum. For halo bias both systematics are negligible since the Lagrangian
radii of halos of interest are $\lesssim10\Mpc$; for power spectrum response
to avoid the systematics we only report results for $k\ge0.05\iMpc$.

\section{Power spectrum response}
\label{sec:pk_resp}
We now calibrate the response of the power spectrum to $\delta_c$ for different
$N_\nu$ and $k_L$. In \refsec{Rgrowth_meas}, we show the measurement of the
power spectrum response in the neutrino SU simulations. In particular, we shall
demonstrate with high statistical significance that the larger the scale of
$k_L$, the smaller the power spectrum response until it is much larger than
the neutrino free-streaming scale. In \refsec{Rgrowth_dep}, we study the
dependence of the growth response on $k_L$ and $N_\nu$.

\subsection{Growth response}
\label{sec:Rgrowth_meas}
In the presence of a long-wavelength $\delta_c$, the locally measured power
spectrum differs from the global one. We can quantify the fractional
difference between the local and global power spectra by
\be
 \frac{\Delta P}{P}\approx\frac{d\ln P}{d\delta_c}\delta_c
 \equiv R_{\rm tot}\delta_c \,,
\ee
where we define $R_{\rm tot}$ as the ``response'' of the local power
spectrum to $\delta_c$. Refs.~\cite{Barreira:2017sqa,Barreira:2017kxd}
provide a rigorous embedding of power spectrum response into the perturbative
framework, and can be generalized to higher-oder responses. At leading
order $R_{\rm tot}$ is independent of the amplitude of $\delta_c$, hence
we can use the SU simulations with $\delta_{c0}=0,\pm0.01$ to calibrate
the response. This effect can also be measured in big $N$-body simulations
by the bispectrum in the squeezed limit with the angle average of the
cosine between the long and short modes (see e.g. Ref.~\cite{Chiang:2014oga})
\be
 \lim_{k_L\to0}B(k,k',k_L) \equiv B^{\rm sq}(k,k_L) = R_{\rm tot}P(k_L)P(k) \,,
\ee
where $k_L$ corresponds to the mode of $\delta_c$ and $k\approx k'$
corresponds to the mode of $P$ averaged over direction. Note that
unless otherwise stated, we denote the CDM+baryon power spectrum
and bispectrum without subscript for compactness.

In the  $\Lambda$CDM cosmology, the evolution of $\delta_c$ is independent 
of the wavenumber of the large-scale mode $k_L$, and so as $R_{\rm tot}$.
As a result, $R_{\rm tot}$ depends only on time and the small-scale wavenumber
$k$. If the universe, however, has additional components that cluster and
remain smooth above and below Jeans scales, such as quintessence or massive
neutrinos, then both $\delta_c$ and $R_{\rm tot}$ would depend on $k_L$ (see
Ref.~\cite{Chiang:2016vxa} for the scale-dependent response in quintessence
SU). Therefore, $R_{\rm tot}=R_{\rm tot}(k,k_L,a)$. This also indicates that
the reduced squeezed-limit bispectrum would depend on $k_L$.

\begin{figure*}[t]
\centering
\includegraphics[width=0.497\textwidth]{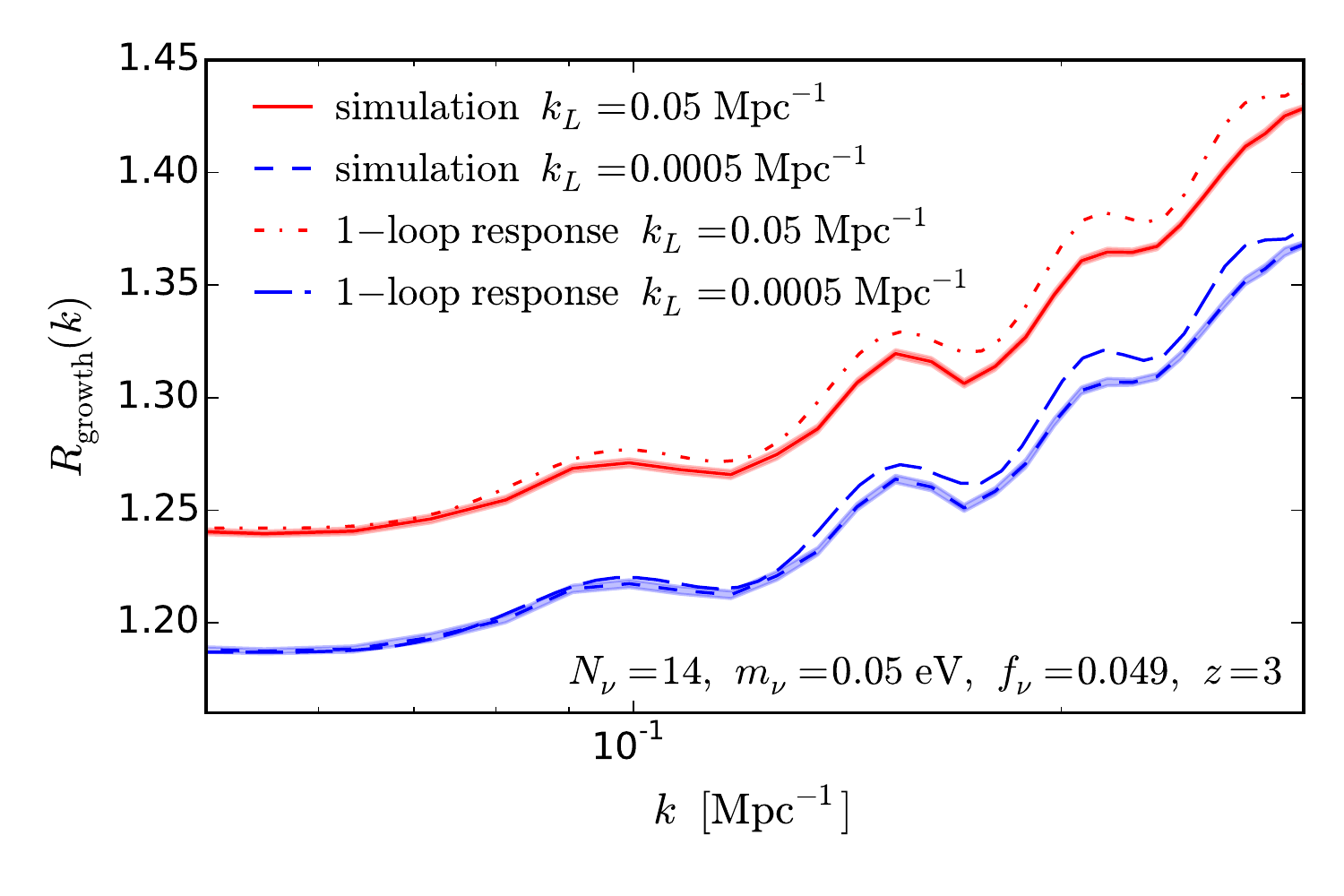}
\includegraphics[width=0.497\textwidth]{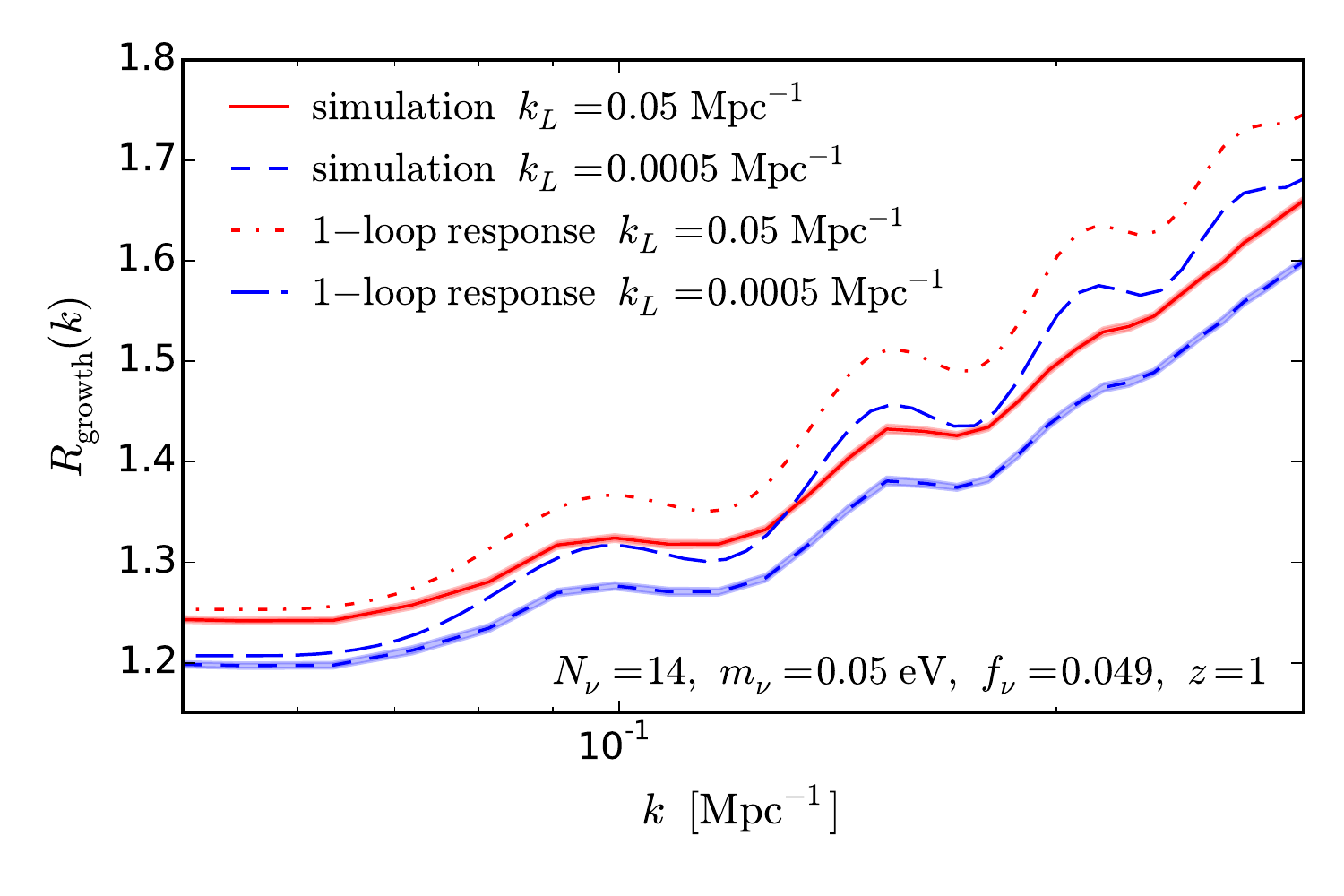} \\ [-3ex]
\includegraphics[width=0.497\textwidth]{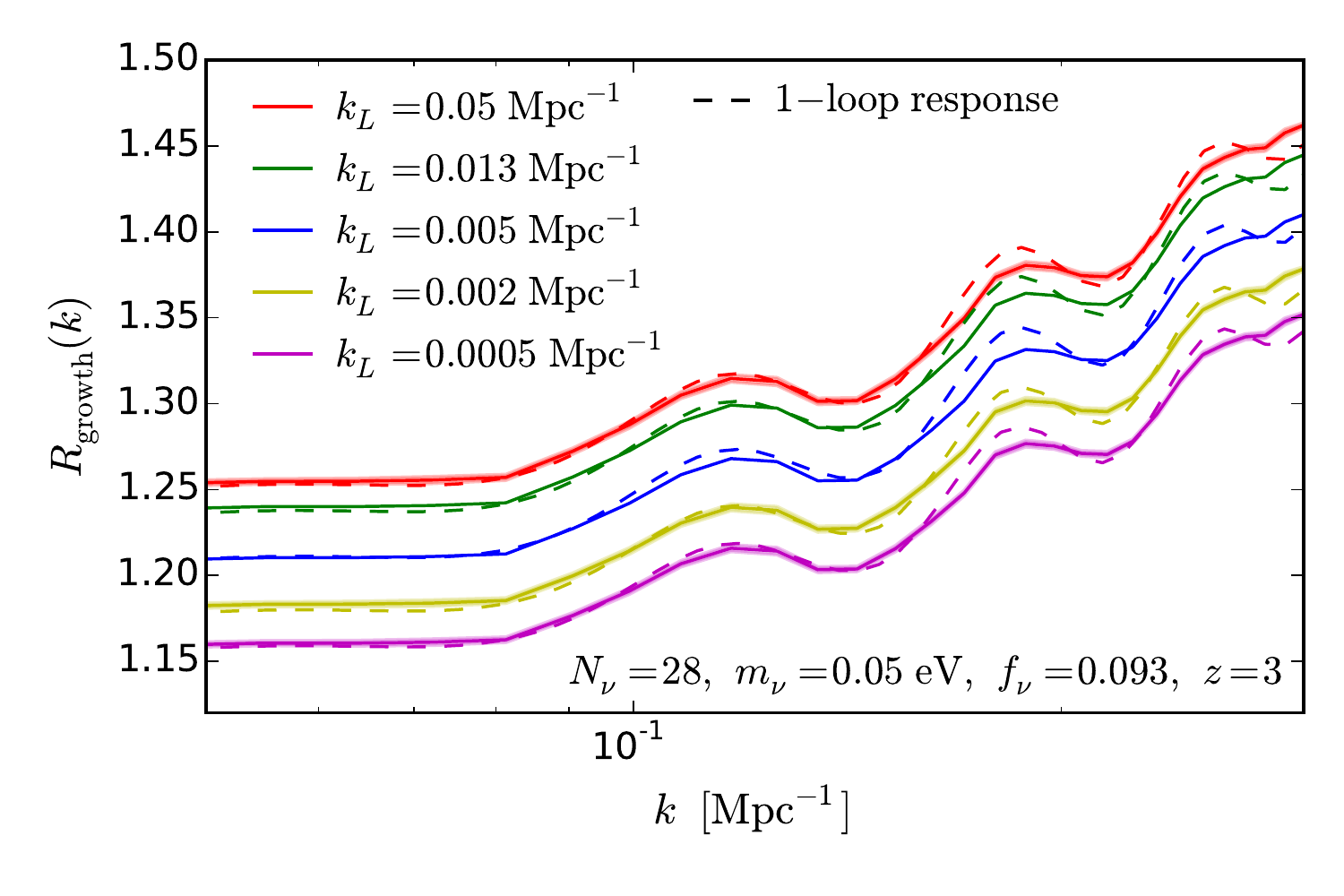}
\includegraphics[width=0.497\textwidth]{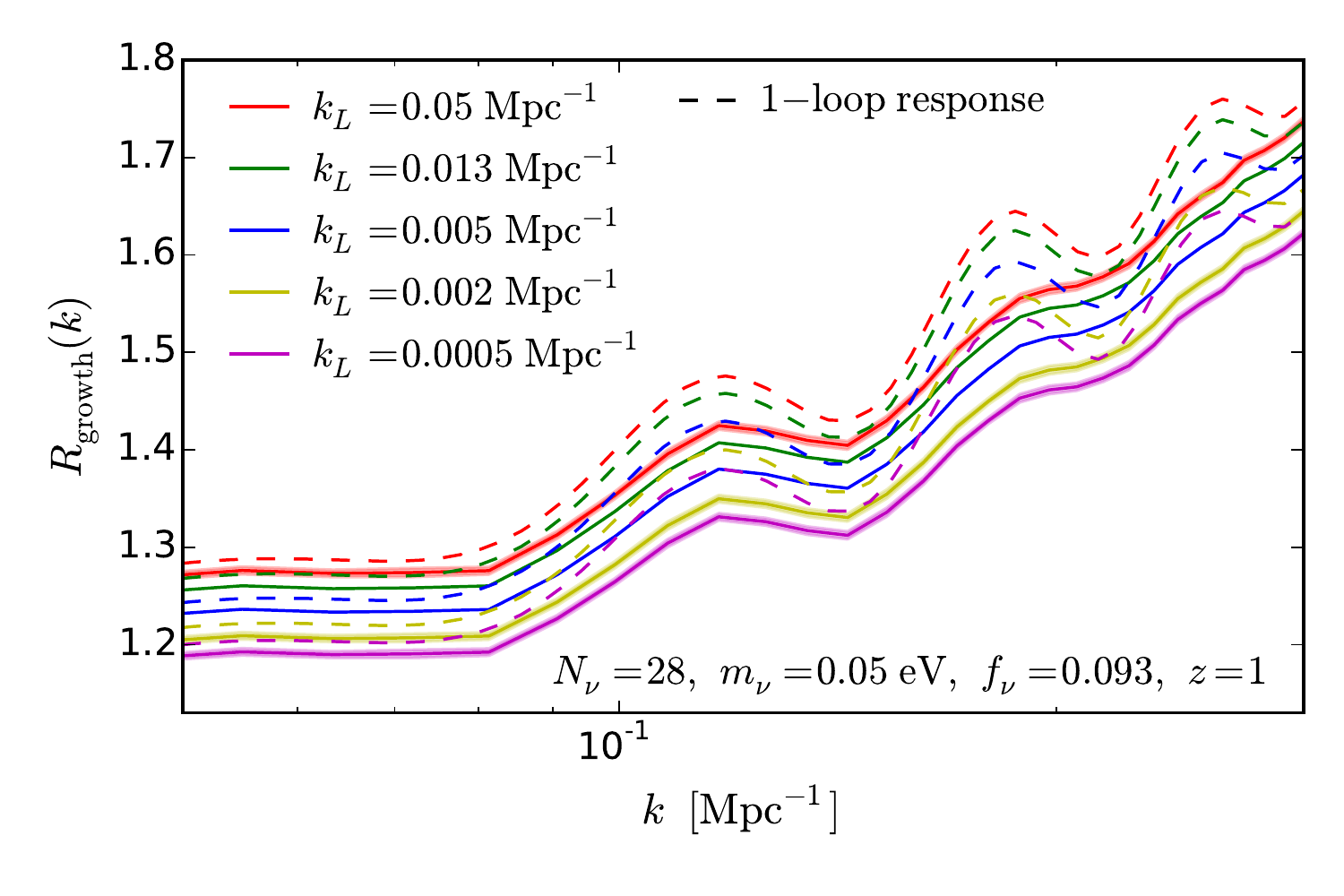}
\caption{Growth response measured from 40 sets of neutrino separate universe
simulations as a function of small-scale $k$ for different large-scale $k_L$
(denoted by various colors). The top and bottom panels show $N_\nu=14$ and 28
($f_\nu=0.049$ and 0.093), and the left and right panels show $z=3$ and 1.
The lines with shaded areas show the measurement from simulations with the error
on the mean, whereas the lines without shaded areas show the analytic calculation
with the 1-loop power spectrum response (see the text for details).}
\label{fig:pk_resp}
\end{figure*}

Physically, the total response can be separated into three pieces \cite{Li:2014sga}:
\be
 R_{\rm tot}=R_{\rm growth}+R_{\rm dilation}+R_{\bar\rho} \,.
\label{eq:Rtot}
\ee
$R_{\rm growth}$ specifies the change due to the growth of the small-scale
density fluctuation between the separate and global universes at a fixed
comoving $k$; $R_{\rm dilation}$ describes the change of comoving $k$ between
separate and global universes due to different expansion histories; $R_{\bar\rho}$
accounts for the different mean densities in separate and global universes
used to define the small-scale density fluctuation. Note that $R_{\rm dilation}$
and $R_{\bar\rho}$ are nondynamical effects and can be computed without
additional simulations, as we shall discuss in detail in \refsec{observables}.
On the other hand, the growth response is dynamical and so requires $N$-body
simulations for an accurate estimate.

In order to measure the growth response from SU simulations, we distribute
the dark matter particles onto a $640^3$ grid using the cloud-in-cell density
assignment to construct the density fluctuation, and Fourier transform the
fluctuation with \texttt{FFTW} \cite{fftw} to estimate the power spectrum
$\hat{P}(k,a)$. We then estimate the growth response by
\be
 \hat{R}_{\rm growth}(k,k_L,a)\equiv
 \frac{\hat{P}(k,a|\delta_{c0}^+(k_L))-\hat{P}(k,a|\delta_{c0}^-(k_L))}{2\hat{P}(k,a|\delta_{c0}^0)\delta_c(k_L,a)} \,,
\ee
where $\delta_{c0}^{0,\pm}=0$, $\pm0.01$.

The lines with shaded areas in \reffig{pk_resp} show the mean of the growth
response measured from neutrino SU simulations as a function of small-scale
$k$ for different large-scale $k_L$ (denoted by various colors), and the shaded
areas represent the error on the mean. The top and bottom panels show $N_\nu=14$
and 28, and the left and right panels show $z=3$ and 1. We show the results at
higher redshifts and on larger scales ($k\le0.3\iMpc$) because it is the regime
in which the 1-loop calculation has the predicting power, which will be discussed
in the following. We find with high statistical significance that the growth
response indeed depends on the large-scale wavenumber $k_L$ for all redshifts
and small scales. Note especially that for $k\gtrsim0.1\iMpc$, where the systematics
due to small-scale neutrino clustering and the curvature of the long-wavelength
mode can be ignored, the dependence of $R_{\rm growth}$ on $k_L$ persists and
is similar in amplitude.

To better understand the measurement from neutrino SU simulations quantitatively,
we compute the growth response using perturbation theory. Specifically,
in perturbation theory the growth response can be modeled as
\be
 R_{\rm growth}(k,k_L,a)=\frac{d\ln P(k,a)}{d\ln D_W(a)}\left[\frac{d\ln D_W}{d\delta_c}(k_L,a)\right] \,,
\label{eq:Rgrowth}
\ee
where the second term in the right-hand side is computed in \reffig{dlnDW}.
Note that in \refeq{Rgrowth} the dependencies on $k$ and $k_L$ are separated
into the first and second terms in the right-hand side. In the linear regime,
$P\approx P_{\rm lin}$ is proportional to $D_W^2$ and so
\be
 \frac{d\ln P_{\rm lin}(k,a)}{d\ln D_W(a)}=2 \,.
\label{eq:plin_growth}
\ee
In the mildly nonlinear regime, we utilize the 1-loop power spectrum in the
standard perturbation theory (see e.g. Ref.~\cite{Jeong:2006xd}):
$P_{\rm 1-loop}=P_{\rm lin}+P_{22}+2P_{13}$, where $P_{22}$ and $P_{13}$
are the nonlinear correction and proportional to $D_W^4$ if $\Omega_{cW}(a_W)/f_W^2(a_W)\approx1$.
As a result,
\be
 \frac{d\ln P_{\rm 1-loop}(k,a)}{d\ln D_W(a)}
 =2\[1+\frac{P_{22}(k,a)+2P_{13}(k,a)}{P_{\rm 1-loop}(k,a)}\] \,.
\label{eq:p1loop_growth}
\ee

The lines without shaded areas in \reffig{pk_resp} show the growth response computed
from the 1-loop power spectrum using \refeq{p1loop_growth} and \reffig{dlnDW}.
On large scale, the nonlinear correction becomes subdominant, and the 1-loop calculation
approaches to the linear prediction and becomes $k$ independent. We find that the 1-loop
calculation is generally in good agreement with the measurement, especially at $z=3$,
as would be expected and hence validates the SU simulations. On smaller scales and at
lower redshift, the nonlinearities are too large to be modeled by the perturbation theory,
hence we find a more significant difference.

\subsection{Dependence on $k_L$ and $N_\nu$}
\label{sec:Rgrowth_dep}
We have shown in \reffig{pk_resp} that the small-scale growth responds differently
to $\delta_c$ with various $k_L$. To better study this feature, let us define the
``growth response step'' as the ratio of the growth response with respect to that
of $k_{L\uparrow}=0.0005\iMpc$, i.e.
\be
 R_{\rm growth}(k,k_L,N_\nu,a)/R_{\rm growth}(k,k_{L\uparrow},N_\nu,a) \,.
\ee

We first examine the dependence of the step on the small-scale $k$ by fixing $N_\nu$
and $k_L$, and we find that it is fairly independent of $k$. More precisely, at $z=0$
the step between $k_{L\uparrow}$ and $k_{L\downarrow}=0.05\iMpc$ departs from a
scale-independent constant at the 10\% level for $k\lesssim0.6\iMpc$. This is not
surprising: in \reffig{pk_resp} $R_{\rm growth}$ of different $k_L$ have similar
scale dependence in $k$. Thus, in the following we fix $k=0.252\iMpc$ to avoid
the systematics from neutrino clustering as well as the curvature of $\delta_c$,
and focus on the dependencies of $k_L$ and the number of massive neutrinos $N_\nu$.

\begin{figure*}[t]
\centering
\includegraphics[width=0.497\textwidth]{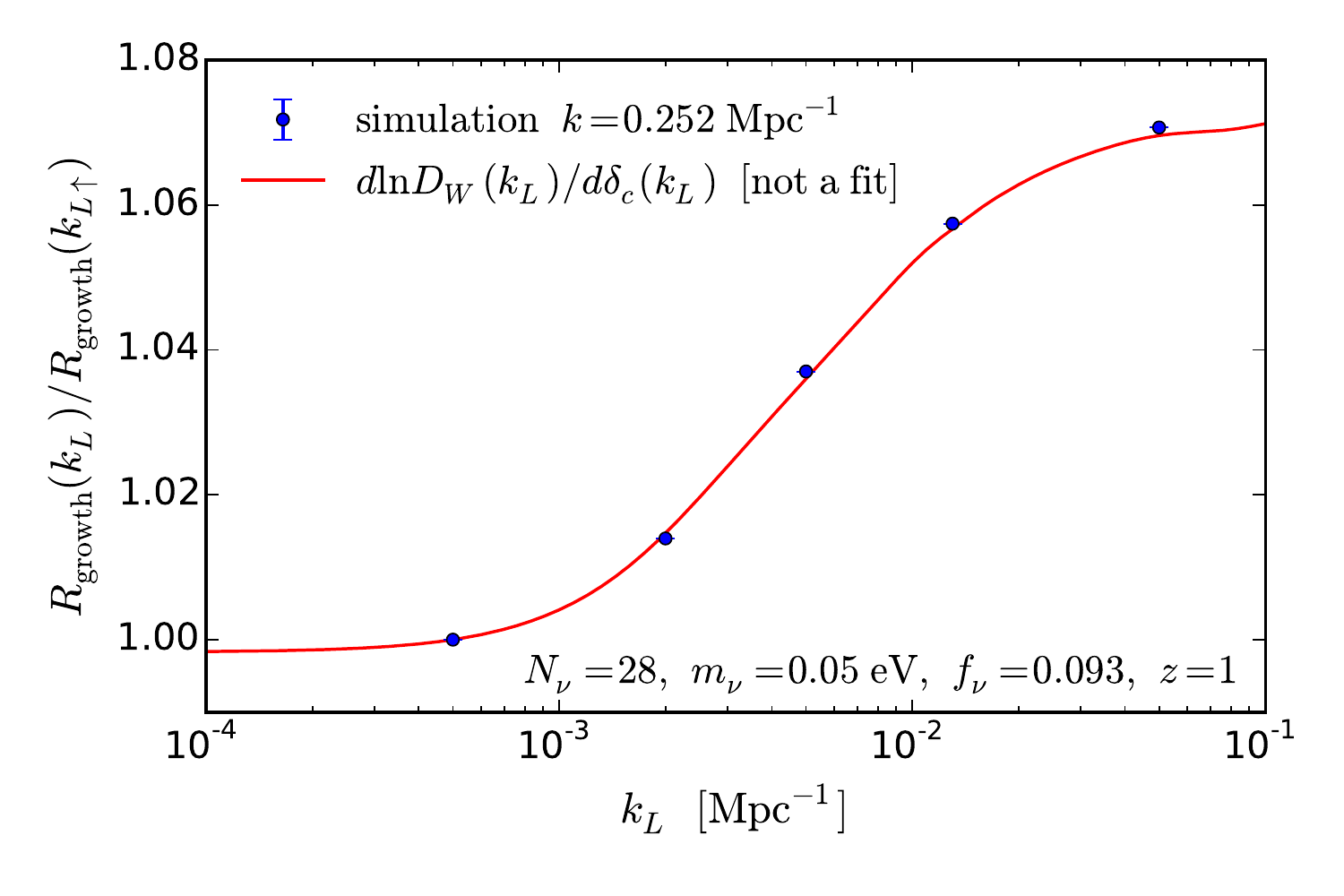}
\includegraphics[width=0.497\textwidth]{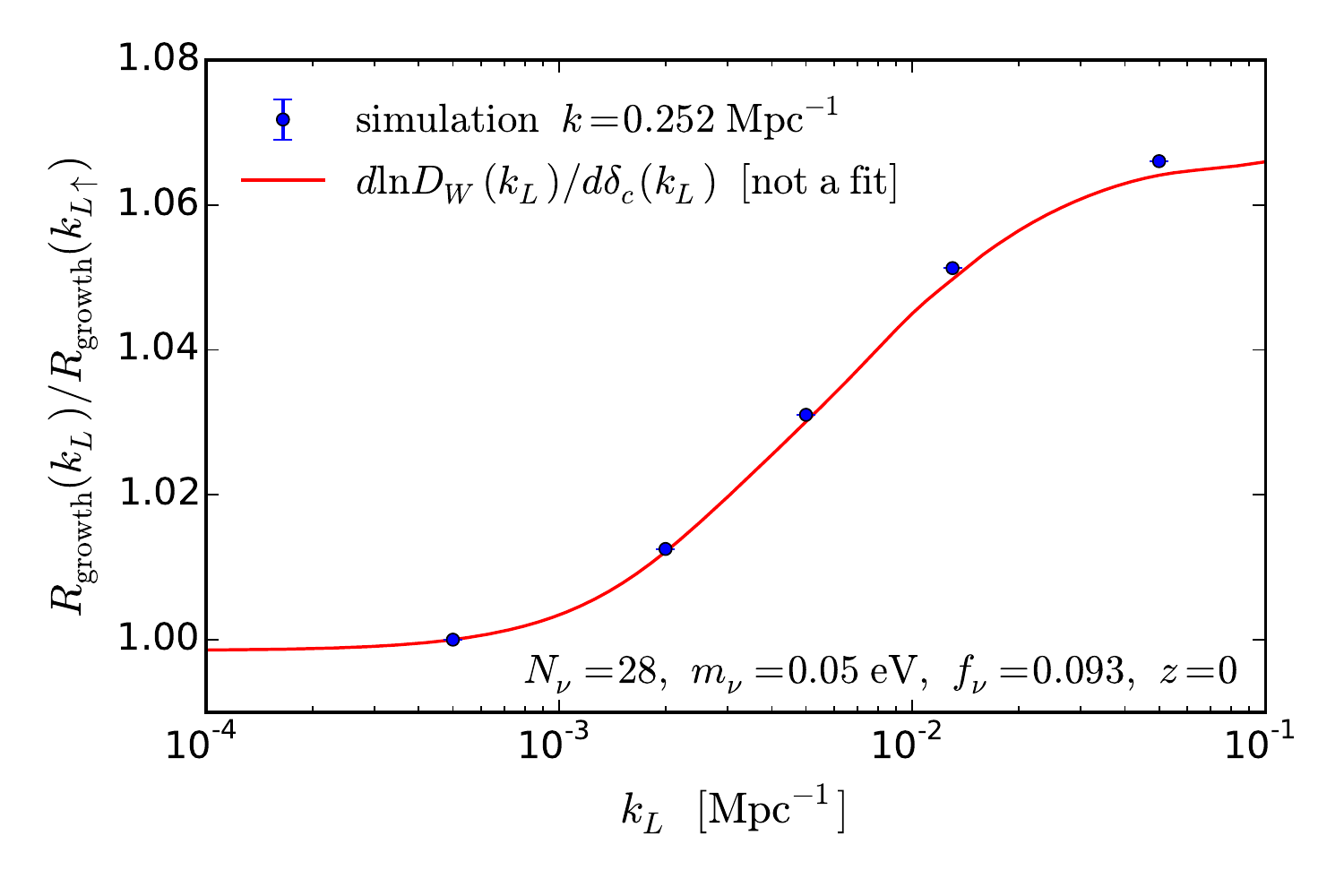}
\caption{Growth response step at $z=1$ (left) and 0 (right) for $N_\nu=28$
as a function of the large-scale mode $k_L$. The blue data points show the
measurement from SU simulations with the small-scale mode $k=0.252\iMpc$,
whereas the red solid line shows the numerical calculation of $d\ln D_W/d\delta_c$
normalized at $k_{L\uparrow}=0.0005\iMpc$. The error bars show the error
on the mean.}
\label{fig:Rgw_step_kL}
\end{figure*}

\refFig{Rgw_step_kL} shows the growth response step for $N_\nu=28$ as a
function of $k_L$. The blue data points with error bars show the measurement
of the step from neutrino SU simulations at $z=1$ (left) and 0 (right).
Note that by definition the error bars of the step are zero at $k_{L\uparrow}$.
For the other $k_L$, the small error bars are the outcome of the highly
correlated growth responses, since we use the same random realizations
for different $k_L$. We find that the dependence of the growth response
step on $k_L$ is statistically significant, indicating that the growth
response is indeed affected by the temporal evolution of $\delta_c$.

Given the independence of the step on the small-scale $k$ and the good
agreement with perturbation theory for linear $k$ demonstrated in the
previous section, the response of the linear growth function should
provide an accurate calibration of its shape and amplitude. We follow
the procedures in \refsec{growth} to numerically compute $d\ln D_W/d\delta_c$
with $N_\nu=28$ for various $k_L$, and normalize it at $k_{L\uparrow}$.
The result is shown as the red solid line in \reffig{Rgw_step_kL}, and
we find it is in excellent agreement with the measurement. Notice that
the red line is slightly less than unity on large scale, because we
normalize the step at $k_{L\uparrow}$. Should we normalize the step
feature with a smaller $k_L$, the red line would approach to unity
on large scale.

\begin{figure*}[t]
\centering
\includegraphics[width=0.497\textwidth]{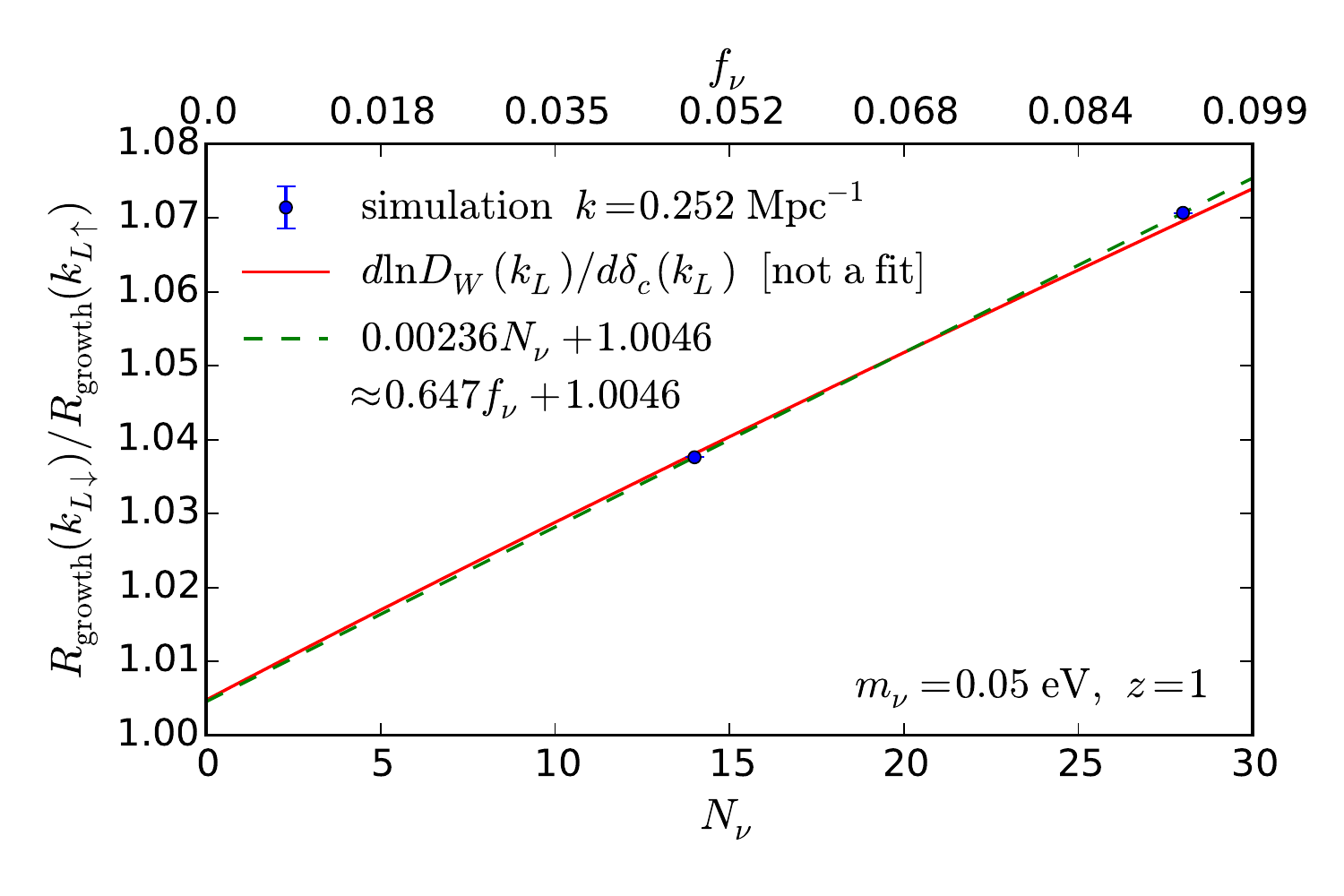}
\includegraphics[width=0.497\textwidth]{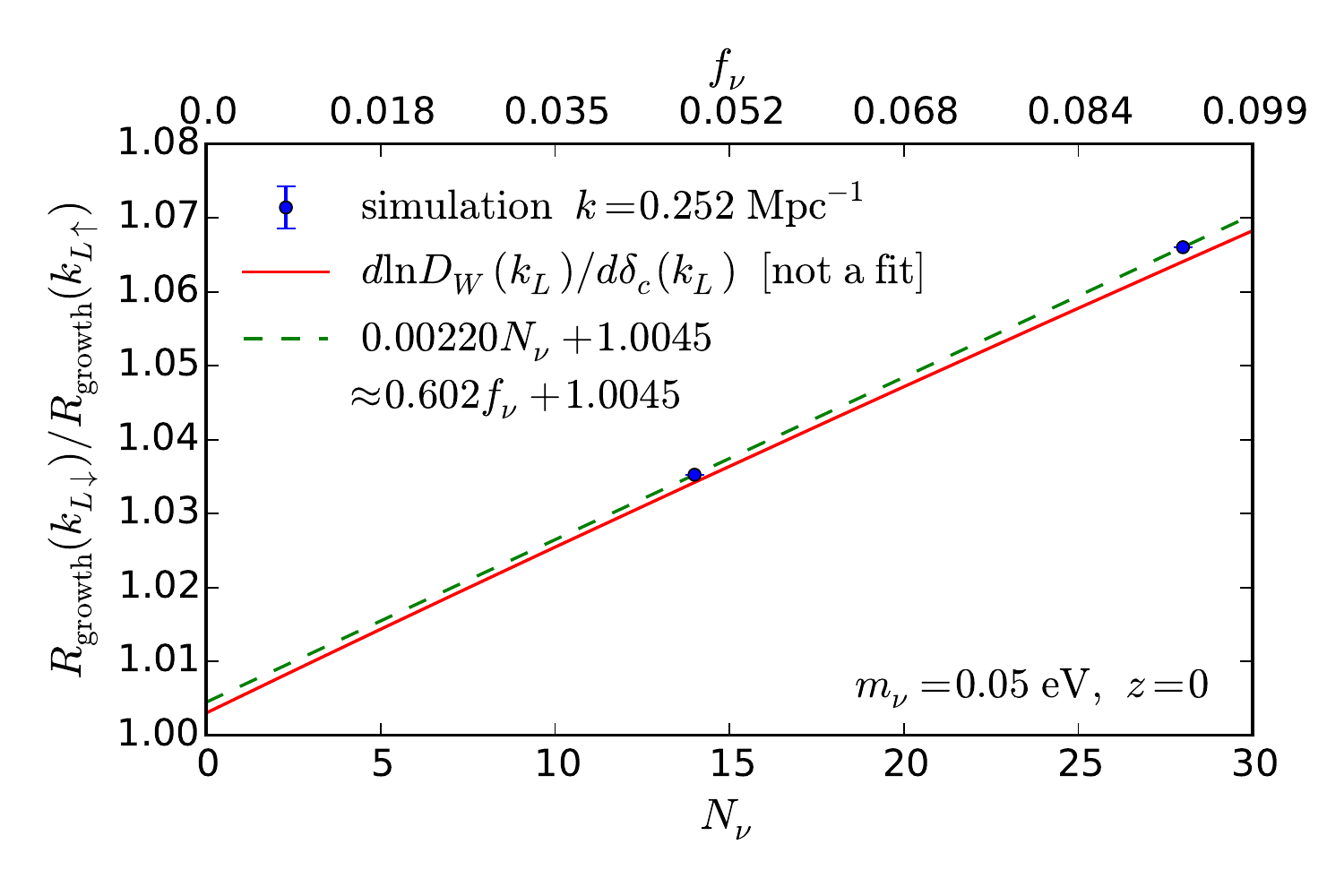}
\caption{Growth response step between $k_{L\uparrow}=0.0005\iMpc$ and
$k_{L\downarrow}=0.05\iMpc$ at $z=1$ (left) and 0 (right) as a function of the
number of massive neutrinos $N_\nu$, with the corresponding $f_\nu$ labeled on
the top $x$-axis. The blue data points show the measurement from SU simulations
with the small-scale mode $k=0.252\iMpc$; the red solid line shows the numerical
calculation of $d\ln D_W/d\delta_c$ normalized at $k_{L\uparrow}$; the green
dashed line shows the linear relation of the two data points. The error bars
show the error on the mean.}
\label{fig:Rgw_step_Nnu}
\end{figure*}

We next examine the growth response step between $k_{L\uparrow}$ and $k_{L\downarrow}$
as a function of $N_\nu$, i.e.\ $R_{\rm growth}(k_{L\downarrow},N_\nu)/R_{\rm growth}(k_{L\uparrow},N_\nu)$.
The blue data points in \reffig{Rgw_step_Nnu} show the measurement from neutrino
SU simulations at $z=1$ (left) and 0 (right), with the corresponding $f_\nu$ labeled
on the top $x$-axis. It is evident that the growth response step is non-zero, and
the larger the $N_\nu$ the larger the step size. Assuming that the measured growth
response step is linear in $N_\nu$, we can solve the slope and the intercept, and
the result is shown as the green dashed line, with the values given in the legend.
To convert $N_\nu$ to the commonly used $f_\nu$, we use \refeq{fnuNnu} and take the
limit that $f_\nu\ll1$, which leads to $N_\nu\approx274f_\nu$.\footnote{Solving
directly for a linear relation in $f_\nu$ leads to slightly different results due
to the nonlinearity in \refeq{fnuNnu} at high $f_\nu$.} We find that the slope of
the measured growth response step for artificially large $N_\nu$ in terms of $f_\nu$
is approximately 0.6. Interestingly the best fit does not go to unity at $N_\nu=0$
though this value is far from the simulated values.

To avoid extrapolation from the simulated values of $N_\nu$, we can also model the
growth response step by numerically evaluating $d\ln D_W/d\delta_c$ at $k_{L\uparrow}$
and $k_{L\downarrow}$ as a function of $N_\nu$. Note that we only vary the number of
massive neutrinos, not their mass, and correspondingly do not include massless neutrinos
when $N_\nu<3$. The result is shown as the red solid line in \reffig{Rgw_step_Nnu},
which is in good agreement with the measurement. If we solve the slope $A$ in terms
of $f_\nu$ and intercept $(1+B)$ for the red solid line in the limit that $f_\nu\ll1$,
then we get $A=0.68,0.63$ and $B=0.0048, 0.0030$ for $z=1,0$.

In this model we can interpret the non-zero $B$ intercept of the step in the $N_\nu=0$
limit as due to photons. Photons possess a Jeans scale of the sound horizon before
recombination and the horizon after and cause their own growth response. If the
calculation is done in the universe without photons, then the growth response step
would go to unity at $N_\nu=0$, since the growth of the long-wavelength modes in
$\Lambda$CDM cosmology is scale independent. Note also that these values are obtained
assuming $m_\nu=0.05\eV$ with the change in $f_\nu$ coming from the number of massive
neutrinos. For a different neutrino mass, we expect the growth response step between
the super- and sub-Jeans limits can still be approximated by the same slope, but the
scale of the transition in $k_L$ would shift due to the a different free-streaming
scale.

\section{Scale-Dependent Bias}
\label{sec:nh_resp}
Let us now turn to the response of the halo mass function, which determines the
relationship between the number density of halos and the long-wavelength density
fluctuation $\delta_c$, i.e.\ the linear density bias of halos. In \refsec{bh_meas},
we discuss the analysis of the response bias from SU simulations. In \refsec{bh_dep},
we present the response bias from the neutrino SU simulations, and study its
dependence on $k_L$ and $N_\nu$. As in the previous section, we shall demonstrate
with high statistical significance that the response bias is scale dependent due
to the presence of the neutrino free-streaming scale. In \refsec{b_model}, we
compare the response bias with different bias models and discuss the importance
of temporal nonlocality in producing scale-dependent bias.

\subsection{Response bias}
\label{sec:bh_meas}
The linear bias of halos can be regarded as the linear response of the halo
abundance of mass $M$ to the long-wavelength density perturbation, i.e.
\be
 b(M)\equiv\frac{d\delta_h(M)}{d\delta_c}=\frac{d\ln n_{\ln M}(M)}{d\delta_c} \,,
\label{eq:bresp}
\ee
where $n_{\ln M}(M)=-dn(M)/d\ln M$ is the differential halo mass function
and $n(M)$ is the cumulative  halo mass function. We  refer to this as
``response bias''. Physically, the enhanced growth in a SU with $\delta_c>0$ 
makes massive halos more abundant compared to their counterparts with
$\delta_c<0$, making their number density fluctuation a biased tracer
of $\delta_c$.  Therefore, by measuring how the halo mass function is affected
by $\delta_c$ in SU simulations, we have a direct calibration of the halo
bias without the standard clustering measurement
\cite{Li:2015jsz,Lazeyras:2015lgp,Baldauf:2015vio}.

While we do not write it explicitly in \refeq{bresp}, $\delta_c$ is a function
of the wavenumber of the long mode $k_L$ when its growth depends on scale.
For example, if there are additional components that possess Jeans scales,
such as quintessence, then the growth becomes scale dependent and so $b(M)$ would
depend on $k_L$ contrary to the expectations of purely spatial and temporally
local bias \cite{Chiang:2016vxa}. The SU simulations thus allow us to study
the scale-dependent bias due to the free-streaming length of massive neutrinos
\cite{LoVerde:2014pxa}.

The most straightforward way of measuring the response of the halo mass function
is to bin the halo abundance in halo mass and compare it for different $\delta_c$.
This, however, is inefficient since we can only measure the effect if the change
in halo mass moves it across the mass bins. Instead, we adopt the abundance matching
technique introduced in Ref.~\cite{Li:2015jsz} to characterize the response bias.
Specifically, for each set of $\delta_{c0}=\pm0.01$ SU simulation we first bootstrap
resample the total 80 realizations and combine the halo catalogs sorted in mass.
We then measure the discrete threshold mass shift by
\be
 s_i(\ln M_i)=\frac{\ln M_i^+-\ln M_i^-}{2|\delta_c|} \,,
\ee
where $M_i^\pm$ are the masses of the $i^{\rm th}$ most massive halo in the $\delta_{c0}=\pm0.01$
SU sample and $M_i=(M_i^+M_i^-)^{1/2}$. Next we use a smoothing spline function with
two knots per dex in halo mass to estimate the continuous threshold mass shift $\hat{s}(\ln M)$
as well as the cumulative halo mass function $\hat{n}(\ln M)$ using all $M_i$ entry
by entry.\footnote{We also estimate the cumulative halo mass function by separately
estimating it in the $\delta_{c0}=\pm0.01$ SUs and then taking the average. The fractional
difference of the Lagrangian bias is less then 0.02\% for the mass and redshift range
of interest, which is much smaller than the bootstrap uncertainty.} The derivative of
the smooth cumulative mass function fit gives the differential mass function
$\hat{n}_{\ln M}=-d\hat{n}/d\ln M$. The cumulative Lagrangian bias of halos with mass
greater than threshold mass $M$, denoted with an overbar, can thus be estimated as
\be
 \hat{\bar{b}}^L(M)=\frac{\hat{n}_{\ln M}(\ln M)\hat{s}(\ln M)}{\hat{n}(\ln M)} \,.
\label{eq:bL_hat}
\ee
\refEq{bL_hat} measures the Lagrangian bias since the SU simulations are performed
with the same comoving instead of physical volume, and we obtain the cumulative
Eulerian bias by transforming from Lagrangian to Eulerian coordinates:
\be
 \hat{\bar{b}}(M)=1+\hat{\bar{b}}^L(M) \,.
\label{eq:b_eulerian}
\ee
Finally, repeating the procedure with different resamplings for 8,000 times,
we have an estimate of the error on the mean.

\subsection{Scale dependence}
\label{sec:bh_dep}
\begin{figure*}[t]
\centering
\includegraphics[width=0.497\textwidth]{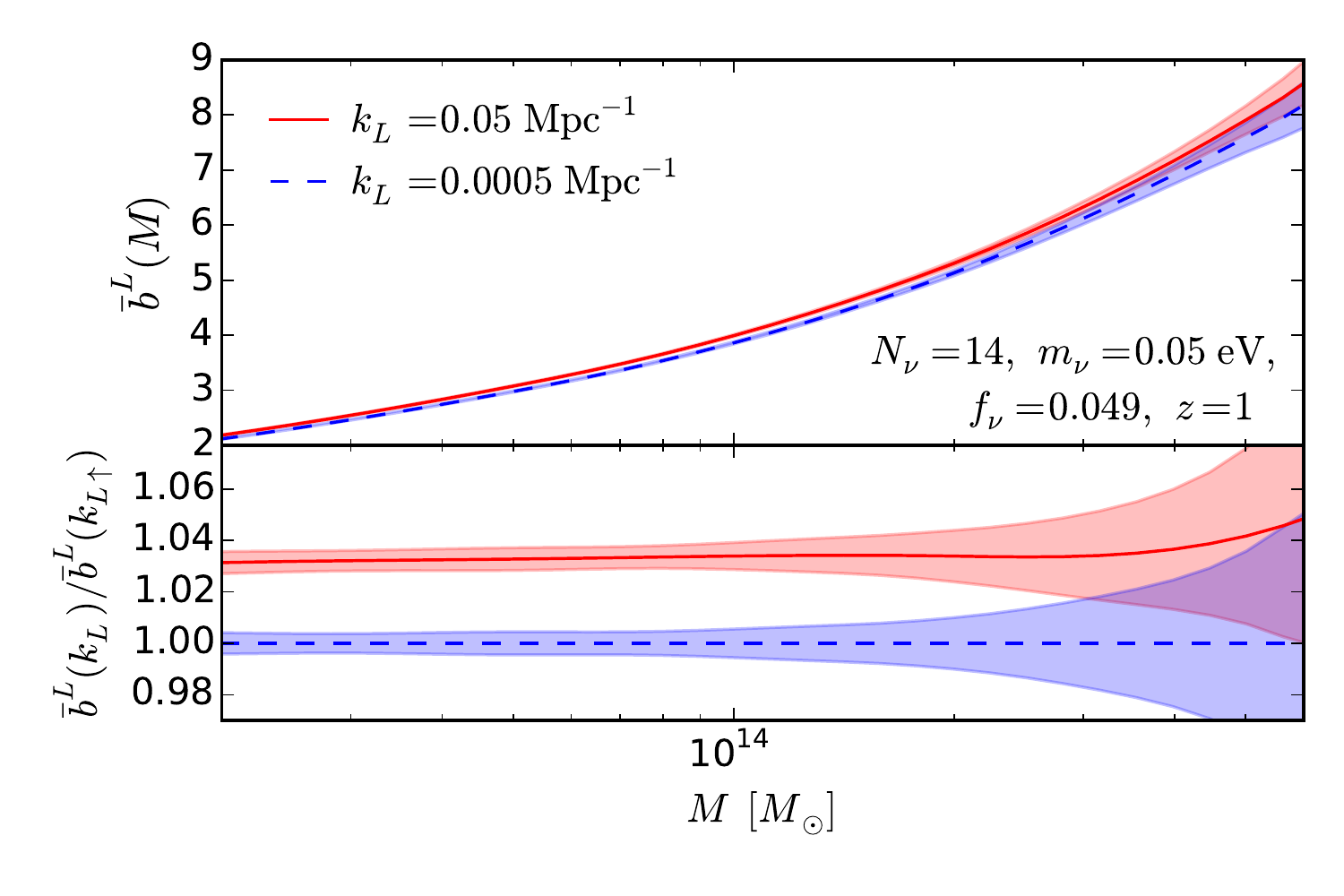}
\includegraphics[width=0.497\textwidth]{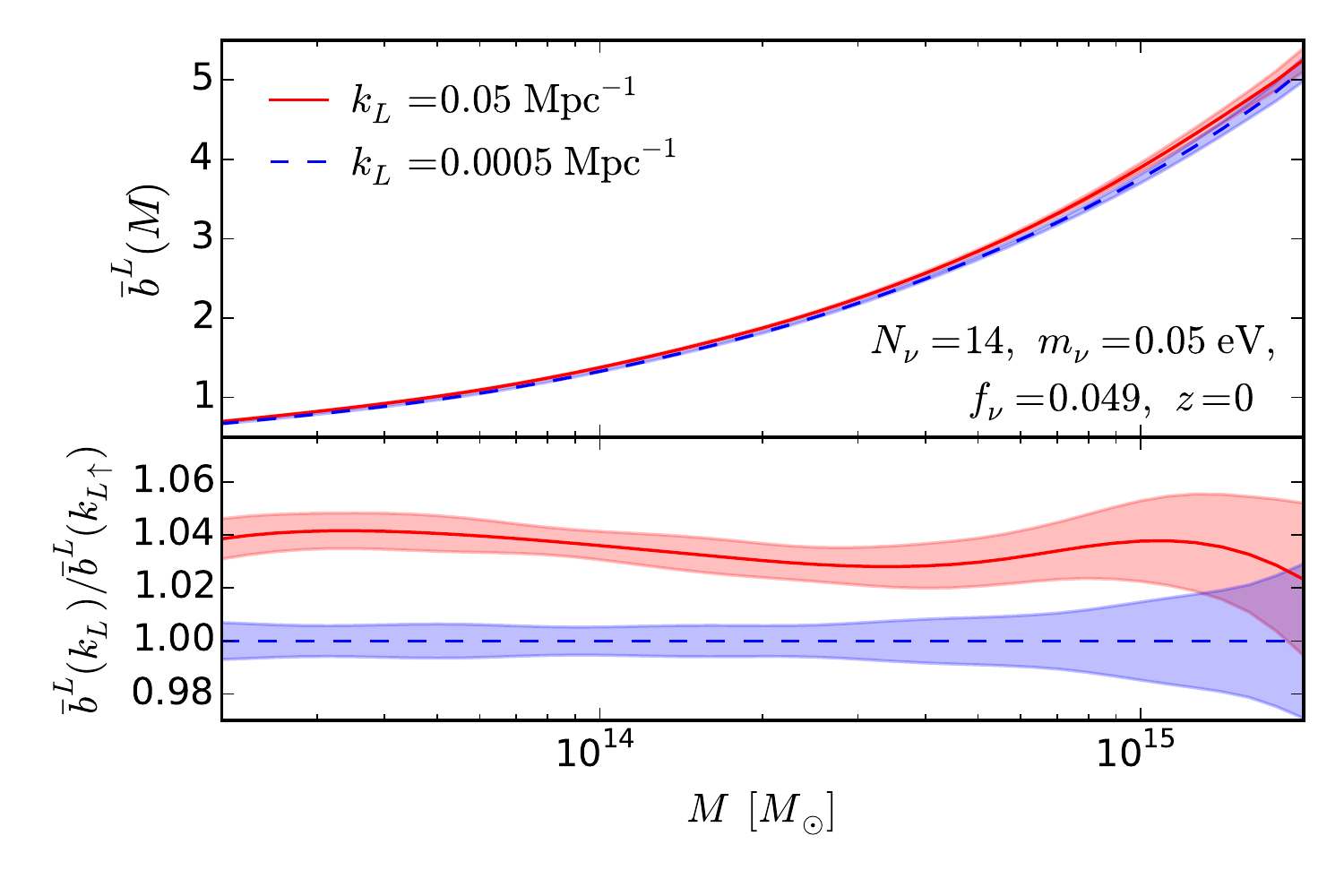} \\ [-3ex]
\includegraphics[width=0.497\textwidth]{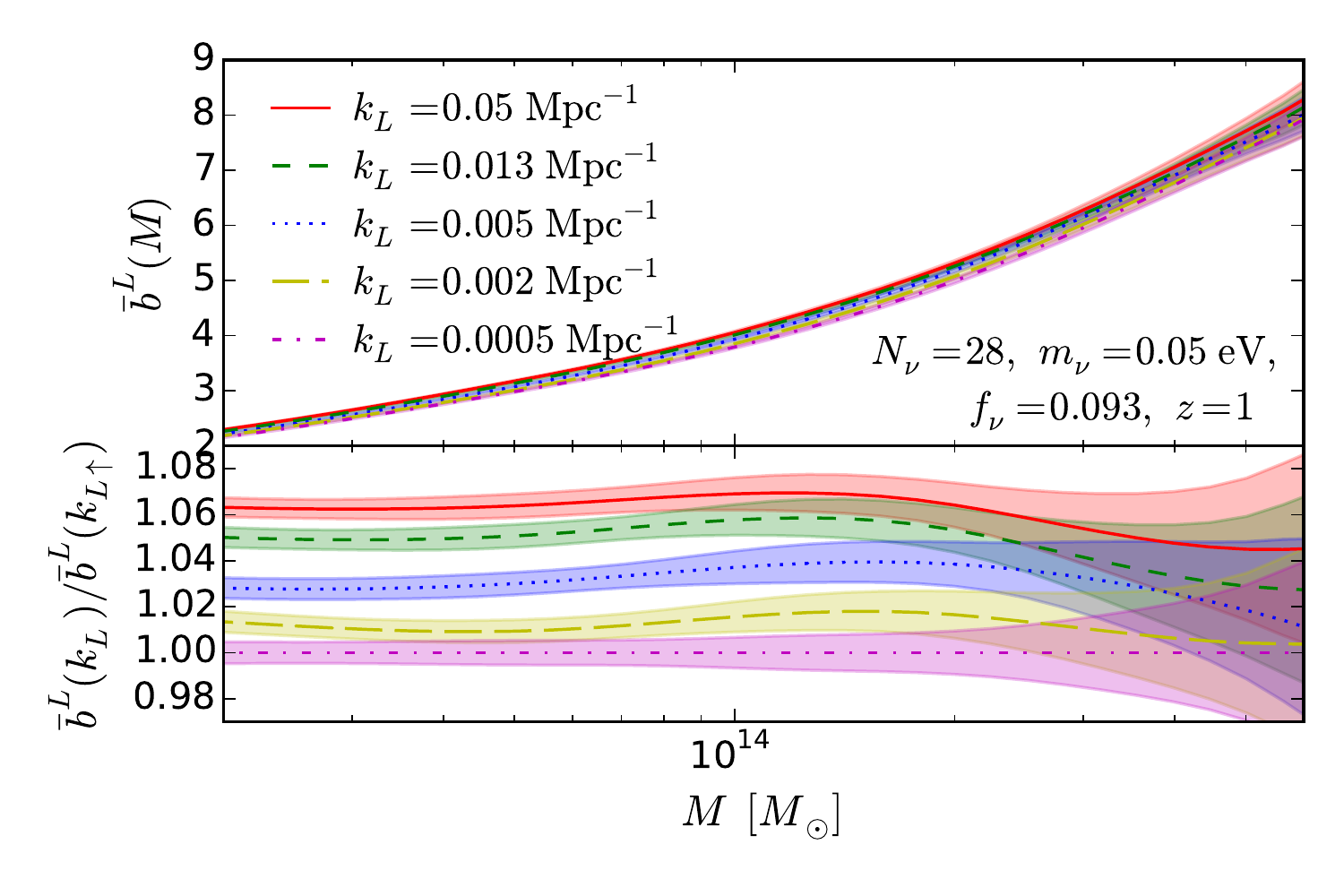}
\includegraphics[width=0.497\textwidth]{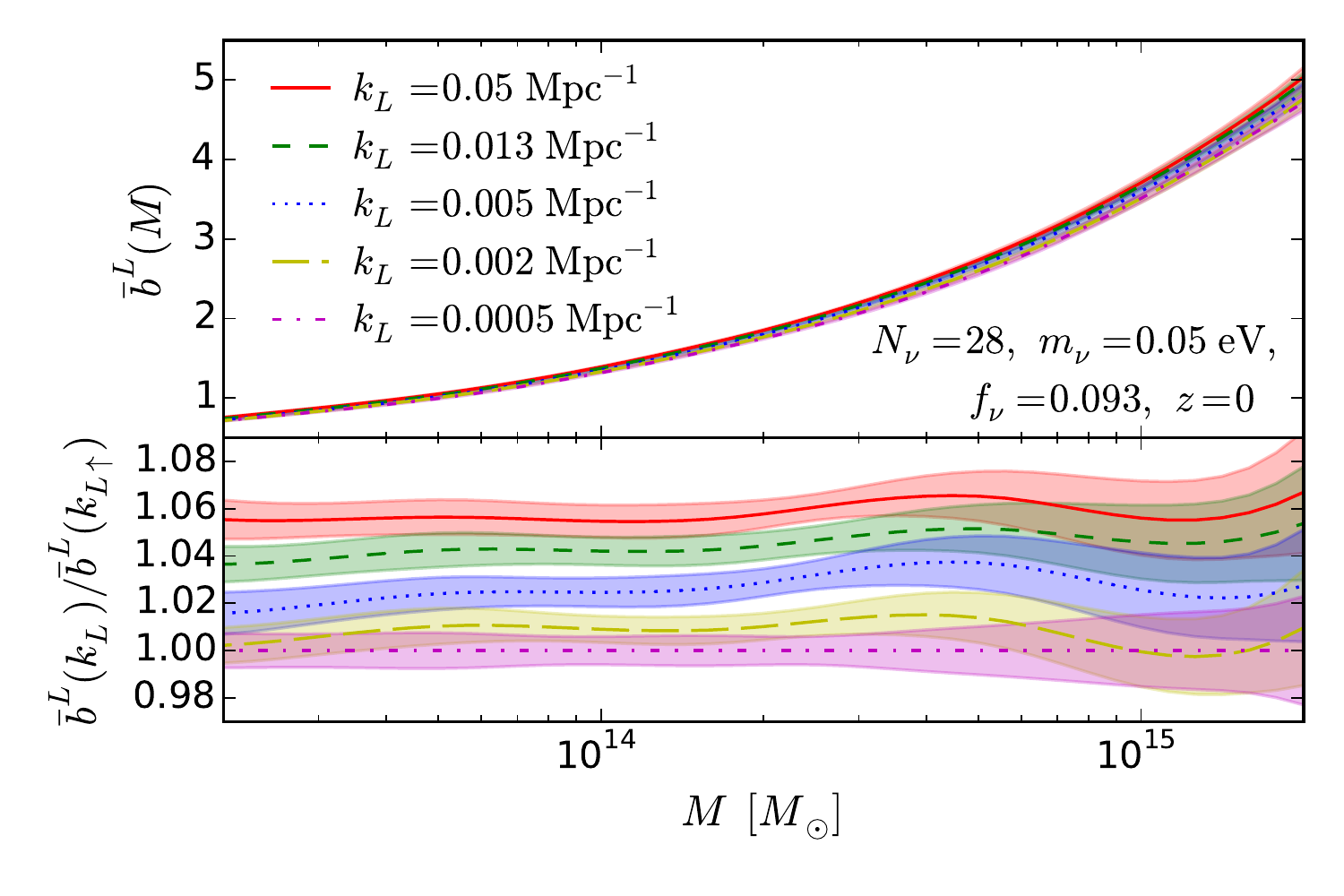}
\caption{Lagrangian response bias measured from 80 sets of neutrino separate
universe simulations as a function of halo mass. The top and bottom panels
show $N_\nu=14$ and 28 ($f_\nu=0.049$ and 0.093); the left and right panels
show $z=1$ and 0. In each panel, the top plot shows the response bias $\bar{b}^L(M)$,
and the bottom plot shows the ratios of the biases with respect to $k_{L\uparrow}=0.0005\iMpc$,
i.e.\ $\bar{b}^L(k_L)/\bar{b}^L(k_{L\uparrow})$. The lines and shaded region
show the smoothed estimate and the bootstrap error on the mean.}
\label{fig:bL_resp}
\end{figure*}

The response bias as measured from the SU simulations shows a statistically
significant scale dependence where the bias increases with $k_L$ and $N_\nu$
in a manner nearly independent of halo mass. \refFig{bL_resp} shows the Lagrangian
response bias measured from the 80 sets of neutrino SU simulations as a function
of halo mass. The top and bottom panels show $N_\nu=14$ and 28, whereas the
left and right panels show $z=1$ and 0. In each panel the top and bottom plots
show the response bias $\bar{b}^L(M)$ and the ratio $\bar{b}^L(k_L)/\bar{b}^L(k_{L\uparrow})$.
The lines and shaded region show the smoothed estimate and the bootstrap error.
We find that for a fixed $N_\nu$, the response bias systematically increases
with $k_L$, apparent especially for $N_\nu=28$ where the five values of $k_L$
more fully map out a larger net effect, indicating that $\bar{b}^L$ is indeed
scale dependent. 

The scale dependence increases with $N_\nu$ and does not evolve much between
$z=1$ and 0 or with mass, although the uncertainty increases at the high-mass
end due to rarity of such objects. Note that the mild oscillations in the ratio
for $N_\nu=14$ at $z=0$ are still statistically consistent with being independent
of halo mass due to the correlations inherent in our estimation technique.

\begin{figure*}[t]
\centering
\includegraphics[width=0.497\textwidth]{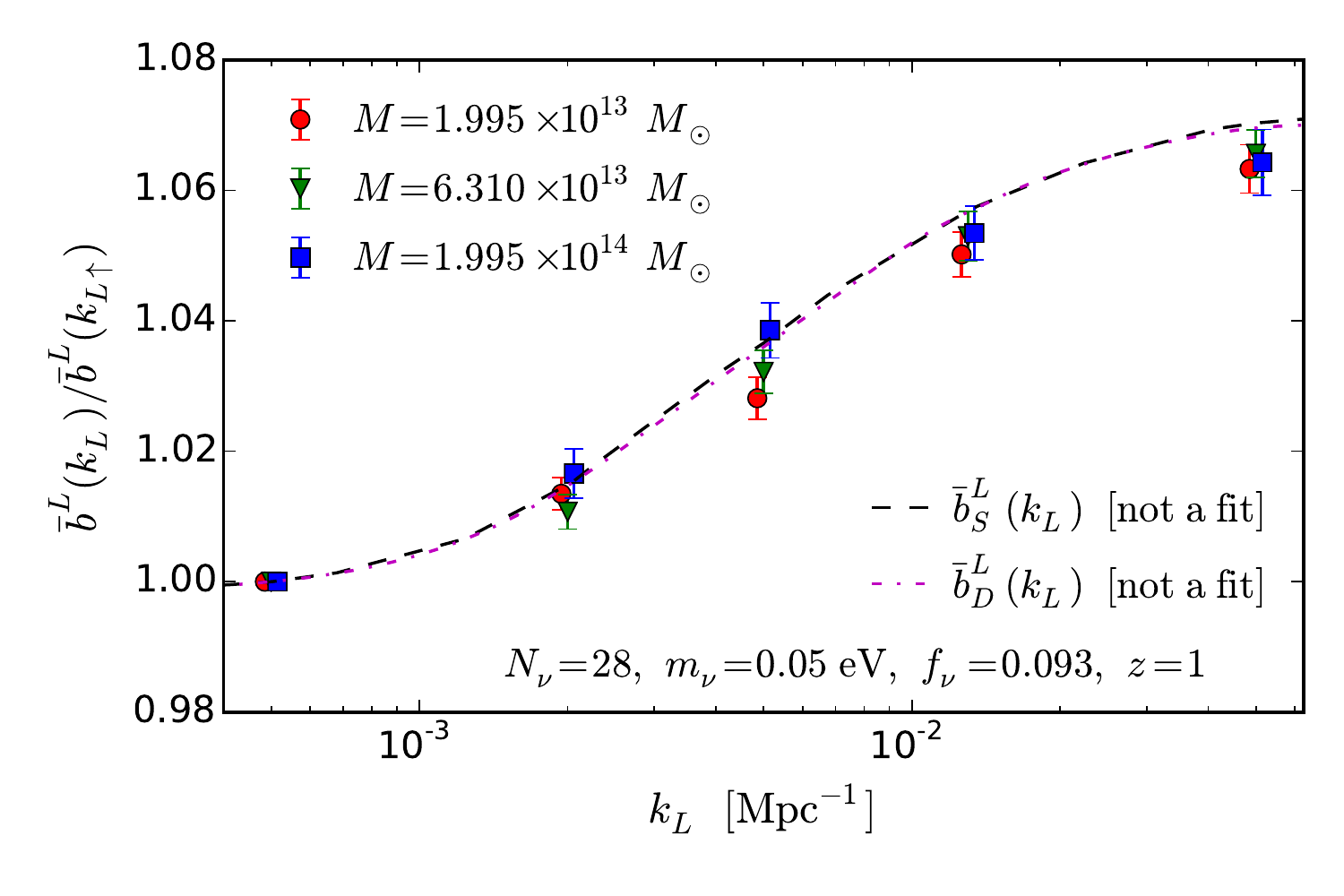}
\includegraphics[width=0.497\textwidth]{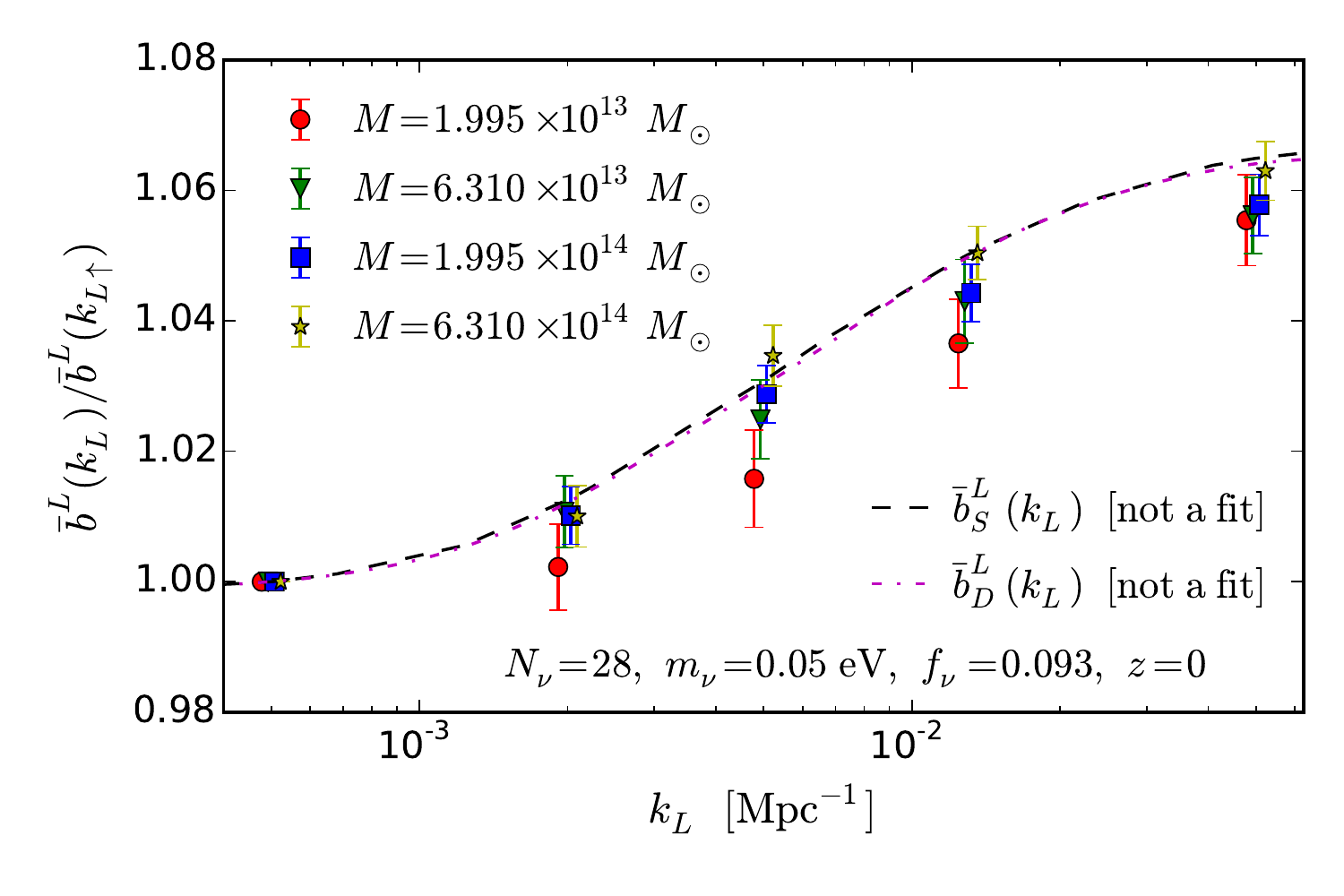}
\caption{Lagrangian response bias step at $z=1$ (left) and 0 (right) for $N_\nu=28$
($f_\nu=0.093$) as a function of the large-scale mode $k_L$. Different colored symbols
represent halo catalogs of different masses, with the error bars showing the bootstrap
error on the mean. The black dashed and magenta dot-dashed lines show the modeling
from spherical collapse ($\bar b_S^L$) and universal mass function ($\bar b_D^L$)
bias models (see detailed discussion in \refsec{b_model}).}
\label{fig:bL_step_kL}
\end{figure*}

\begin{figure*}[t]
\centering
\includegraphics[width=0.497\textwidth]{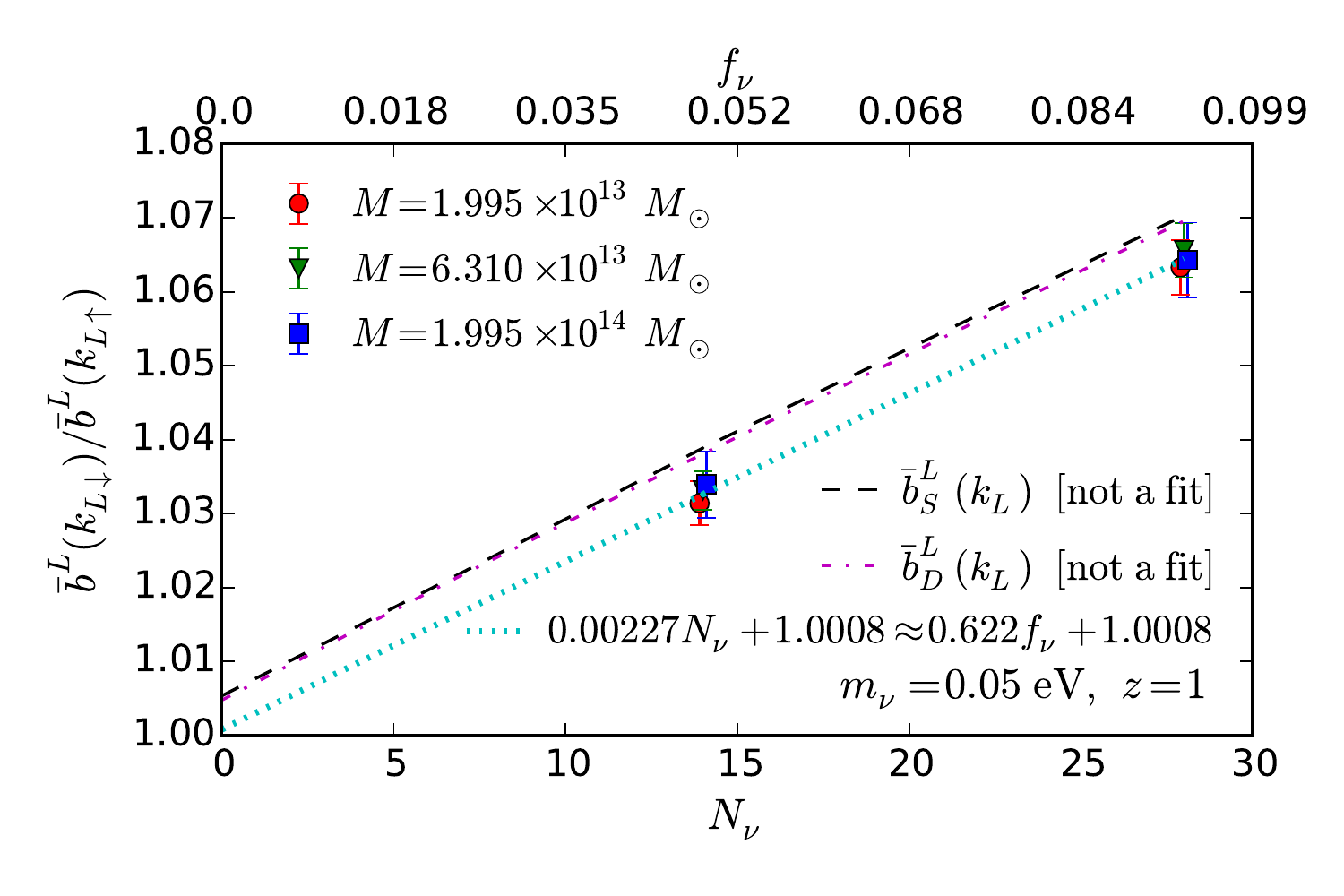}
\includegraphics[width=0.497\textwidth]{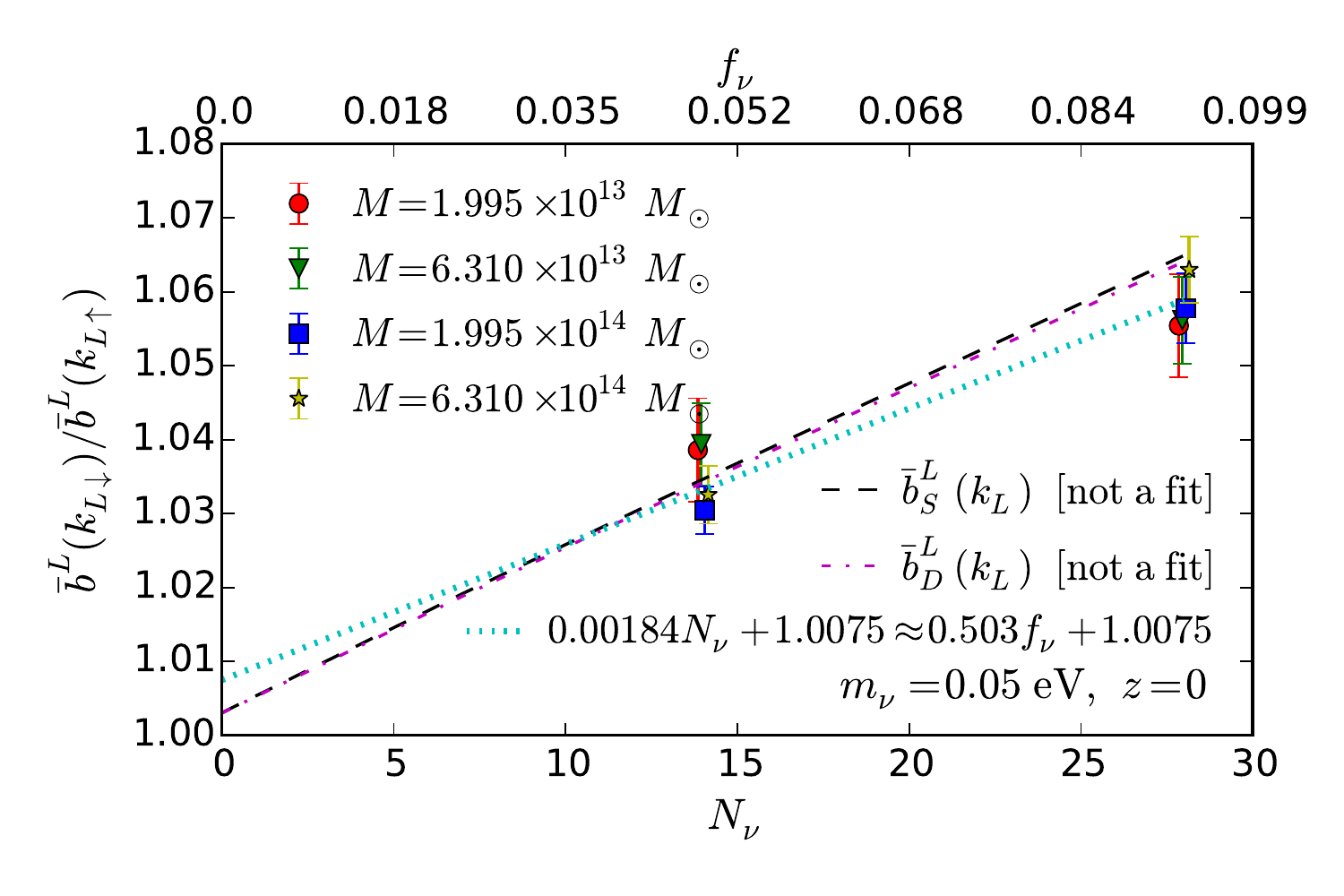}
\caption{Lagrangian response bias step between $k_{L\uparrow}=0.0005\iMpc$
and $k_{L\downarrow}=0.05\iMpc$ at $z=1$ (left) and 0 (right) as a function
of the number of massive neutrinos $N_\nu$, with the corresponding $f_\nu$
labeled on the top $x$-axis. Different colored symbols represent halo catalogs
of different masses, and the error bars show the bootstrap error on the mean.
The cyan dotted line is a fit to all the colored symbols with the slope and
the intercept shown in the legend, which provides a rough estimate of the
magnitude of the effect. The black dashed and magenta dot-dashed lines show
the predictions from spherical collapse ($\bar b_S^L$) and universal mass
function ($\bar b_D^L$) bias models (see detailed discussion in \refsec{b_model}).}
\label{fig:bL_step_Nnu}
\end{figure*}

In order to quantify the scale dependence of bias further, we first isolate the
scale dependence by taking the ratio with respect to the smallest wavenumber
in the simulations $k_{L\uparrow}=0.0005\iMpc$, i.e.
\be
 \bar{b}^L(M,k_L,N_\nu,a)/\bar{b}^L(M,k_{L\uparrow},N_\nu,a) \,,
\ee
for various choices of $M$. The uncertainty in this ratio becomes large for
high-mass halos, again due to their rarity, and for low-mass halos, likely
due to the inability of mass ordering to identifying the ``same'' halos in
the $\delta_{c0}=\pm0.01$ SU halo catalogs due to mergers and accretion. For
these reasons we take $M=2\times10^{13}$, $6.3\times10^{13}$, and $2\times10^{14}~M_\odot$
at $z=1$, and the same for $z=0$ with the addition of $6.3\times10^{14}~M_\odot$.
Note that since we measure the cumulative bias, $\bar{b}^L(M)$ of different
masses are correlated.

The colored symbols in \reffig{bL_step_kL} show the response bias step for $N_\nu=28$
as a function of the long mode $k_L$ for halos with different masses. The uncertainty
on the response bias is larger than that of the growth response, but a similar step-like
feature is still evident. Note that the error bar is for the ratio with respect to
$k_{L\uparrow}$, so by definition the uncertainly is zero at $k_{L\uparrow}$. This
step feature is at most weakly dependent on mass, but the dependence can also be a
consequence of the estimation technique, i.e.\ the mild oscillations of the ratio
of the response bias shown in \reffig{bL_resp}.

We next examine the $N_\nu$ dependence of the full amplitude of the step, as
quantified by the bias ratio between the largest and smallest wavenumbers i.e.
$\bar{b}^L(k_{L\downarrow},N_\nu)/\bar{b}^L(k_{L\uparrow},N_\nu)$, and the
result is shown in \reffig{bL_step_Nnu}. The results are consistent with a
linear dependence on $N_\nu$ that is weakly dependent on mass. To guide the
eye, we fit all the data points in \reffig{bL_step_Nnu} to a linear relation.
Note that this is not strictly correct due to the correlation of the cumulative
biases of halos with different mass cuts, but it suffices for a rough estimation.
The cyan dotted line in \reffig{bL_step_Nnu} shows the fit from all the colored
symbols, with the slope and intercept shown in the legend. Note that as in the
power spectrum response, the intercept at $N_\nu=0$ needs not vanish since the
photons can also produce scale-dependent bias. We shall discuss the interpretation
of these results next.

\subsection{Bias models}
\label{sec:b_model}
The behavior of scale-dependent bias uncovered in the SU simulations above
both illuminate the assumptions behind, and are illuminated by the predictions
of, various bias models. Here we consider how three types of models commonly
found in the literature can be extended to accommodate scale-dependent bias:
spherical collapse, universal mass function, and scale-free linear bias with
respect to multiple tracers. We call these $\bar{b}_S^L$, $\bar{b}_D^L$, and
$\bar{b}_T$ respectively.   

Scale-dependent spherical collapse bias is based on calculating the effect of
$\delta_c$ on the collapse of a spherical tophat overdensity in the SU and we implement
it here by assuming that the collapse depends on the CDM alone. Specifically,
the Lagrangian bias with respect to CDM is given by
\be
\bar b^L_S(M,k_L)=\frac{d\ln n(M)}{d\delta_c}
 =\frac{d\ln n(M)}{d\delta_{\rm crit}}\frac{d\delta_{\rm crit}}{d\delta_c}(k_L) \,,
\ee
where $\delta_{\rm crit}$ is the linearly extrapolated density threshold
for collapse and $d\ln n/d\delta_{\rm crit}$ depends only on the halo mass.
In the presence of massive neutrinos, the response of $\delta_{\rm crit}$
with respect to $\delta_c$ becomes scale dependent \cite{LoVerde:2014rxa,LoVerde:2014pxa},
and the scale dependence of the halo bias is entirely characterized by
$d\delta_{\rm crit}/d\delta_c$. In \refapp{sphericalcoll} we outline in
detail the setup for numerically evaluating $d\delta_{\rm crit}/d\delta_c$.
The resulting prediction for the step in the Lagrangian bias
\be
 \frac{\bar b^L_S(M,k_L)}{\bar b^L_S(M,k_{L\uparrow})}
 =\frac{\[d\delta_{\rm crit}/d\delta_c\](k_L)}{\[d\delta_{\rm crit}/d\delta_c\](k_{L\uparrow})} \,,
\label{eq:bS_step}
\ee
is shown as the black dashed line in \reffig{bL_step_kL}. Note that with
no free parameters, the spherical collapse model captures the main trends
of scale-dependent bias quite well. In particular this model predicts that
the scale dependence of the bias is independent of halo mass and helps
explain the weak dependence seen in the SU simulations.

Another simple model of bias which we call the universal mass function bias,
$\bar{b}_D^L$, comes from assuming that the mass function is a universal
functional of the power spectrum. In the SU picture, this should be the
local power spectrum $P_W$ so that
\be
 \bar{b}_D^L(M,k_L) = \frac{d \ln n(M;P_W)}{d \delta_c} (k_L)\propto \frac{d\ln D_W}{d \delta_c}(k_L) \,,
\label{eq:bD}
\ee
where the mass-dependent proportionality approximately holds regardless of whether
$P_W$ is considered to be the linear or nonlinear power spectrum as shown in
\refsec{pk_resp}. Like the spherical collapse model, the step of $\bar{b}_D^L$
does not contain free parameters, i.e.
\be
 \frac{\bar{b}_D^L(M,k_L)}{\bar{b}_D^L(M,k_{L\uparrow})}
 =\frac{\[d\ln D_W/d\delta_c\](k_L)}{\[d\ln D_W/d\delta_c\](k_{L\uparrow})} \,.
\label{eq:bD_step}
\ee
In this model, the scale dependence of bias is directly inherited from the
scale dependence of the power spectrum response, which itself is a proxy
for the response of small-scale structure responsible for halo formation.
It therefore plays a role similar to $\delta_{\rm crit}$ in the spherical
collapse model and indeed we find in \reffig{bL_step_kL} that the growth
response (magenta dot-dashed line) and spherical collapse predictions are
similar
\be
 \frac{d\ln D_W}{d \delta_c}(k_L) \approx \frac{d\delta_{\rm crit}}{d\delta_c}(k_L) \,,
\ee
and describe the SU simulations results equally well.

\begin{figure*}[t]
\centering
\includegraphics[width=0.497\textwidth]{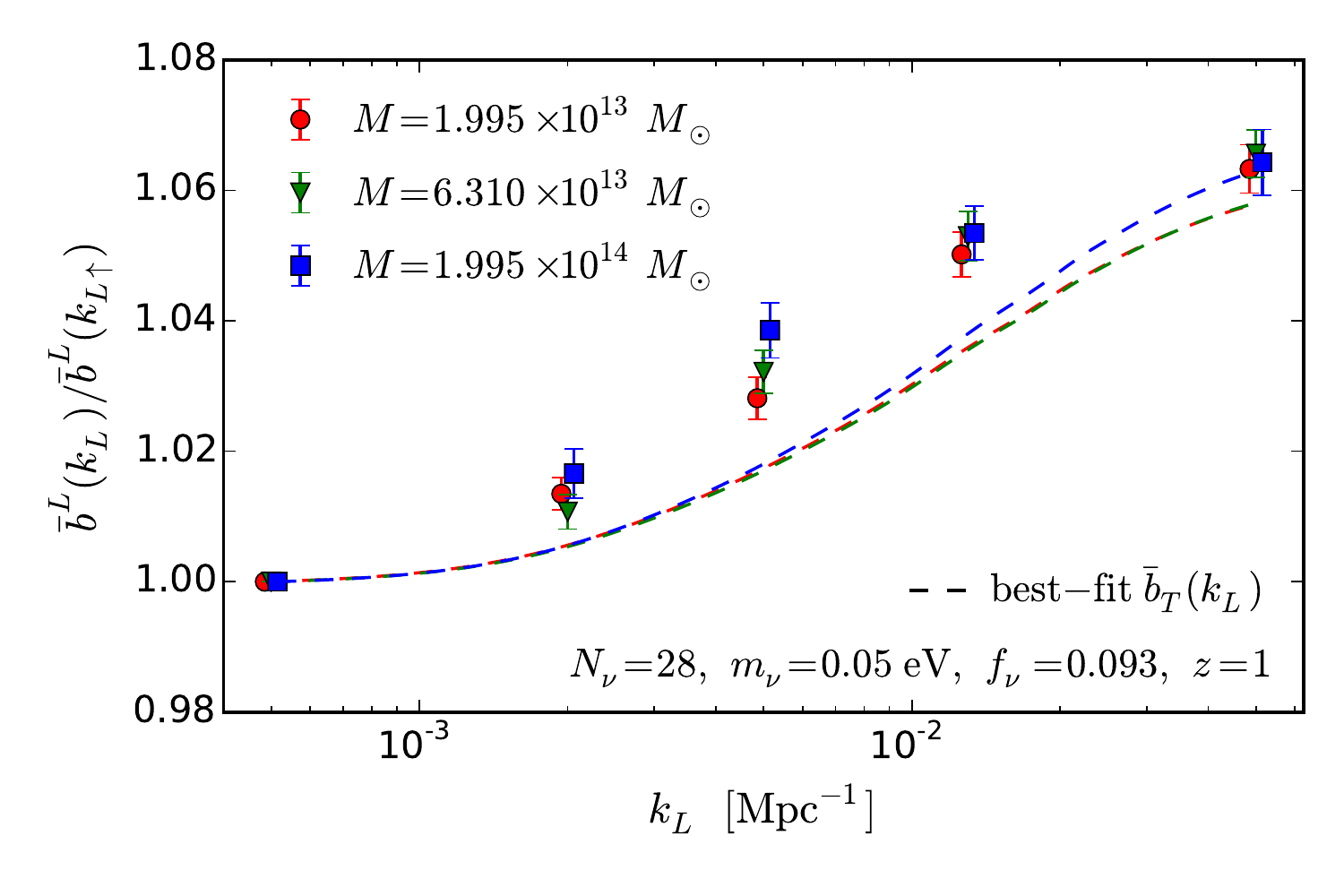}
\includegraphics[width=0.497\textwidth]{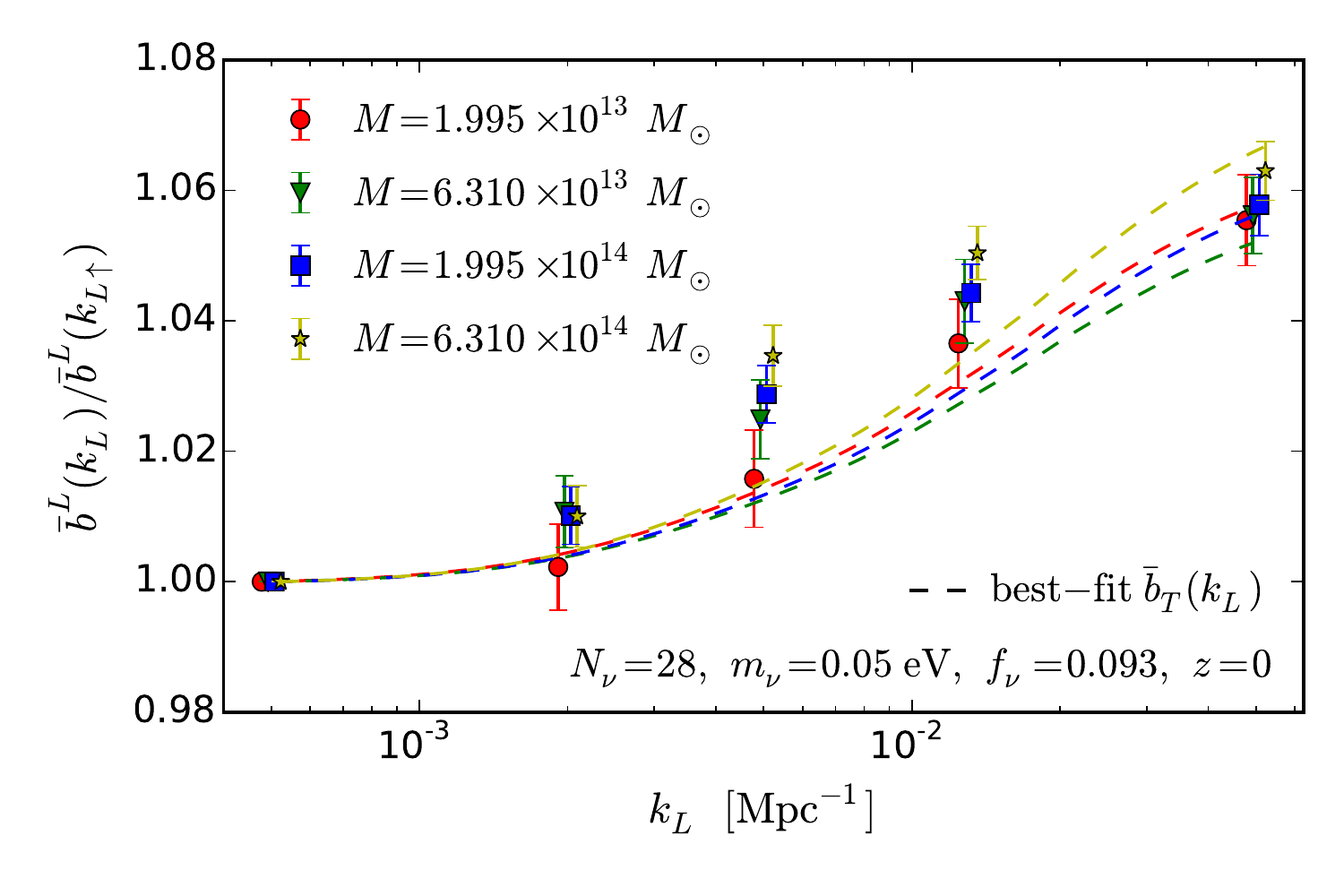}
\caption{Comparison of the best-fit transfer function bias model (dashed lines;
see the text for detailed description of the model) at $z=1$ (left) and 0 (right)
to the measured Lagrangian response bias step from neutrino separate universe
simulations as in \reffig{bL_resp}. Note that the data points of a given mass
are highly correlated in $k_L$ due to the same random realizations used in separate
universe simulations which forbids the large change in slope that would be required
to move the curves towards the central points. See \reffig{bL_step3}, which displays
the same fitting result with less correlation by taking the difference of neighboring
data points, for a better visual representation.}
\label{fig:bL_step_bT}
\end{figure*}

The third type is based on the assumption that bias is temporally and spatially
local, and hence scale free, but that halos can be biased with respect to multiple
species of matter separately, in this case CDM and neutrinos $\delta_h=b_c\delta_c+b_\nu\delta_\nu$.
Therefore, the Eulerian bias with respect to CDM, i.e.\ $\delta_h(M, k_L)=\bar{b}_T(M,k_L)\delta_c(k_L)$,
can be written as
\be
 \bar{b}_T(M,k_L)=b_c(M)+b_\nu(M)\frac{T_\nu(k_L)}{T_c(k_L)} \,,
\label{eq:bT}
\ee
where $b_c$ and $b_\nu$ are bias parameters depending on halo mass, and the
scale dependence is encoded in the neutrino and CDM+baryon transfer functions,
which are computed using \texttt{CLASS}. To compare with the step of the Lagrangian
bias, the transfer function bias can be written as
\be
 \frac{\bar b_T(M,k_L)-1}{\bar b_T(M,k_{L\uparrow})-1}
 =\frac{\tilde{b}_c+T_\nu(k_L)/T_c(k_L)}{\tilde{b}_c+T_\nu(k_{L\uparrow})/T_c(k_{L\uparrow})} \,,
\label{eq:bT_step}
\ee
where $\tilde{b}_c=(b_c-1)/b_\nu$. Therefore, the step of the transfer
function bias model has one fitting parameter $\tilde{b}_c$ which controls
the amplitude of the scale dependence but crucially the shape cannot be
adjusted. The $\bar{b}_T$ model with the best-fit $\tilde{b}_c$ is shown
in \reffig{bL_step_bT} and is a poor fit to the simulations especially at
$z=1$. Note that despite having an adjustable amplitude parameter the best
fit systematically undershoots the step results because of their strong constraint
on the shape of the step. The large correlation in the data points means
that the slope of the step is much better constrained than $\chi$-by-eye
would suggest.

\begin{figure*}[t]
\centering
\includegraphics[width=0.497\textwidth]{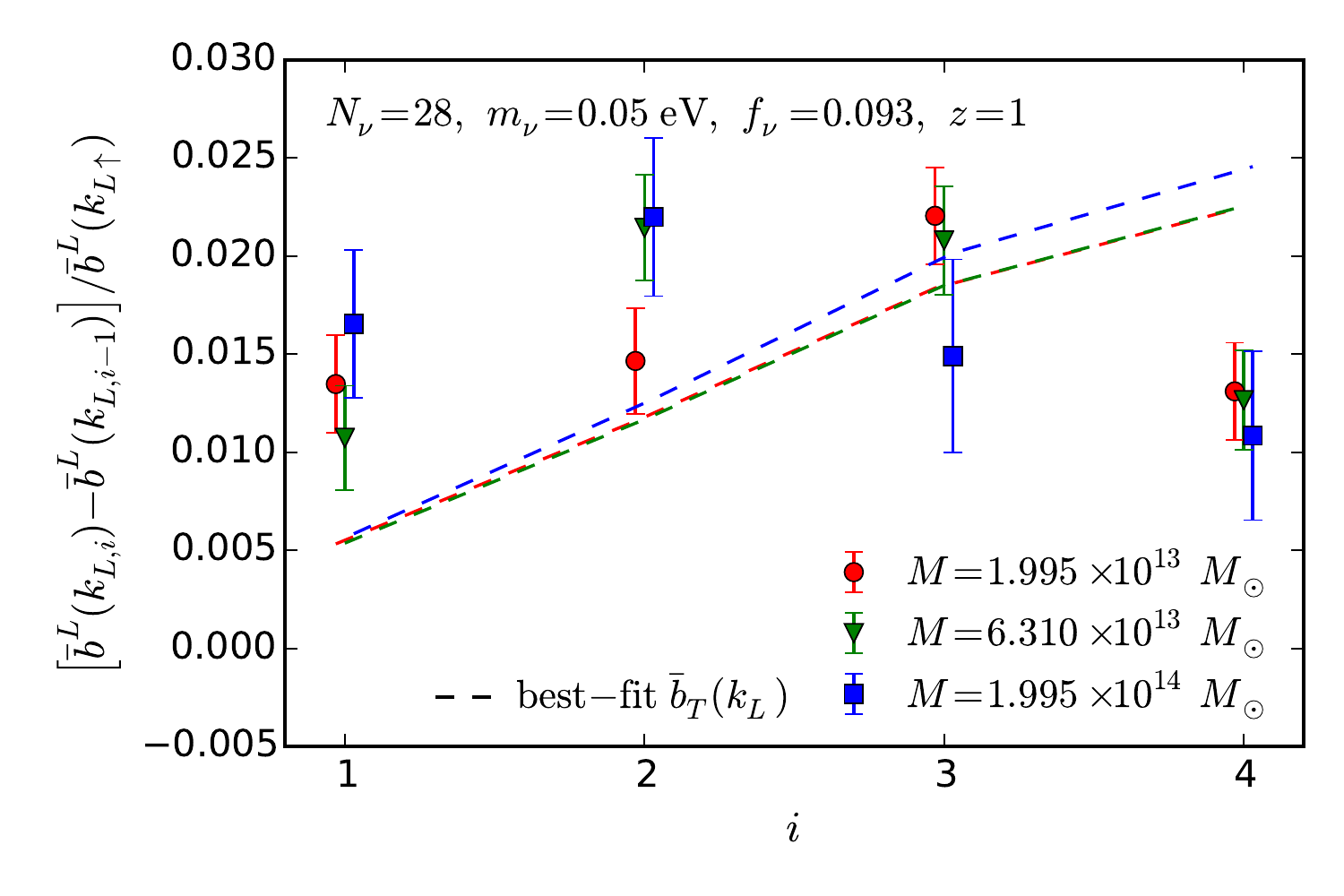}
\includegraphics[width=0.497\textwidth]{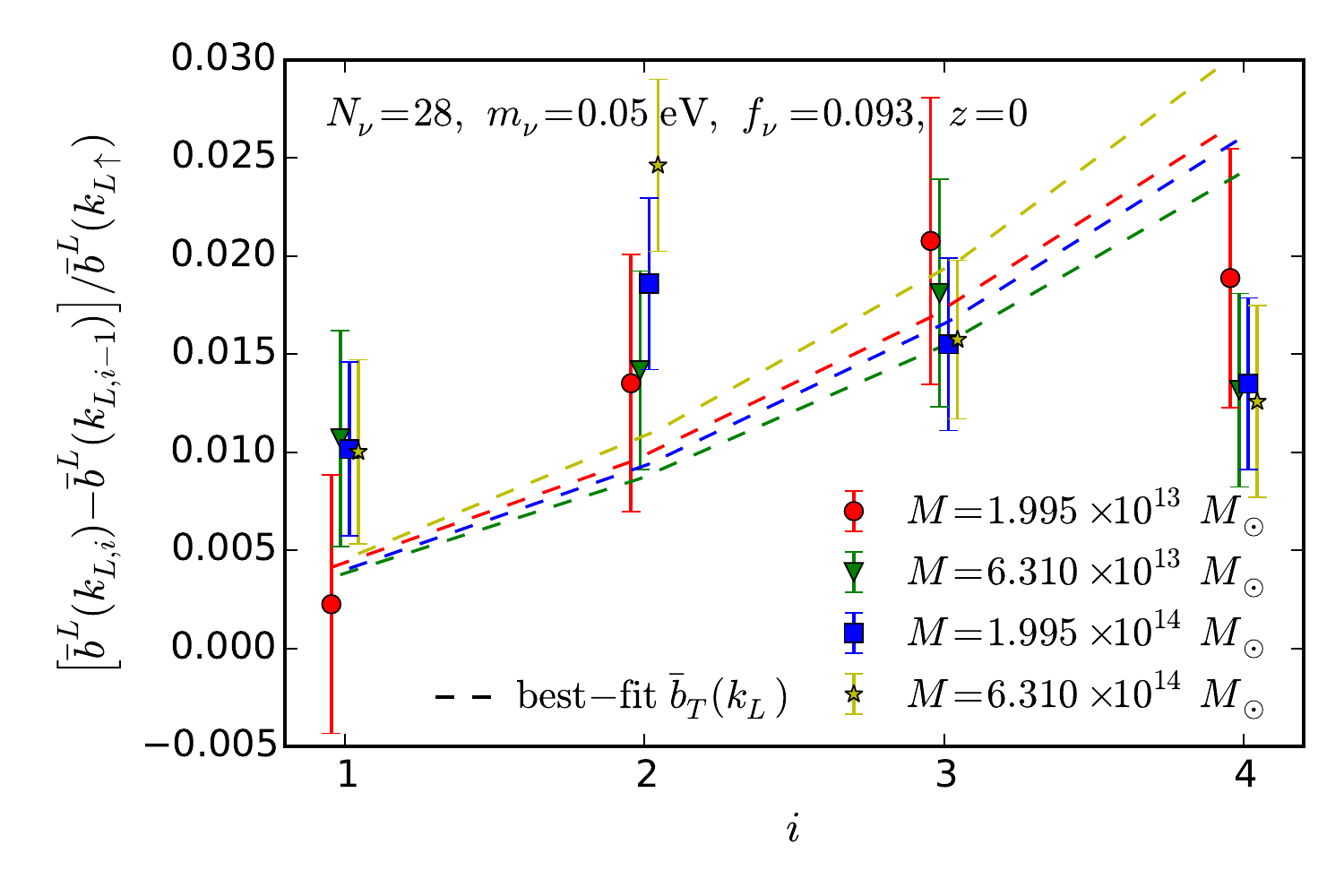} \\ [-3ex]
\includegraphics[width=0.497\textwidth]{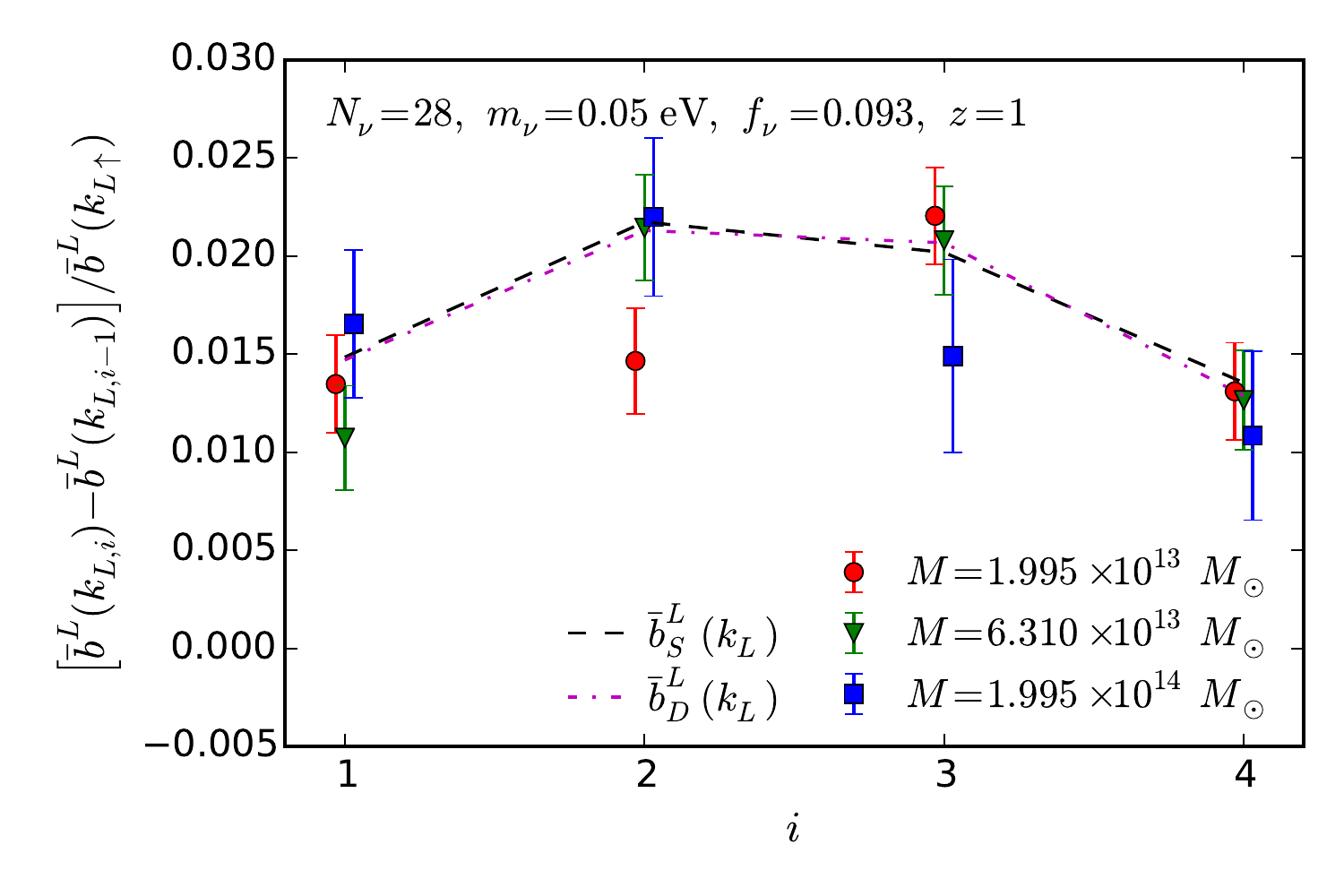}
\includegraphics[width=0.497\textwidth]{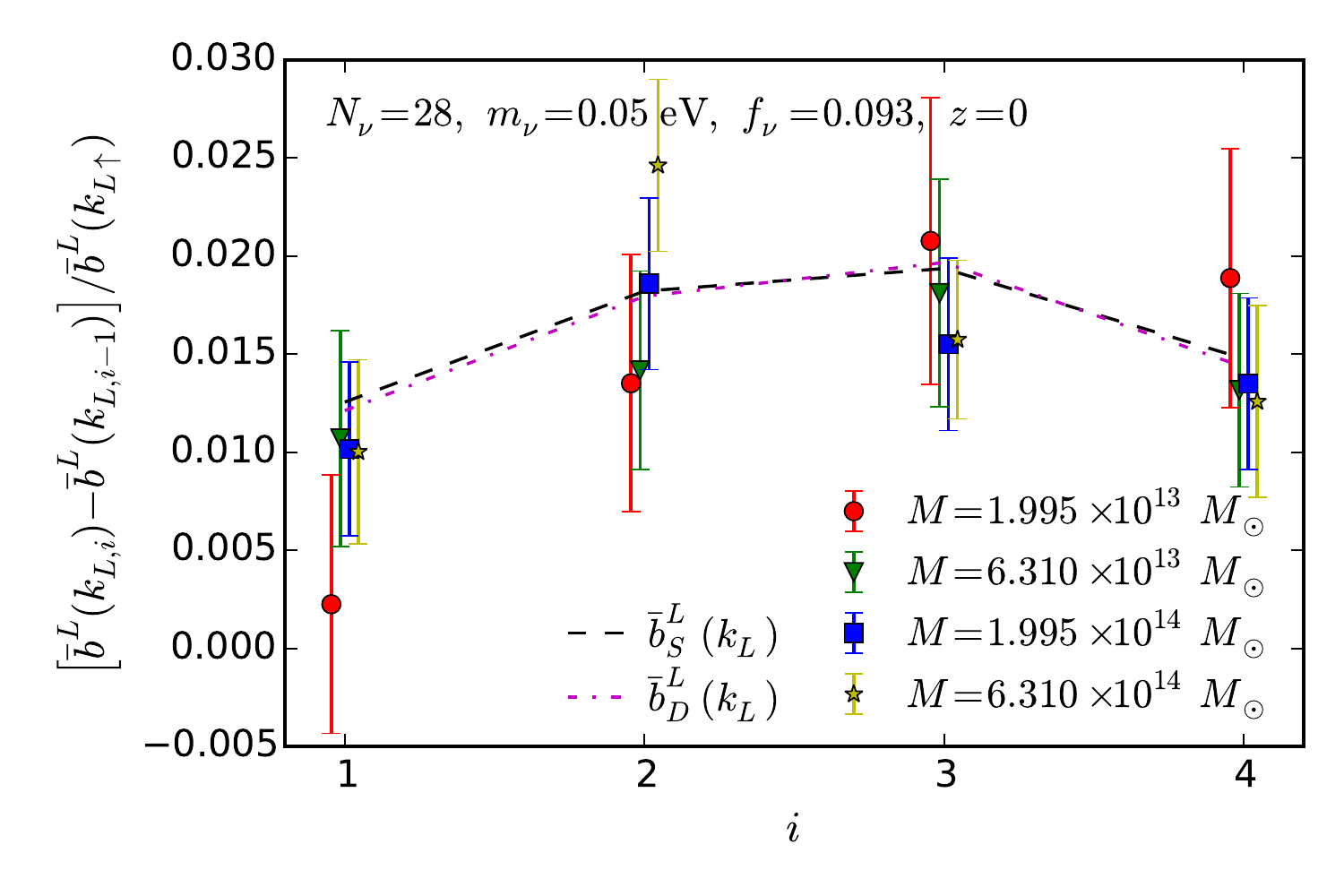}
\caption{Difference of the Lagrangian response bias step between neighboring
points, i.e.\ $\[\bar{b}^L(k_{L,i})-\bar{b}^L(k_{L,i-1})\]/\bar{b}^L(k_{L\uparrow})$,
where $i=0,\cdots,4$ corresponds to $k_{L,i}=0.0005$, 0.002, 0.005, 0.013, and
$0.05\iMpc$, at $z=1$ (left) and 0 (right). The top panel show the comparison
between the measurement from the neutrino SU simulations and the best-fit transfer
function bias model ($\bar{b}_T$); the bottom panel shows the comparison with
the spherical collapse ($\bar{b}^L_S$) and universal mass function bias models
($\bar{b}^L_D$). Taking the difference of the neighboring points largely reduces
the correlation between data points and exposes, e.g.\ the strong constraint
at $i=4$ and the cause of the poor $\bar{b}_T$ fit in \reffig{bL_step_bT}.}
\label{fig:bL_step3}
\end{figure*}

To obtain a better visual representation of the problem, we take the the difference
of the Lagrangian response bias step between neighboring points, i.e.\
$\[\bar{b}^L(k_{L,i})-\bar{b}^L(k_{L,i-1})\]/\bar{b}^L(k_{L\uparrow})$, where
$i=0\cdots4$ corresponds to $k_{L,i}=0.0005$, 0.002, 0.005, 0.013, and $0.05\iMpc$.
This largely reduces the correlation between data points (the typical absolute
values of the correlation drops from 0.6 to 0.2), while keeping the fitting
unchanged. The top panel in \reffig{bL_step3} shows the comparison at $z=1$
(left) and 0 (right) between the measurement and the $\bar{b}_T(k_L)$ model,
and the bottom panel shows the comparison with the $\bar{b}_S^L(k_L)$ and
$\bar{b}_D^L(k_L)$ models. We can see that the problem with $\bar{b}_T$ is
that the slope monotonically rises with $k_L$ whereas the data prefer a decline
by the highest point for most masses and redshifts. This problem does not
occur for $\bar{b}_S^L$ and $\bar{b}_D^L$ which have the right shape for
the scale dependence.   

We can further quantify how well the models fit the measurement including the
covariance between points by computing
\begin{eqnarray}
 \chi^2&=&\sum_{ij}\[C^{-1}\]_{ij}\[D(k_{L,i})-M(k_{L,i})\]\nonumber\\
 && \times \[D(k_{L,i})-M(k_{L,i})\] \,,
\end{eqnarray}
where $i,j$ run over the five values of $k_L$,
$D(k_{L,i})=\langle \hat{\bar{b}}^L (k_{L,i})/\hat{\bar{b}}^L (k_{L\uparrow}) \rangle$
is the mean of the measured Lagrangian response bias step, $C^{-1}$ is the inverse
covariance matrix of $D$, and $M$ is the model of $\bar{b}_S^L$, $\bar{b}_D^L$,
and $\bar{b}_T$. Since we adopt the same Gaussian random realizations for setting
up the initial conditions of SU simulations with different $k_L$, $D$ of different
$k_L$ are highly correlated and so it is necessary to use the full covariance matrix
to unbiasedly compute $\chi^2$. We estimate the covariance matrix by bootstrap
resampling the identical realizations for all five $k_L$ and repeating the procedure
8,000 times. By definition the first data point of the Lagrangian response bias
step has zero variance and no contribution to $\chi^2$  (see \reffig{bL_step_kL}),
so we only consider the rest of the four $k_L$. For $\bar{b}_T(k_L)$, there is one
fitting parameter so the degrees of freedom (d.o.f.) is three; $\bar{b}_S^L(k_L)$
and $\bar{b}_D^L(k_L)$ do not contain fitting parameters so d.o.f. is four.

\begin{table}
\centering
\begin{tabular}{c | c | c | c | c }
\hhline{=====}
$~z~$ & $~M\ [M_\odot]~$ & $~\bar{b}_S^L(k)~$ & $~\bar{b}_D^L(k)~$ & $~\bar{b}_T(k)~$ \\
\hline
1 & $2.0\times10^{13}$ & 2.17 & 1.89 & 9.92 \\
  & $6.3\times10^{13}$ & 0.74 & 0.61 & 12.6 \\
  & $2.0\times10^{14}$ & 0.86 & 0.81 & 14.7 \\
\hline
0 & $2.0\times10^{13}$ & 1.37 & 1.28 & 0.77 \\
  & $6.3\times10^{13}$ & 0.78 & 0.65 & 2.64 \\
  & $2.0\times10^{14}$ & 0.85 & 0.75 & 7.19 \\
  & $6.3\times10^{14}$ & 0.64 & 0.66 & 9.80 \\
\hhline{=====}
\end{tabular}
\caption{Summary of the reduced $\chi^2$ ($\chi^2$/d.o.f.) for the three bias
models. Both $\bar{b}_S^L(k)$ and $\bar{b}_D^L(k)$ have d.o.f.=4, whereas
$\bar{b}_T(k)$ has d.o.f.=3.}
\label{tab:fit}
\end{table}

\refTab{fit} summarizes the reduced $\chi^2$ ($\chi^2$/d.o.f.) for the three
bias models. We find that even with one additional fitting parameter, $\bar{b}_T(k_L)$
is generally a very poor fit, especially for halos with high bias, i.e.\ at high
redshift and with high mass. It only works well for halos with the lowest mass
at $z=0$. We can therefore rule out transfer function bias as a model for the
scale-dependent bias of the SU simulations with high confidence. On the other
hand, $\bar{b}_S^L(k_L)$ and $\bar{b}_D^L(k_L)$ give reasonable values of reduced
$\chi^2$ except for halos with mass $2\times10^{13}~M_\odot$ at $z=1$. Note that
the bootstrap construction of the covariance is a noisy estimate so even this
case is not necessarily significant. In \reffig{bL_step_Nnu}, we also show that
$\bar{b}_S^L$ and $\bar{b}_D^L$ predict fairly well the linear trend of the step
amplitude with $N_\nu$.

The transfer function model fails because it has a transition centered at higher
$k_L$ than the data requires. The transfer function ratio represents the ratio
of CDM and neutrino linear density fluctuations at the observed redshift. Since
the free-streaming scale, above which the two are the same, decreases with time
this model underestimates the effect of scale-dependent growth at earlier times.
On the other hand, both the spherical collapse and universal mass function models
automatically incorporate the whole past growth history of the long-wavelength mode.
This distinction is even more dramatic at $a \gg 1$ when the transfer functions
coincide across the whole linear regime so that $\bar{b}_T$ predicts no scale-dependent
bias. In the spherical collapse and universal mass function models, once the scale
dependence is imprinted in the Lagrangian bias, it remains although in the Eulerian
bias it can only be measured for sufficiently rare, high-mass halos where the Lagrangian
bias is large compared to unity.

More generally, the failure of the transfer function model indicates that halo bias
is {\it temporally nonlocal} \cite{Hui:2007zh,Parfrey:2010uy,LoVerde:2014pxa,LoVerde:2014rxa,Senatore:2014eva}.
It is not sufficient to know the properties of the long-wavelength perturbation at
just the observation epoch. To make a precise prediction of the scale-dependent bias,
one needs to know the whole growth history of $\delta_c(a)$.

\section{Effects on observables}
\label{sec:observables}

Because of the scale-dependent responses, the small-scale observables are affected
correspondingly. In this section we study how the linear halo power spectrum and
squeezed-limit bispectrum are influenced.

Let us start with the linear halo power spectrum. In \refsec{nh_resp}, we have shown
that the Lagrangian bias follows to good approximation the growth response $d\ln D_W/d\delta_c$
as a function of $k_L$ which in turn takes the form of a step across the neutrino
free-streaming scale of amplitude
\be
 \frac{\bar{b}^L(k_{L\downarrow})}{\bar{b}^L(k_{L\uparrow})}= 1+ A f_\nu +B \,,
 \label{eq:fit}
\ee
with $A \approx 0.6$ and $B\approx0.003$ calculated from $d\ln D_W/d\delta_c$ at
$z=0$. This allows us to extrapolate the simulation results which have an unrealistic
$N_\nu=14$ and 28 to more relevant lower values. In this section we choose $m_\nu=0.05\eV$,
$N_\nu=3$, $\Omega_b=0.05$, and $\Omega_c=0.25$, which corresponds to $f_\nu=0.011$.
Note that the crude empirical fit to the response bias results give slightly different
values of $A\approx 0.5$ and $B\approx 0.0075$, predicting a larger effect in the
untested but cosmologically relevant region of $N_\nu\lesssim 10$. A finite $B$
represents the scale-dependent bias from the photons rather than neutrinos. To be
conservative, when displaying results in this section we set $A=0.6$ and $B=0$ to
focus on the neutrino induced bias effects, though we leave our expressions general.
Also note that the scaling with $f_\nu$ in \refeq{fit} is at fixed $m_\nu$ and varying
$N_\nu$. This should not be conflated with fixed $N_\nu$ and varying $m_{\nu}$, which
would change the free-streaming scale.

\begin{figure}[h]
\centering
\includegraphics[width=0.498\textwidth]{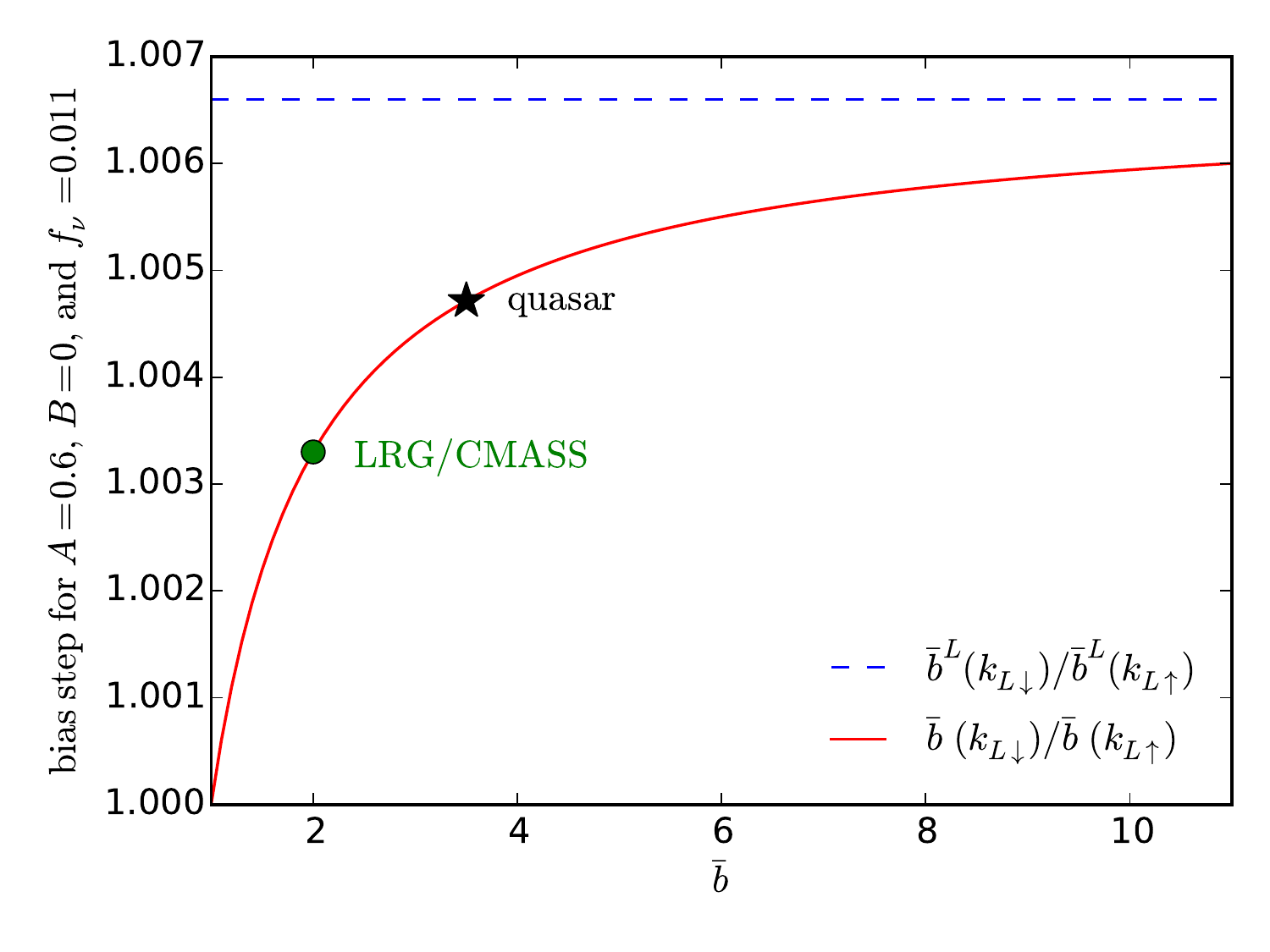}
\caption{Step in bias between $k_{L\uparrow}=0.0005\iMpc$ and $k_{L\downarrow}=0.05\iMpc$
as a function of Eulerian bias $\bar{b}$ (evaluated at $k_{L\uparrow}$) for $A=0.6$,
$B=0$, and $f_\nu=0.011$. The blue dashed and red solid lines show the Lagrangian
and Eulerian bias, respectively. The green circle and black star correspond respectively
to $\bar{b}=2$ (LRG/CMASS) and 3.5 (quasar).}
\label{fig:bE_step}
\end{figure}

The observed halo bias is in Eulerian space, so we convert the step amplitude  in
the Lagrangian bias to equivalent in the Eulerian bias
\ba
 \frac{\bar{b}(k_{L\downarrow})}{\bar{b}(k_{L\uparrow})}\:&=\frac{\bar{b}^L(k_{L\downarrow})+1}{\bar{b}^L(k_{L\uparrow})+1}
 =\left(1-\frac{1}{\bar{b}}\right)\frac{\bar{b}^L(k_{L\downarrow})}{\bar{b}^L(k_{L\uparrow})}+\frac{1}{\bar{b}} \vs
 \:&=\left(1-\frac{1}{\bar{b}}\right)(1+ A f_\nu +B)+\frac{1}{\bar{b}} \,,
\label{eq:bE_step}
\ea
where $\bar{b}$ is the Eulerian bias evaluated as $k_{L\uparrow}$. \refFig{bE_step}
shows the Eulerian (red solid) and Lagrangian (blue dashed) bias step as a function
of $\bar{b}$ for $A=0.6$, $B=0$, and $f_\nu=0.011$. Unlike the Lagrangian bias step,
the Eulerian bias step shows a strong dependence on the value of the bias and hence
halo mass. For $\bar{b}=1$, the Lagrangian bias is zero, so the scale dependence
vanishes; for $\bar{b}\gg1$, the Eulerian bias step approaches to the Lagrangian
bias step, meaning the impact of the scale-dependent bias on the clustering of
halos is most significant in high bias, high mass objects for $\bar{b} \ge 1$.
The green circle and black star correspond to objects of $\bar{b}=2$ (LRG/CMASS
\cite{Marin:2010iv,Gil-Marin:2014sta}) and 3.5 (quasar \cite{Eftekharzadeh:2015ywa}).
While the net effect for $f_\nu=0.011$ is quite small, we shall see next that
it has a significant effect on the linear halo power spectrum relative to the
also small step in the linear CDM power spectrum.

\begin{figure}[h]
\centering
\includegraphics[width=0.498\textwidth]{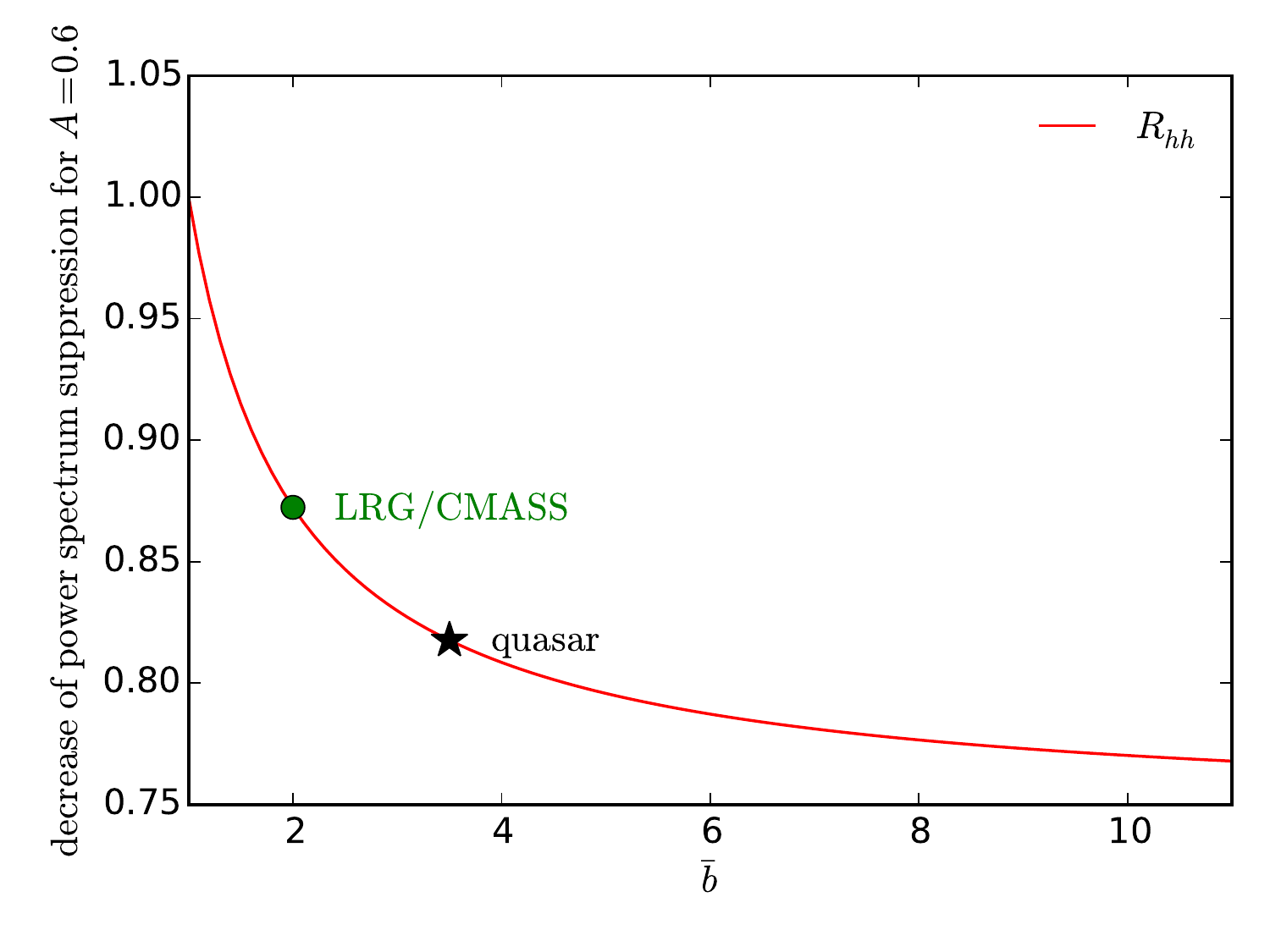}
\caption{Decrease of linear halo power spectrum suppression due to neutrino scale-dependent
bias at $k_{L\downarrow}$, i.e. $R_{hh}$, as a function of $\bar{b}$ for $A=0.6$. The
green circle and black star correspond to $\bar{b}=2$ (LRG/CMASS) and 3.5 (quasar).}
\label{fig:Rhh}
\end{figure}

Since our bias is defined with respect to the underlying CDM(+baryon) field
instead of total matter (including massive neutrinos), the linear halo power
spectrum is given by $P_{hh}=\bar b^2P_{cc}$, where $P_{cc}(k)$ is the linear
CDM power spectrum. Neutrinos suppress the growth of CDM perturbations and
produce a downward step in its power spectrum with respect to the $m_\nu=0$
model with the same total nonrelativistic matter. The amplitude of the full
step is approximately $\Delta P_{cc}/P_{cc} \approx-6f_\nu$\footnote{The commonly
quoted empirical relation $-8f_\nu$ \cite{Hu:1997mj} is for the total matter,
i.e.\ $\Delta P_{mm}/P_{mm}$. Since below the free-streaming scale $P_{mm}\approx(1-2f_\nu)P_{cc}$,
we have $\Delta P_{cc}/P_{cc} \approx-6f_\nu$.}, but as we have seen in \refsec{nh_resp}
the step in the CDM transfer function is not fully complete by $k_{L\downarrow}$
where the step in bias is essentially complete. For an accurate comparison, we
use \texttt{CLASS} to compute the linear power spectrum for $N_\nu=3$ massless
neutrinos with $\Omega_c=0.253$, and find
\be
 \frac{P_{cc}(k_{L\downarrow})}{P_{cc}^{m_\nu=0}(k_{L\downarrow})}\approx 1-4.7f_\nu \,.
\ee
Using \refeq{bE_step} and linearizing in $f_\nu$, we find the suppression in
linear halo power spectrum becomes
\be
 \frac{P_{hh}(k_{L\downarrow})}{P_{hh}^{m_\nu=0}(k_{L\downarrow})}
 \approx 1 - 4.7 R_{hh}(\bar{b}) f_\nu+2\left( 1-\frac{1}{\bar{b}} \right) B \,,
\ee
where
\be
 R_{hh}(\bar{b}) = 1 - \frac{2 A}{4.7} \left(1-\frac{1}{\bar{b}}\right) \,.
\ee
The decrease of the linear halo power spectrum suppression relative to CDM due to 
neutrino induced scale-dependent bias is captured by $R_{hh}(\bar{b})$. \refFig{Rhh}
shows $R_{hh}(\bar{b})$ for $A=0.6$. We find that for LRG/CMASS and quasars the
decrease of linear halo power spectrum suppression is 13\% and 18\%, respectively.
In the limit that $\bar b\gg1$, the reduction is 26\%. Of course the observed halo
and mass power spectra also involve nonlinear corrections with their own scale dependence.
Nonetheless the scale-dependent linear bias thus should be taken into account whenever
neutrino growth suppression is considered in galaxy survey data for any $f_\nu$ if
the free-streaming scale is deep in the linear regime as it is for $m_\nu=0.05\eV$.

Let us now turn to the squeezed-limit bispectrum. Unlike the linear halo power
spectrum, the halo bispectrum at the leading order contains the contribution from
the nonlinear bias, which can be regarded as the response of the linear bias to
$\delta_c$ (e.g. Ref.~\cite{Chiang:2017jnm}). Accurate calibration of higher-order
responses requires SU simulations with larger $|\delta_{c0}|$ \cite{Wagner:2015gva,Lazeyras:2015lgp}.
Lacking such simulations, we thus consider only one piece of the halo bispectrum
and set $B_{hhh}^{\rm sq} (k,k_L)=\bar{b}^2(k) \bar{b}(k_L)B_{ccc}^{\rm sq} (k,k_L)$
with $B_{ccc}$ being the CDM bispectrum. Furthermore, to highlight the effect from
the scale-dependent growth response, we only show the result for the CDM squeezed-limit
bispectrum, i.e.
\ba
 B^{\rm sq}_{ccc}(k,k_L)\:&=\langle P_{cc}(k|\tilde \delta_c(k_L))\tilde{\delta}_c(k_L)\rangle \vs
 \:&=R_{\rm tot}(k,k_L)P_{cc}(k)P_{cc}(k_L) \,,
\ea
where $k$ and $k_L$ are the small- and large-scale modes, $P_{cc}(k|\tilde \delta_c(k_L))$
is the CDM power spectrum in the presence of a single long-wavelength mode of Fourier
amplitude $\tilde\delta_c$ and wavenumber $k_L$.\footnote{Here we distinguish between
$\delta_c$, the dimensionless real-space average value of the mode, and $\tilde\delta_c$
the dimensionful Fourier space amplitude of the mode for clarity.} As pointed out in
\refsec{Rgrowth_meas}, the total power spectrum response $R_{\rm tot}$ contains the
contributions from the growth response $R_{\rm growth}$, that can be measured from the
SU simulations, as well as dilation and reference-density effects. To the leading order,
$R_{\rm dilation}$ and $R_{\bar\rho}$ are given by
\be
 R_{\rm dilation}(k)=-\frac{1}{3}\frac{d\ln k^3P(k)}{d\ln k} \,, \quad
 R_{\bar\rho}=2 \,,
\label{eq:RdilRrho}
\ee
which are independent of $k_L$. Therefore, the step feature in the squeezed-limit
bispectrum due to the scale-dependent growth response is diluted by
$R_{\rm dilation}$ and $R_{\bar\rho}$.

\begin{figure}[h]
\centering
\includegraphics[width=0.498\textwidth]{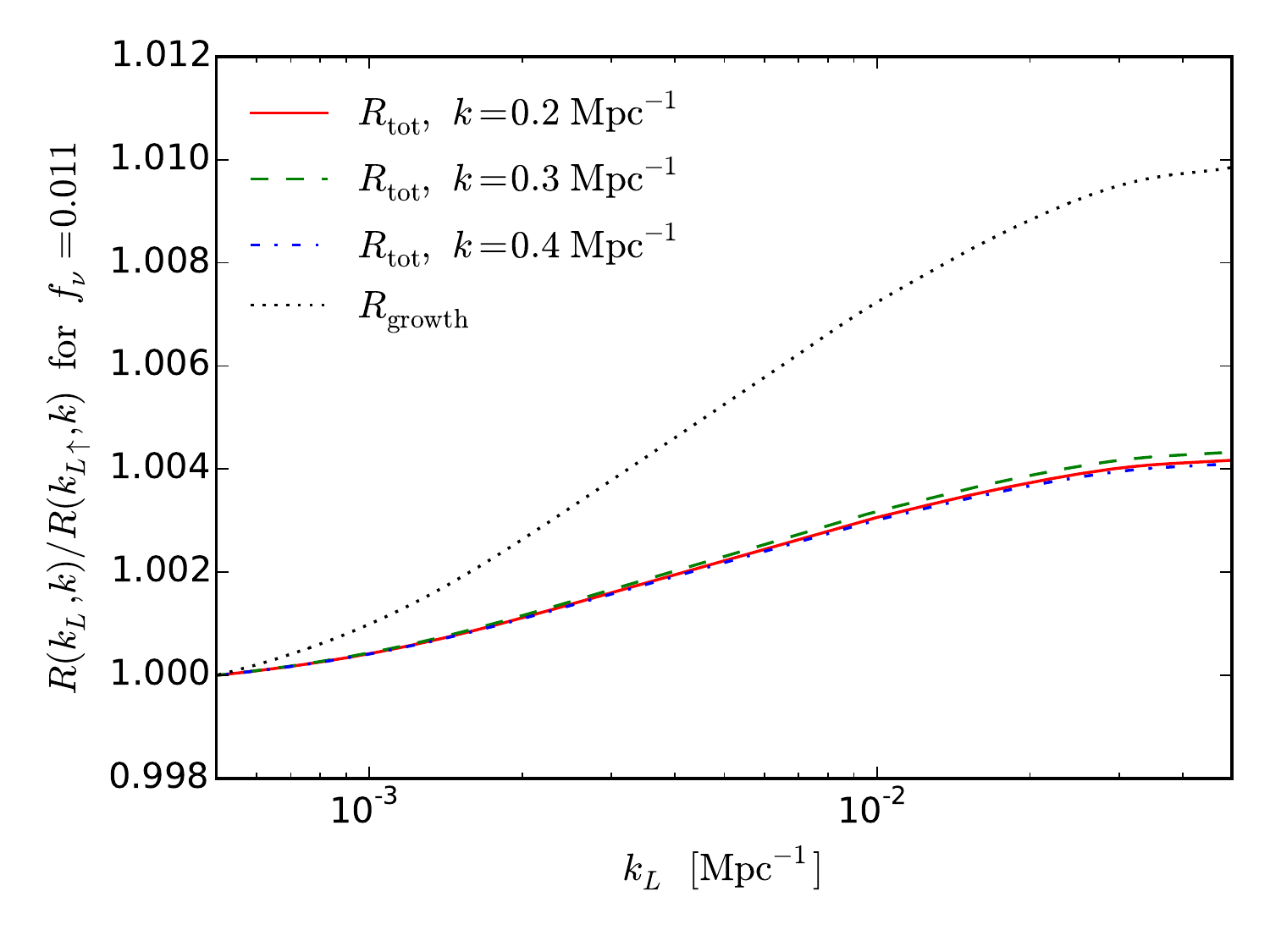}
\caption{Step of CDM power spectrum response as a function of the long mode
$k_L$. The red solid, green dashed, and blue dot-dashed lines show the step
of the total power spectrum response, equivalent to the reduced squeezed-limit
bispectrum, with the short mode $k=0.2$, 0.3, and 0.4 Mpc, whereas the black
dotted line shows the growth response alone.}
\label{fig:Rtot_3nu}
\end{figure}

In \refsec{Rgrowth_dep} we have shown that the growth responses measured from the
SU simulations of $N_\nu=14$ and 28 are in excellent agreement with the analytic
calculation. Thus, for $f_\nu=0.011$ cosmology we assume $R_{\rm growth}=2(d\ln D_W/d\delta_c)$,
which depends only on $k_L$. For $R_{\rm dilation}$, we use the spectral index of
the CDM power spectrum, which depends on $k$. \refFig{Rtot_3nu} shows the power
spectrum response as a function of the long mode $k_L$. The red solid, green dashed,
and blue dot-dashed lines show the step of the total power spectrum response,
equivalent to the reduced squeezed-limit bispectrum, with the short mode $k=0.2$,
0.3, and $0.4\iMpc$, whereas the black dotted line shows the growth response alone.
As expected, $R_{\rm dilation}$ and $R_{\bar\rho}$ dilute the dependence on $k_L$,
and compared to $R_{\rm growth}$ the step size is reduced by 60\%. Albeit small,
the scale dependence is a distinct feature due to massive neutrinos, hence can
serve as an independent probe. On the other hand, we find that $R_{\rm tot}$
depends only weakly on $k$. This is because for $k\gtrsim0.2\iMpc$ the baryonic
acoustic oscillation is less prominent, and so the spectral index is closer to
a constant.

Our discussion focuses on the galaxy squeezed-limit bispectrum, but the derivation
is similar for the galaxy-galaxy-lensing bispectrum (see Ref.~\cite{Chiang:2017qoh}
for the cross-correlation between Lyman-$\alpha$ power spectrum and lensing
convergence), where the lensing convergence provides the long mode. Since
lensing measures the total matter fluctuation along the line-of-sight, the
bispectrum is schematically given by $B^{\rm sq}_{hhm}(k,k_L)=\bar{b}^2(k)B^{\rm sq}_{ccm}(k,k_L)$,
where the CDM-CDM-matter squeezed-limit bispectrum is given by
\ba
 B^{\rm sq}_{ccm}(k,k_L)\:&\equiv\langle P_{cc}(k|\tilde \delta_m(k_L))\tilde \delta_m(k_L)\rangle \vs
 \:&=R_{\rm tot}(k,k_L)P_{cc}(k)P_{cm}(k_L) \,. 
\ea
Here we have assumed that since the perturbations are adiabatic, there is a one-to-one
relation between the long-wavelength $\tilde \delta_m$ and $\tilde \delta_c$ given by
the linear transfer functions so that
\ba
 \frac{d\ln P_{cc}(k)}{d\delta_m}(k_L)\:&= R_{\rm tot}(k,k_L)\frac{T_c(k_L)}{T_m(k_L)} \,, \vs
 P_{cm}(k_L)\:&=\frac{T_c(k_L)}{T_m(k_L)}P_{mm}(k_L) \,.
\ea
We find that the only difference between the $B^{\rm sq}_{ccc}$ and $B^{\rm sq}_{ccm}$
is the large-scale power spectra, i.e.\ $P_{cc}(k_L)$ versus $P_{cm}(k_L)$.

\section{Discussion}
\label{sec:discussion}
Massive neutrinos provide an ideal arena to explore the scale dependence
of the response of small-scale structure formation to the large-scale
density environment. Due to their large thermal velocities, massive neutrino
fluctuations track the CDM only on scales larger than the free-streaming
scale with smaller-scale fluctuations washed out. As a result, the growth
of CDM perturbations becomes scale dependent. The response of small-scale
observables to these long-wavelength perturbations consequently becomes
scale dependent as well. These effects can be captured in the separate
universe (SU) technique by absorbing the long-wavelength perturbations
into the local expansion.

Using the SU technique, we perform $N$-body simulations in overdense
and underdense SUs with different wavelengths of the large-scale CDM density
perturbations in a universe with massive neutrinos. By differencing pairs
of overdense and underdense SU simulations with the same Gaussian realizations
of initial phases, we measure how the power spectrum and halo mass
function respond to large-scale CDM density perturbations of different
wavelengths, which give rise to the squeezed-limit bispectrum and the halo
bias, respectively. Due to the cancellation of the cosmic variance, the SU
simulations yield a precise characterization of the scale dependence of these
responses.

Specifically, for the cosmology with $m_\nu=0.05\eV$ but an artificially
high number of neutrinos totaling a fraction $f_\nu=0.093$ of the matter,
we perform SU simulations for five long-wavelength perturbations (from
$k_{L\uparrow}=0.0005\iMpc$ to $k_{L\downarrow}=0.05\iMpc$) spanning the
free-streaming transition. Scale dependence in the responses for the power
spectrum and Lagrangian bias are detected with high significance (see
\reffig{Rgw_step_kL} and \reffig{bL_step_kL}). Interestingly, we find
that the scale dependence in both cases can be described well with just
the linear growth response, which can be calculated without simulations
and with no free parameters. For Lagrangian bias, this result follows if
the mass function is universal in the local power spectrum but is also
equally consistent with the spherical collapse model. To further confirm
this result, we also perform SU simulations with $f_\nu=0.049$ for $k_{L\uparrow}$
and $k_{L\downarrow}$ (see \refFig{Rgw_step_Nnu} and \reffig{bL_step_Nnu})
and show that they are consistent with the linear growth response as well,
implying a scaling of the steps across the free streaming scale of $\sim 0.6 f_\nu$.
This scaling allows us to extrapolate our results to a more realistic case
of three massive neutrinos of $0.05\eV$ with $f_\nu=0.011$.
 
There are two important implications of our results. First, the scale-dependent
responses due to massive neutrinos produce new features in the halo power spectrum
and squeezed-limit bispectrum, and the effects are shown in \reffigs{Rhh}{Rtot_3nu}
with $f_\nu=0.011$. For the linear halo power spectrum, we find that the scale-dependent
bias reduces the difference between linear power spectra of massive and massless
neutrinos by 13 and 26\% for objects of $\bar{b}=2$ and $\bar{b}\gg1$, respectively
independently of $f_\nu$. The larger the halo mass (hence the halo bias), the larger
effect due to the scale-dependent bias. This effect must be taken into account for
future surveys that use the halo power spectrum to constrain neutrino mass \cite{LoVerde:2016ahu}.
For the CDM-CDM-CDM reduced squeezed-limit bispectrum, we find that the step size
is around 4\% with a weak dependence on the small-scale mode $k$. The effect is
small because the dilation and reference-density responses dilute the scale dependence
from the growth response. Albeit small, the scale dependence is a distinct characteristic
due to massive neutrinos can be used as an independent probe.

Second, we find that halo bias is temporally nonlocal. For the same value of the
CDM density fluctuations at two different wavenumbers, the bias differs due to the 
evolutionary histories of the modes. A local model of density bias, even one that
allows for local bias with respect to the different density components cannot describe
this effect. Specifically, we demonstrate in \reffigs{bL_step_bT}{bL_step3} that
the transfer function bias $\bar{b}_T(k)=b_c+b_\nu\[T_\nu(k)/T_c(k)\]$, where $b_c$
and $b_\nu$ are the CDM and neutrino bias parameters and $T_c$ and $T_\nu$ are the
CDM+baryon and neutrinos transfer functions, is a poor fit to the scale-dependent
bias measured from our SU simulations for any $b_c$ and $b_\nu$. Therefore, the
standard Lagrangian picture that the halo statistics at any time are determined
entirely by the linear density field at a single epoch is falsified in this case
where the free-streaming or Jeans scale is deeply in the linear regime.   

Our SU simulations assume a fixed and low value of the individual neutrino masses $m_\nu=0.05\eV$ 
and should not be naively extrapolated to higher values. Since the SU ignores small-scale
neutrino clustering, the free-streaming scale must be much larger than the scale
of the observables. For $m_\nu=0.05\eV$ we argue in \refapp{caveat} that neutrino
clustering can be neglected for $k\gtrsim 0.05\iMpc$. Furthermore, since we approximate
the long-wavelength mode as spatially constant, it must also be much larger than
the region that encompasses the small-scale observables. Specifically for halo bias,
corrections will enter at $\mathcal{O}(k_L^2R_M^2)$ with $R_M$ being the Lagrangian
radius of halo with mass $M$ and for the power spectrum response $\mathcal{O}(k_L^2/k^2)$
with $k_L$ and $k$ being the wavenumbers of the large- and small-scale modes respectively.   

With too small a free-streaming scale, these limitations would make it impossible
to track responses across the free-streaming scale with the SU technique. Even
for $m_\nu=0.05\eV$, we expect there to be some correction to our results from
the clustering of slow neutrinos in the tail of the neutrino distribution function
(see discussion in Ref.~\cite{Senatore:2017hyk}) and the scale dependence of
power spectra from nonlinearity.
In \refapp{clusteringbias} we compare our prediction to one of the best suites
of $N$-body simulations with massive neutrino particles, and find that it is equally
consistent with the neutrino particle simulations as the scale-independent linear
bias. As discussed in more detail there, to distinguish
our novel scale dependence  it is necessary to run neutrino particle simulations
of a few Gpc, to see the full scale dependence at large scales, and with high enough
$f_\nu$ or a large enough number of simulations to robustly constrain the $\mathcal{O}(f_\nu)$ 
corrections discussed here. Note that $f_\nu$ should be increased while keeping
$m_\nu \sim 0.05$ eV so that the free-streaming scale and nonlinear clustering of
neutrinos are not qualitatively changed.
We leave the detailed comparison between our
SU results with other techniques of simulating massive neutrinos for future work.

\acknowledgements{
We would like to thank Yu Feng, Neal Dalal, Emanuele Castorina, Masahiro Takada,
Fabian Schmidt, and Shun Saito for useful discussion.
We would further like to thank Francisco Villaescusa-Navarro for providing
the simulation data in Ref.~\cite{Villaescusa-Navarro:2017mfx}.
Results in this paper were obtained using the high-performance computing 
system at the Institute for Advanced Computational Science at Stony Brook
University and with the computation and storage resources provided by the
University of Chicago Research Computing Center.
CC and ML are supported by grant NSF PHY-1620628 and DOE DE-SC0017848.
WH is supported by NASA ATP NNX15AK22G, DOE DE-FG02-13ER41958, the Simons Foundation,
and the Kavli Institute for Cosmological Physics at the University of Chicago
through grant NSF PHY-1125897 and an endowment from the Kavli Foundation and its founder Fred Kavli.

\appendix
\section{Robustness to initial conditions}
\label{app:setup}
In this appendix we test the choice of initial and horizon-entry scale
factors, $a_i$ and $a_H$, for solving the differential equation of the
small-scale growth, i.e. \refeqs{D}{epsiloneom}. We shall particularly
focus on the growth response at $a\ge0.02$, as it corresponds to the
quantity observable and time of interest.

\begin{figure}[h]
\centering
\includegraphics[width=0.498\textwidth]{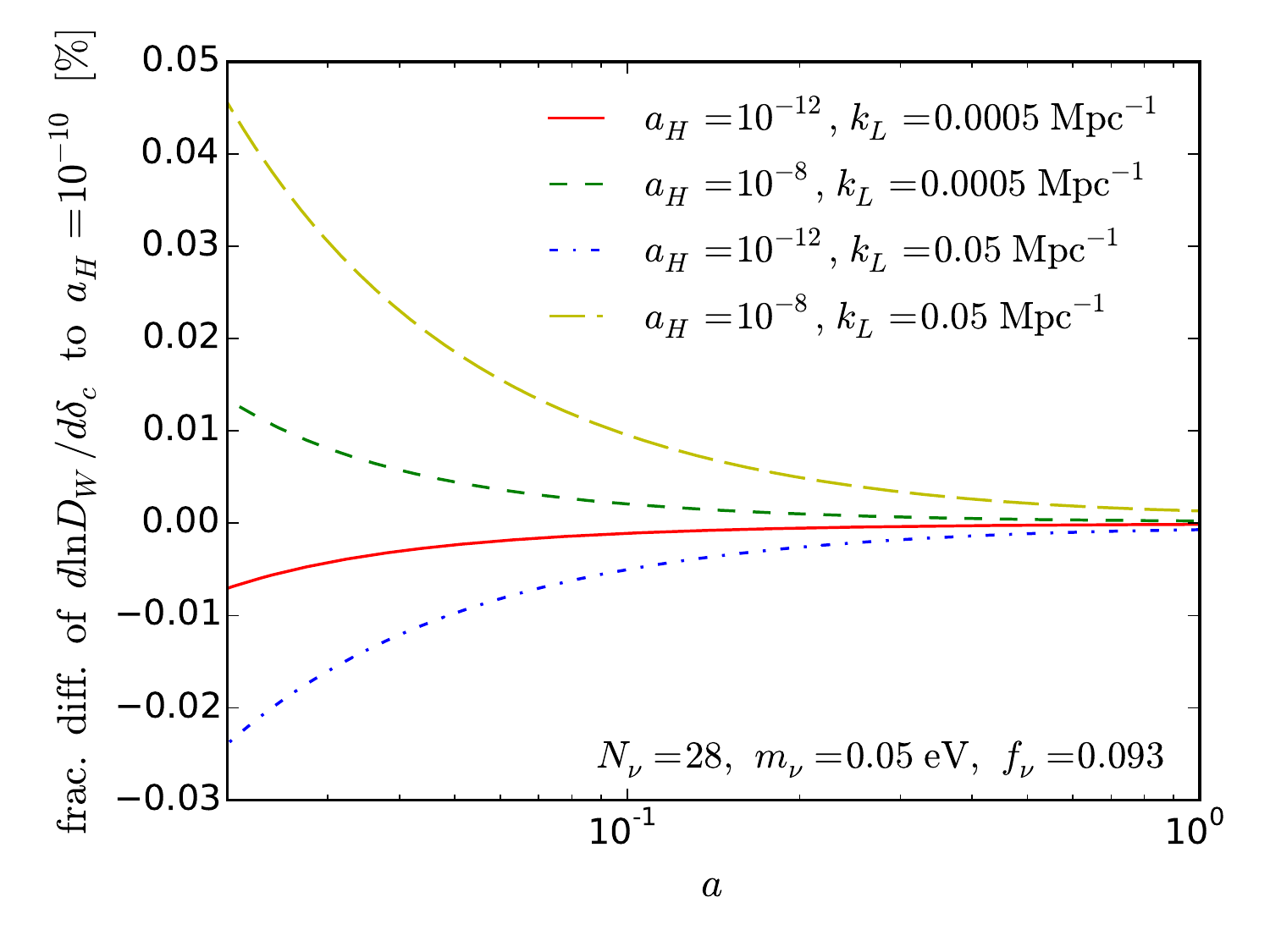}
\caption{Fractional difference of the growth response of various $a_H$
to that with $a_H=10^{-10}$, for fixed $a_i=10^{-6}$ and two limiting
$k_L$.}
\label{fig:dlnDW_aH}
\end{figure}

Following the discussion in \refsec{growth}, we have the constraint
$a_{\rm eq}\gg a_i\gg a_H$. Let us start by fixing $a_i=10^{-6}$ and
varying $a_H=10^{-12}$, $10^{-10}$, and $10^{-8}$. \refFig{dlnDW_aH}
shows the fractional difference of the growth response of various $a_H$
to that with $a_H=10^{-10}$, for fixed $a_i=10^{-6}$ and two limiting
$k_L$. We find that for all cases the differences are less than 0.05\%,
and the agreement is better for $a$ approaching to unity. This demonstrates
that the difference due to $a_H$ can be safely neglected and justifies our 
choice of $a_H=10^{-10}$ in the main text.

\begin{figure}[h]
\centering
\includegraphics[width=0.498\textwidth]{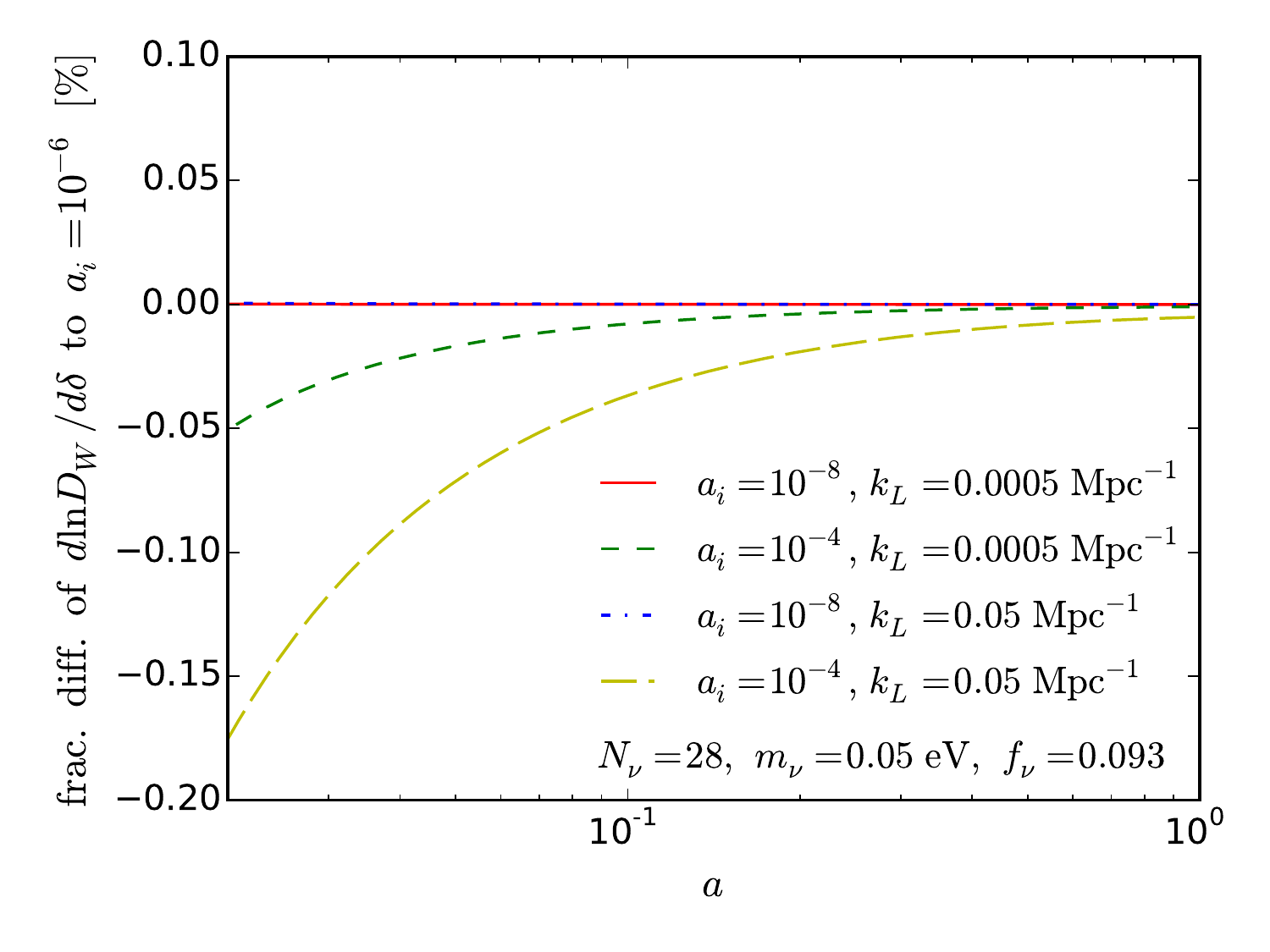}
\caption{Same as \reffig{dlnDW_aH}, but for various $a_i$ to that with
$a_i=10^{-6}$, for fixed $a_H=10^{-10}$.}
\label{fig:dlnDW_ai}
\end{figure}

We next fix $a_H=10^{-10}$, and compute the growth response with $a_i=10^{-4}$,
$10^{-6}$, and $10^{-8}$. Note that unlike the $a_H$ dependence, which corresponds
to a true dependence on the short-wavelength $k$ that cannot be captured in our
 SU implementation, this tests a purely computational error from assuming
$a_i \ll a_{\rm eq}$ in the derivation. \refFig{dlnDW_ai} shows the fractional
difference of the growth response of various $a_i$ to that with $a_i=10^{-6}$,
for fixed $a_H=10^{-10}$ and two limiting $k_L$. We first find that the agreement
is good between $a_i=10^{-6}$ and $10^{-8}$, indicating the convergence for even
earlier $a_i$. Between $a_i=10^{-6}$ and $10^{-4}$, the differences at $a=0.02$
(initial redshift of the SU simulations) are apparent. This justifies our choice
of $a_i=10^{-6}$ used in the main text.

\section{2LPT in the separate universe}
\label{app:2lpt}
In this appendix, we derive the small-scale growth of the displacement field
under the framework of 2LPT, assuming that CDM is the only component that
clusters (e.g. below the Jeans scales of all other clustering components),
in order to set up the initial conditions of the SU simulations. Following
the standard convention in App.~D of Ref.~\cite{Scoccimarro:1997gr}, we
define $D_1$ and $D_2$ to be the growths of the first and second order
perturbations of the displacement field. In the absence of the long-wavelength
perturbation $\delta_c$, the evolution of $D_1$ and $D_2$ are given by
\ba
 \:&D''_1+\(2+\frac{H'}{H}\)D'_1-\frac{3}{2}\Omega_c(a)D_1=0 \,, \\
 \:&D''_2+\(2+\frac{H'}{H}\)D'_2-\frac{3}{2}\Omega_c(a)D_2=-\frac{3}{2}\Omega_c(a)D_1^2 \,.
\ea
Note that the equation of the first order growth is identical to \refeq{D}.
If we set the initial conditions of the differential equations at $a_i$
during the matter-dominated epoch, then $H'/H=-3/2$ and if CDM dominates
the matter density then $\Omega_c(a)=1$; hence we obtain the standard results
{(e.g. \cite{Bouchet:1994xp,Scoccimarro:1997gr,Crocce:2006ve})}:
\ba
 \:&D_1(a_i)=a_i \,, \quad D'_1(a_i)=a_i \,, \vs
 \:&D_2(a_i)=-\frac{3}{7}a_i^2 \,, \quad D'_2(a_i)=-\frac{6}{7}a_i^2 \,.
\label{eq:2lpt_md}
\ea
On the other hand, if $a_i$ is in the radiation-dominated epoch, then
$(2+H'/H)\propto a_i/a_{\rm eq}\to 0$ and $\Omega_c(a)\propto a_i/a_{\rm eq}\to 0$,
and the solutions to the small-scale growths become
\ba
 \:&D_1(a_i)=\ln\frac{a_i}{a_H} \,, \quad D'_1(a_i)=1 \,, \vs
 \:&D_2(a_i)=-\frac{3}{2}\Omega_c(a)\[\(\ln\frac{a_i}{a_H}\)^2-4\ln\frac{a_i}{a_H}+6\] \,, \vs
 \:&D'_2(a_i)=-\frac{3}{2}\Omega_c(a)\[\(\ln\frac{a_i}{a_H}\)^2-2\ln\frac{a_i}{a_H}+2\] \,.
\label{eq:2lpt_rd}
\ea

Let us now turn to the universe with a long-wavelength perturbation $\delta_c$.
Within the SUs the small-scale growths of 2LPT follow
\ba
 \:&\frac{d^2D_{W1}}{d\ln a_W^2}+\(2+\frac{d\ln H_W}{d\ln a_W}\)\frac{dD_{W1}}{d\ln a_W} \vs
 \:&\hspace{0.5cm}-\frac{3}{2}\frac{H_{0W}^2}{H_W^2}\frac{\Omega_{cW}}{a_W^3}D_{W1}=0 \,, \\
 \:&\frac{d^2D_{W2}}{d\ln a_W^2}+\(2+\frac{d\ln H_W}{d\ln a_W}\)\frac{dD_{W2}}{d\ln a_W} \vs
 \:&\hspace{0.5cm}-\frac{3}{2}\frac{H_{0W}^2}{H_W^2}\frac{\Omega_{cW}}{a_W^3}D_{W2}
 =-\frac{3}{2}\frac{H_{0W}^2}{H_W^2}\frac{\Omega_{cW}}{a_W^3}D_{W1}^2 \,.
\ea
Rewriting the differential equations in terms of the global coordinate
and linearizing in $\delta_c$, we have the perturbations to the growths
$\epsilon_1=D_{W1}-D_1$ and $\epsilon_2=D_{W2}-D_2$ to be
\ba
 \:&\epsilon''_1+\(2+\frac{H'}{H}\)\epsilon'_1-\frac{3}{2}\Omega_c(a)\epsilon_1 \vs
 \:&\hspace{0.5cm}=\frac{2}{3}\delta_c'D'_1+\frac{3}{2}\Omega_c(a)\delta_cD_1 \,, \\
 \:&\epsilon''_2+\(2+\frac{H'}{H}\)\epsilon'_2-\frac{3}{2}\Omega_c(a)\epsilon_2 \vs
 \:&\hspace{0.5cm}=\frac{2}{3}\delta_c'D'_2+\frac{3}{2}\Omega_c(a)\[\delta_c(D_2-D_1^2)-2D_1\epsilon_1\] \,.
\ea
To solve $\epsilon_1$ and $\epsilon_2$ in the matter-dominated universe,
we assume that $\delta_c$ is sub-horizon and so proportional to $a$, which
leads to
\ba
 \:&\epsilon_1(a_i)=\frac{13}{21}\delta_c(a_i)D_1(a_i) \,, \quad
 \epsilon'_1(a_i)=\frac{26}{21}\delta_c(a_i)D_1(a_i) \,, \vs
 \:&\epsilon_2(a_i)=\frac{32}{27}\delta_c(a_i)D_2(a_i) \,, \quad
 \epsilon'_2(a_i)=\frac{32}{9}\delta_c(a_i)D_2(a_i) \,.
\ea
On the other hand, for the radiation-dominated universe we assume that
$\delta_c$ is super-horizon and so proportional to $a^2$. This then
leads to
\ba
 \:&\epsilon_1(a_i)=\frac{1}{3}\delta_c(a_i)D'_1(a_i) \,, \quad
 \epsilon'_1(a_i)=\frac{2}{3}\delta_c(a_i)D'_1(a_i) \,, \vs
 \:&\epsilon_2(a_i)=-\frac{3}{2}\Omega_c(a_i)\delta_c(a_i)\[D'_1(a_i)\]^2 \vs
 \:&\hspace{1.3cm}\[\frac{7}{27}\ln\(\frac{a_i}{a_H}\)^2-\frac{46}{81}\ln\frac{a_i}{a_H}+\frac{50}{81}\] \,, \vs
 \:&\epsilon'_2(a_i)=-\frac{3}{2}\Omega_c(a_i)\delta_c(a_i)\[D'_1(a_i)\]^2 \vs
 \:&\hspace{1.3cm}\[\frac{7}{9}\ln\(\frac{a_i}{a_H}\)^2-\frac{32}{27}\ln\frac{a_i}{a_H}+\frac{104}{81}\] \,.
\ea
Note that the results of $D_1(a_i)$ and $\epsilon_1(a_i)$ are identical
to \refeq{Di_ei_RD}.

For our SU simulations, we first set the initial conditions of the
small-scale growths at $a_i=10^{-6}$, with $a_H=10^{-10}$. We then
evolve the differential equations to the initial redshift of the
simulations at $a_{Wi}(\delta_{c0})=0.02$. Since usually the code
for setting up the initial conditions of $N$-body simulations uses
growth rate for the velocities of particles, we rewrite
\ba
 f_{W1}\:&=\frac{d\ln D_{W1}}{d\ln a_W}=\(1+\frac{1}{3}\delta'_c\)
 \frac{D'_1+\epsilon'_1}{D_1+\epsilon_1} \,, \vs
 f_{W2}\:&=\frac{d\ln D_{W2}}{d\ln a_W}=\(1+\frac{1}{3}\delta'_c\)
 \frac{D'_2+\epsilon'_2}{D_2+\epsilon_2} \,.
\ea

\section{Separate universe systematics}
\label{app:caveat}

\subsection{Neutrino clustering}
\begin{figure}[h]
\centering
\includegraphics[width=0.498\textwidth]{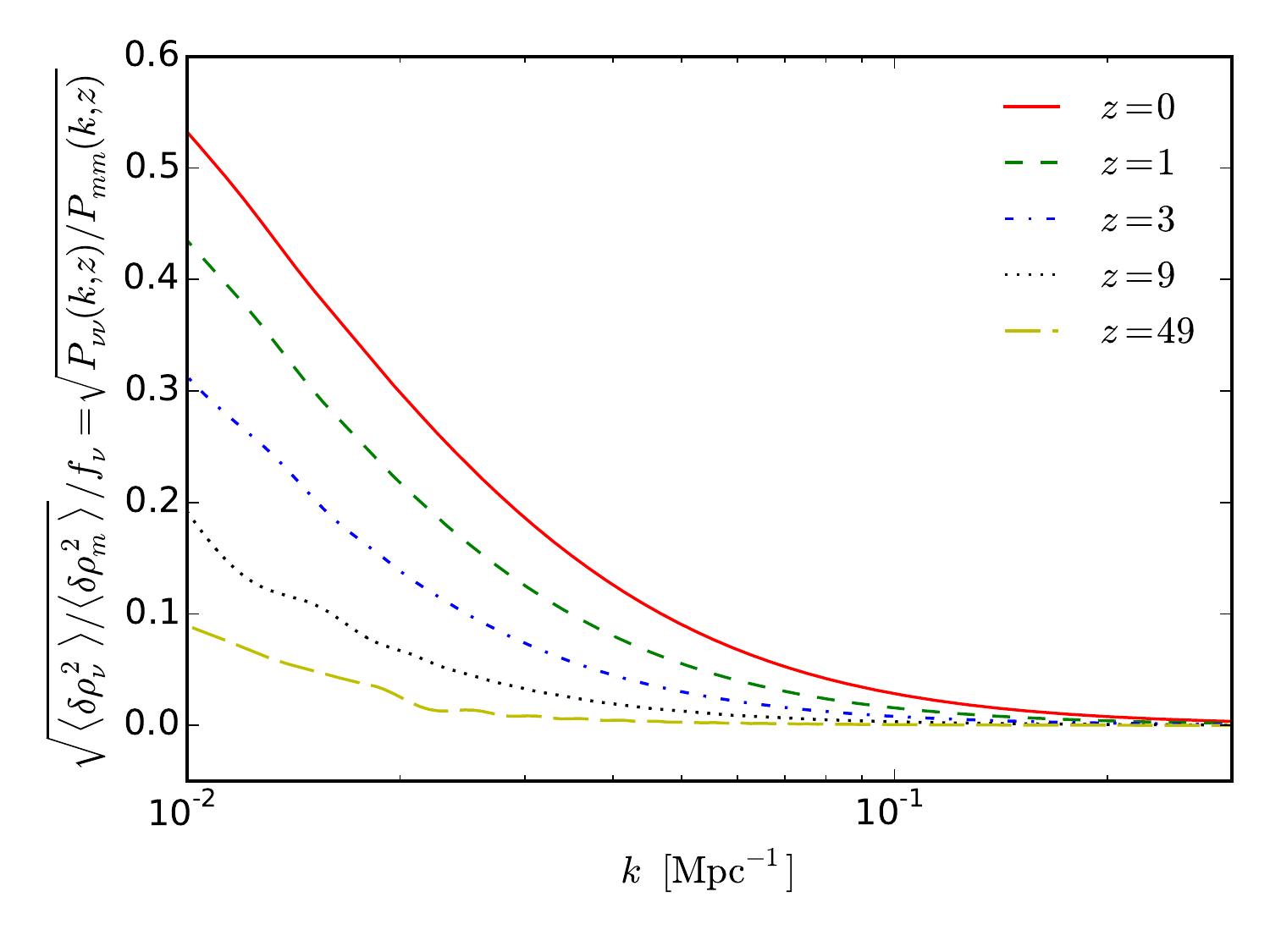}
\caption{Ratio of the typical amplitude of (linear) neutrino perturbations to total matter perturbations. 
This quantifies the error (in units of $f_{\nu}$) induced on the density field on scale $k$ from assuming
that neutrinos are completely smooth in our simulations. This plot uses $N_\nu = 28$ neutrinos $(f_\nu=0.093)$
but we find that the results are only weakly dependent on the number of neutrinos.}
\label{fig:PnunuOverPmm}
\end{figure}

The separate universe approach used in this paper assumes that the neutrino energy density
is completely smooth within the simulations box. For our choice of box size $L = 700$ Mpc, 
neutrinos are clustered at late times on the largest scales so our simulations are missing
some neutrino perturbations on the largest scales. These scales are in the linear regime so
we can straightforwardly estimate the scale at which neutrino perturbations are important
from their fractional contribution to the total gravitational potential as a function of $k$
using the linear power spectrum
\be
 \frac{\Delta \Phi_\nu(k)}{\Phi}\approx\sqrt{\frac{\langle\delta\rho_\nu(k)^2\rangle}{\langle \delta\rho_m(k)^2\rangle}}
 =f_\nu\sqrt{\frac{P_{\nu\nu}(k)}{P_{mm}(k)}} \,.
\ee
This quantity is plotted in \reffig{PnunuOverPmm}. We can see that so long as we restrict
our attention to $k\gtrsim 0.05\iMpc$ the error due to neglecting neutrino clustering should
be $\lesssim 0.1f_\nu$. Note that the ``missing" neutrino perturbations are absent in each
of the SU simulations ($\delta_{c0} = \pm 0.01, 0$ for each $k_L$) so the absolute error on
the response quantities determined from differences between simulations should be even smaller.
In this paper we only report the power spectrum response with $k\ge 0.05\iMpc$.

\begin{figure}[h]
\centering
\includegraphics[width=0.498\textwidth]{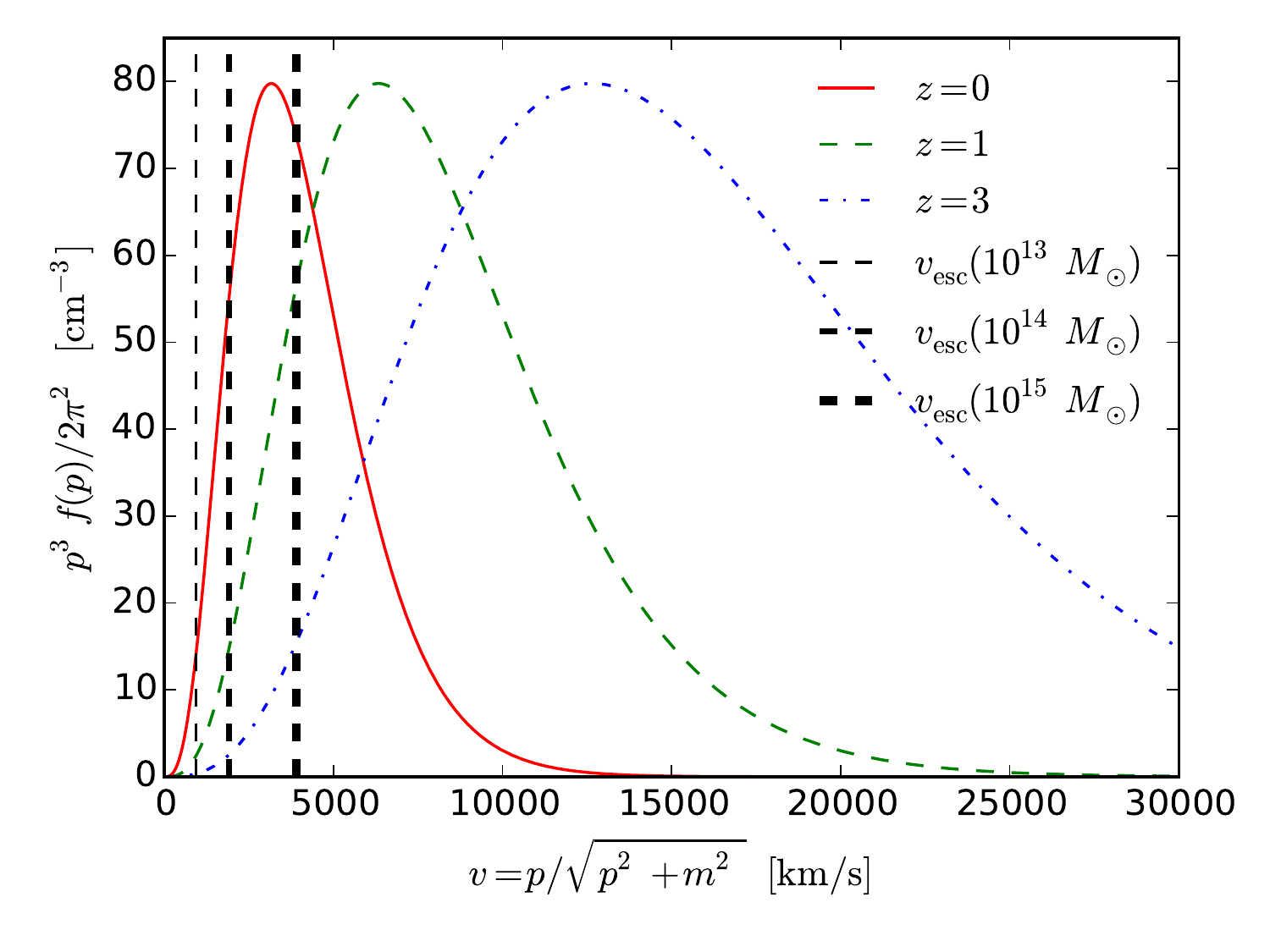}
\caption{The velocity distribution of a single species of neutrinos with mass $0.05\eV$
at several redshifts. Vertical dashed lines show the escape velocities of dark matter
halos of several masses.}
\label{fig:velocitydist}
\end{figure}

We have seen that the error from neglecting the large-scale linear theory neutrino
perturbations should be small ($\mathcal{O}(0.1 f_{\nu})$). On smaller scales where
the CDM perturbations have become nonlinear, however, one could worry that the
gravitational potentials are strong enough to cause significant clustering of
slow-moving neutrinos. 

In \reffig{velocitydist}, we show the mean velocity distribution of neutrinos at
several epochs in comparison with the escape velocities of dark matter halos. We
define the escape velocities as $v_{\rm sec} \equiv \sqrt{2[\Phi(r = \infty) - \Phi(r = R_s)]}$
where $\Phi$ is the gravitational potential of a Navarro, Frenk, and White dark
matter halo \cite{Navarro:1996gj} and $R_s$ is the scale radius of the halo. At
the latest times, a significant fraction of neutrinos can indeed become bound,
or at least significantly deflected, by nonlinear structure. On the other hand,
the relative contribution of these neutrinos to the total gravitational potential
remains small since the CDM perturbations on the same scale are large. For instance,
for halos with virial masses $M = 10^{13}$, $10^{14}$, and $10^{15} M_\odot$, the
local overdensity of neutrino mass for each mass state is approximately $10^8$,
$3\times 10^9$, and $10^{11}~M_\odot$ respectively \cite{LoVerde:2013lta}. Hence,
even for our simulations with $N_\nu = 28$, the fractional correction to the
gravitational potential, and therefore the evolution the CDM particles is $\sim 0.003 f_{\nu}$,
$0.01 f_{\nu}$, and $0.03f_{\nu}$ for the halo masses $M = 10^{13}$, $10^{14}$,
and $10^{15} M_\odot$. This estimate is consistent with Ref.~\cite{LoVerde:2014rxa},
which found that the effects of neutrino clustering around CDM halos was negligible
in spherical collapse calculations so long as the individual neutrino mass $m_\nu\lesssim 0.2\eV$.
We thus consider it a safe assumption to ignore the effect of neutrino clustering
on halo response bias.

\subsection{Large-scale averaging}
In the SU approximation we take the wavelength of the large-scale density perturbation
to be sufficiently larger than the scale of interest for small-scale observables that
its spatially varying amplitude can be replaced by its locally averaged value. In other
words we conflate $\delta_c$ with $\delta_{cW}$ where
\be
 \delta_{cW}({\bf x}_0) = \int d^3 x W_R({\bf x}_0-{\bf x}) \delta_c({\bf x}) \,,
\ee
and $W_R$ is some spherically symmetric window with support on $|{\bf x}_0-{\bf x}| \lesssim R$
which is normalized to integrate to unity. In Fourier space
\be
 \tilde \delta_{cW}({\bf k}) = \tilde W_R({\bf k}) \tilde \delta_c ({\bf k}) \,,
\ee
and for wavelengths that are long compared to $R$
\be
 \lim_{kR \rightarrow 0} \tilde \delta_{cW}({\bf k})
 = \left[ 1 + {\cal O}(k^2 R^2)  \right]  \tilde \delta_c ({\bf k}) \,.
\ee
At a fixed $k=k_L$, the SU approximation is limited to observables that are
sensitive to only a region $R \ll 1/k_L$. For the halo mass function a rough
indication of the region of influence is the Lagrangian radius of the halo $R_M$.
For the power spectrum the typical scale is $R \sim 1/k$. We thus conclude that
averaging errors on the response bias scale as  $\mathcal{O}(k_L^2R_M^2)$, and
those on the squeezed bispectrum as $\mathcal{O}(k_L^2/k^2)$. Note that these
scalings apply to both neutrino and $\Lambda$CDM separate universe simulations
but for the former we take $k_L \gtrsim k_{\rm fs}$ in order to measure the
scale dependence of the responses.

\section{Spherical collapse bias}
\label{app:sphericalcoll}
We follow the method of Ref.~\cite{LoVerde:2014pxa} to make spherical collapse
calculations of the Lagrangian response bias. A region of size $R$ containing
a constant mass $M$ of CDM will evolve according to
\ba
\label{eq:ddotRLW}
 \ddot{R}=\:&-\frac{GM(<R)}{R^2} \\
 \:& -\frac{4\pi G}{3} \sum_x\[\rho_x(t) + \delta\rho _{x} + 3P_x(t)+3\delta P_x\]R \,, \nonumber
\ea
where $\dot{\,}\equiv d/dt$, $\rho_x$ and $P_x$ are the energy density and pressure
of any non-CDM components of the universe and $\delta\rho_x$, and $\delta P_x$
are long-wavelength perturbations in the energy density and pressure of $x$.
Using $M = \frac{4}{3}\pi R^3\rho_c(1+\delta)$, and factoring out a long-wavelength
CDM perturbation from $\delta$ allows this equation to be rewritten in terms of
the fluctuation from the local mean $\delta_S$, defined via
$(1+\delta) = [1+\delta_c(a)](1+\delta_S)$ as
\ba
 \:&\frac{d^2\delta_S}{d\ln a_W^2}+\left(2+\frac{d\ln H_W}{d\ln a_W}\right)\frac{d\delta_S}{d\ln a_W}
 -\frac{4}{3}\frac{\left(\frac{d\delta_S}{d\ln a_W}\right)^2}{1+\delta_S} \vs
 \:&\hspace{0.5cm}\qquad = \frac{3}{2}\frac{\Omega_{cW}H_{0W}^2}{a_W^3H_W^2}\delta_S(1+\delta_S)  \,,
\label{eq:deltaSNL}
\ea
where $a_W = a[1-\delta_c(a)/3]$, as before. This equation is equivalent to
\refeq{ddotRLW} and can be viewed as a nonlinear generalization of \refeq{DW}
(for spherical density perturbations). \refEq{deltaSNL} is identical to the
usual nonlinear equation for a spherical density perturbation and the solutions
are independent of $M$. For given initial conditions $\delta_{Si}=\delta_S(a_{Wi})$
and $\left[d\ln \delta_S/d\ln a_W\right](a_{Wi})$, it can be solved to determine
$a_{{\rm coll},W}$, the SU scale factor at which $\delta_S\rightarrow \infty$.

Following our calculations for $D_W$ and \refapp{setup}, we start at $a_i = 10^{-6}$
with initial velocity
\be
 \frac{d\ln \delta_S}{d\ln a_W}(a_i) = \frac{d\ln D_W}{d\ln a_W}(a_i) \,.
\ee
We iteratively solve \refeq{deltaSNL} to determine the initial density perturbation
$\delta_{Si}$ that will ``collapse'' at global scale factor $a_{\rm coll}$ for each
$\delta_c(a, k_L)$. Our criteria for collapse is that $d\ln \delta_S/d\ln a_W = 100$.
The linearly extrapolated threshold for collapse is given by
$\delta_{\rm crit} \equiv \left[D(a_{\rm coll})/D(a_{i})\right]\delta_{Si}$, where
the initial value of $\delta_{Si}$ that produces collapse at $a_{\rm coll}$ is a
function of the long-wavelength mode $\delta_c(a)$. From this we compute
\be
 \frac{\delta \delta_{\rm crit}}{\delta \delta_c}(a_{\rm coll})
 = \frac{\delta_{\rm crit}(\delta_c)  - \delta_{\rm crit}(-\delta_c)}{2\delta_c(a_{\rm coll})} \,,
\ee
and define the Lagrangian bias with respect to CDM as 
\be
 \bar{b}_S^L = \frac{\partial\ln n(M)}{\partial \delta_{\rm crit}}\frac{\delta \delta_{\rm crit}}{\delta \delta_c}(k_L) \,.
\ee

\section{Comparison of scale-dependent bias with neutrino particle simulations}
\label{app:clusteringbias}
In this appendix we compare the scale-dependent bias model based on the response
of the halo mass function in SU simulations to the clustering bias measured from $N$-body
simulations with massive neutrino particles in Ref.~\cite{Villaescusa-Navarro:2017mfx}.
This simulation suite is one of the largest to date and contains 100 realizations
of $1~h^{-1}~{\rm Gpc}$ boxes. The cosmological parameters are $h=0.6711$, $\Omega_b=0.049$,
and $\Omega_c=0.2649$, with three massive neutrinos of 0.05 eV, corresponding to $f_\nu=0.011$.
The halos are identified with the Friends-of-Friends algorithm \cite{Davis:1985rj},
and to be conservative we only consider halos with more than 100 dark matter particles,
corresponding to a minimum halo mass of $9.673\times10^{13}~M_\odot$. The clustering
bias is measured with respect to CDM,
i.e. $\bar{q}(k)=P_{ch}(k)/P_{cc}(k)$.
We use a Fourier bin of $0.002~h~{\rm Mpc}^{-1}$ so that the data point are dense enough
while we still have enough realizations to estimate the covariance matrix.

We fit the mean of the measured clustering bias to two linear bias models. The first
one is that the linear bias is scale independent, hence the total bias is given by
\be
 \bar{b}(k)=\bar{b}_1+\bar{b}_{k^2}k^2 \,,
\ee
where the $k^2$ term absorbs the loop corrections in the large-scale limit \cite{Assassi:2014fva}.
We also explore the inclusion of additional $\bar{b}_{k^4}k^4$ but find no significant
difference, so we present the simpler model. The second one is our predicted scale
dependence, and the model is given by
\be
 \bar{b}(k)=1+\bar{b}^L_1f(k)+\bar{b}_{k^2}k^2 \,,
\ee
where we set $f(k)=[d\ln D_W/d\delta_c](k)/[d\ln D_W/d\delta_c](k_\uparrow)$. To compute
$f(k)$ we use the same cosmology as in Ref.~\cite{Villaescusa-Navarro:2017mfx}. Note
that this model is identical to $\bar{b}^L_D$ defined in \refeq{bD} with an additional
$\bar{b}_{k^2}$ accounting for the nonlinear correction. Both models contain two free
parameters, and we find the parameters by minimizing
\be
 \chi^2=\sum_{ij}[\bar{q}(k_i)-\bar{b}(k_i)][\bar{q}(k_j)-\bar{b}(k_j)][C^{-1}(\bar{q})]_{ij} \,,
\ee
where $C(\bar{q})$ is the covariance of $\bar{q}$ estimated from 100 realizations.
We set $k_{\rm max}=0.05~h~{\rm Mpc}^{-1}$, and confirm that the conclusion is
insensitive to the choice of fitting range.

\begin{figure*}[t]
\centering
\includegraphics[width=0.497\textwidth]{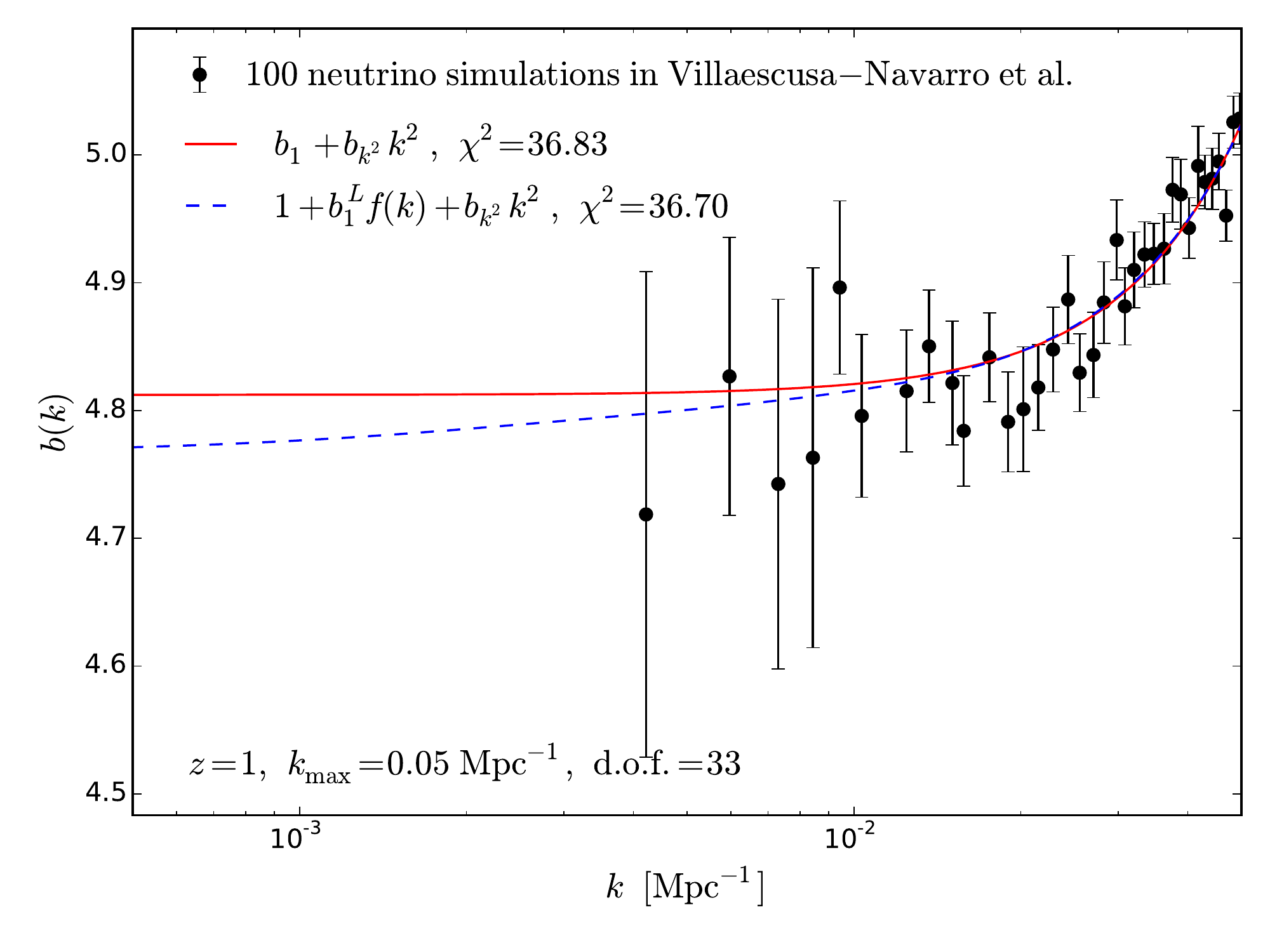}
\includegraphics[width=0.497\textwidth]{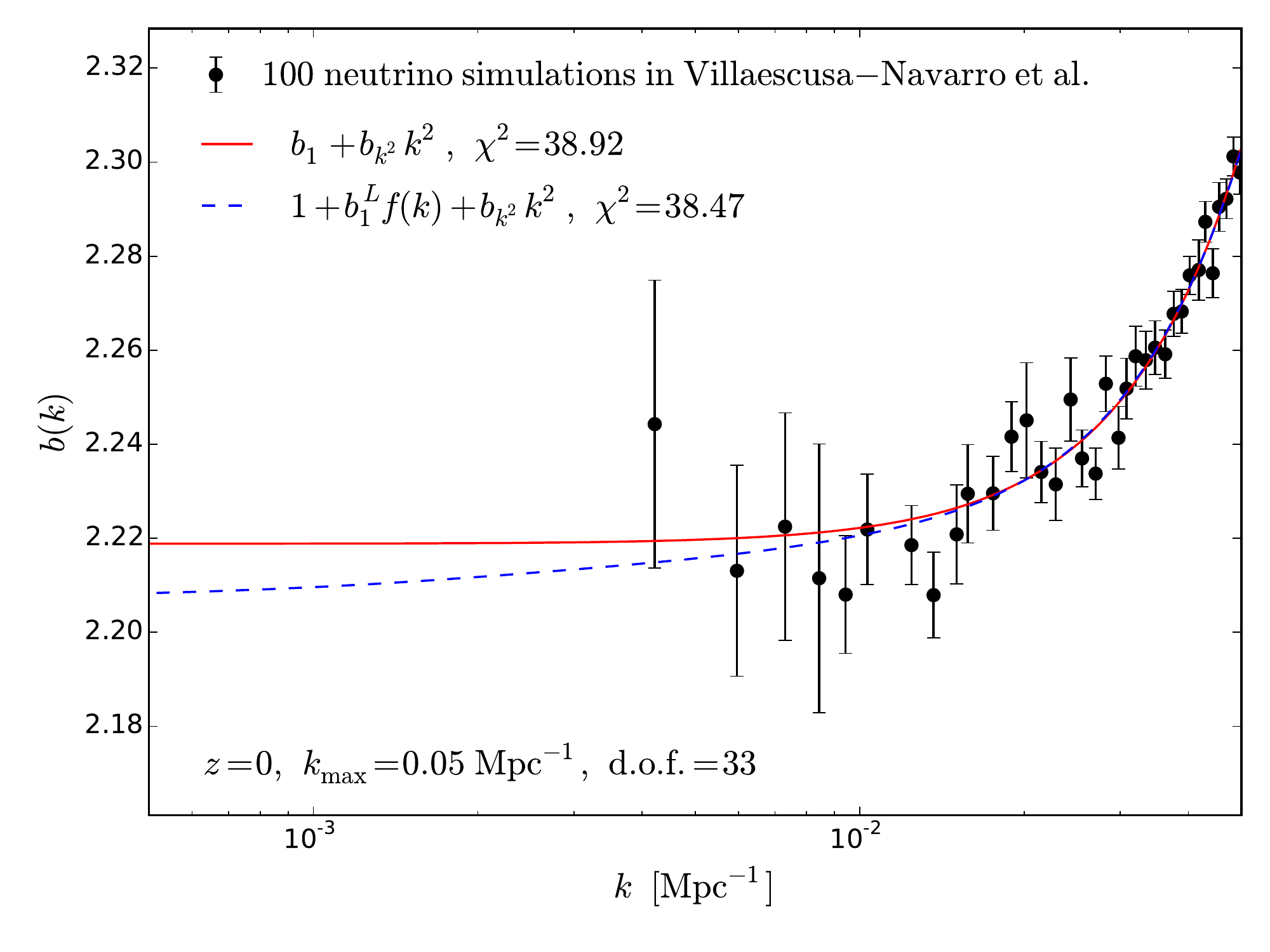}
\caption{Comparison of the clustering bias measured from $N$-body simulations
with massive neutrino particles and the two bias models at $z=1$ (left)
and 0 (right). The data points show the mean of simulations with error
bar showing the error on the mean, whereas the red solid and blue dashed
lines show the best-fit scale-independent and scale-dependent linear bias
models with $k_{\rm max}=0.05~{\rm Mpc}^{-1}$. The corresponding $\chi^2$
values are shown in the legend, with d.o.f.=33.}
\label{fig:bclustering}
\end{figure*}

\refFig{bclustering} shows the results at $z=1$ (left) and 0 (right). The data point
show the mean of the measurement from 100 realizations with the error on the mean,
and the red solid and blue dashed lines show the two models correspondingly. There
are 35 data points with two fitting parameters, so the number of degrees of freedom (d.o.f.) is 33. The $\chi^2$ values
from fitting the mean of the clustering bias to the models are shown in the legend.
First we find that the reduced $\chi^2$ is close to unity, indicating that the models
describe simulations results well. More importantly, we find that the $\chi^2$ values
between the two models are close, implying that the simulations do not have enough
statistical power to distinguish one from another. Namely, both scale-independent
and scale-dependent linear bias models are equally consistent with the simulations.

There are two main reasons that the simulations cannot distinguish the two models.
First, the scale dependence we predict is on large scale, and even the fundamental
mode of the simulation box of $1~h^{-1}~{\rm Gpc}$ cannot probe the full effect.
This can clearly be seen in \reffig{bclustering}: at $k\gtrsim0.01~{\rm Mpc}^{-1}$
the two models are almost identical. Note that the large-scale difference between
the two bias models will approach to a constant because the response becomes scale
independent when $k\ll k_{\rm fs}$. Therefore, to probe this effect we need simulations
with box size of a few Gpc. Second, people usually increase $f_\nu$ by increasing
the neutrino mass instead of number. For larger neutrino mass, the free-streaming
length approach to nonlinear scale, so it is challenging to separate the scale-dependent
linear bias and nonlinear bias. Moreover, the neutrino clustering becomes important
for larger neutrino mass, hence the approximation breaks down and our prediction
is invalid. Therefore, an ideal set of simulations to detect this effect is to have
a few Gpc box size and $0.05\eV$ of massive neutrinos mass with $f_\nu=5-10\%$. This study
is beyond the scope of this paper and is work in progress.

\bibliography{fsunu}

\end{document}